\documentclass[aps, prb, reprint, amsmath, amssymb, groupedaddress]{revtex4-2}
\usepackage[colorlinks=true, allcolors=blue]{hyperref}
\usepackage{graphicx}
\usepackage{overpic}
\usepackage{mathrsfs}
\usepackage{bm}
\newcommand{\lt}{\left}
\newcommand{\rt}{\right}
\newcommand{\pa}{\partial}

\newcommand{\bS}{\mathbf{S}}
\newcommand{\bb}{\mathbf{b}}

\newcommand{\bF}{\mathbf{F}}

\newcommand{\bnu}{\bm{\nu}}
\newcommand{\bzeta}{\bm{\zeta}}
\newcommand{\btheta}{\bm{\theta}}
\newcommand{\bN}{\mathbf{N}}

\newcommand{\bE}{\mathbf{E}}
\newcommand{\bu}{\mathbf{u}}
\newcommand{\bk}{\mathbf{k}}
\newcommand{\bx}{\mathbf{x}}
\newcommand{\bX}{\mathbf{X}}

\newcommand{\ve}{\varepsilon}
\newcommand{\bbeta}{\bm{\eta}}
\newcommand{\bdelta}{\bm{\zeta}}
\newcommand{\uu}[1]{\underline{\underline{#1}}}

\makeatletter
\newsavebox{\@brx}
\newcommand{\llangle}[1][]{\savebox{\@brx}{\(\m@th{#1\langle}\)}%
  \mathopen{\copy\@brx\kern-0.5\wd\@brx\usebox{\@brx}}}
\newcommand{\rrangle}[1][]{\savebox{\@brx}{\(\m@th{#1\rangle}\)}%
  \mathclose{\copy\@brx\kern-0.5\wd\@brx\usebox{\@brx}}}
\makeatother

\begin{document} 
\title{Frustrated magnets in the limit of infinite dimensions: dynamics and disorder-free 
glass transition}
\author{Achille Mauri}
\altaffiliation{Present address: Institute of Physics, Ecole Polytechnique Fédérale de 
Lausanne (EPFL), CH-1015 Lausanne, Switzerland}
\email[Email: ]{achille.mauri@epfl.ch}

\author{Mikhail I. Katsnelson}
\email[Email: ]{m.katsnelson@science.ru.nl}
\affiliation{Institute for Molecules and Materials, Radboud University, Heijendaalseweg 
135, 6525AJ Nijmegen, The Netherlands}

\date{\today}

\begin{abstract}
We study the statistical mechanics and the equilibrium dynamics of a system of classical 
Heisenberg spins with frustrated interactions on a $d$-dimensional simple hypercubic 
lattice, in the limit of infinite dimensionality $d \to \infty$.
In the analysis we consider a class of models in which the matrix of exchange constants is 
a linear combination of powers of the adjacency matrix.
This choice leads to a special property: the Fourier transform of the exchange coupling 
$J(\bk)$ presents a $(d-1)$-dimensional surface of degenerate maxima in momentum space.
Using the cavity method, we find that the statistical mechanics of the system presents 
for $d \to \infty$ a paramagnetic solution which remains locally stable at all 
temperatures down to $T = 0$.
To investigate whether the system undergoes a glass transition we study its dynamical 
properties assuming a purely dissipative Langevin equation, and mapping the system to an 
effective single-spin problem subject to a colored Gaussian noise.
The conditions under which a glass transition occurs are discussed including the 
possibility of a local anisotropy and a simple type of anisotropic exchange.
The general results are applied explicitly to a simple model, equivalent to the isotropic 
Heisenberg antiferromagnet on the $d$-dimensional fcc lattice with first and second 
nearest-neighbour interactions tuned to the point $J_{1} = 2J_{2}$.
In this model, we find a dynamical glass transition at a temperature $T_{\rm g}$ 
separating a high-temperature liquid phase and a low temperature vitrified phase.
At the dynamical transition, the Edwards-Anderson order parameter presents a jump 
demonstrating a first-order phase transition.
\end{abstract}

\maketitle

\section{Introduction}
\label{s:introduction}

Many important phenomena in magnetism are controlled by frustration.
Striking examples are spin systems on triangular, kagome, pyrochlore, fcc, and other 
geometrically-frustrated lattices, which present an extremely rich and complex 
phenomenology~\cite{balents_nat_2010}.
In many of these systems, frustration forces every plaquette to present multiple 
degenerate minima.
In the lattice, when plaquettes share corners or edges, this leads often to striking 
collective behaviors.

At the same time, frustration is a key ingredient in systems displaying rugged energy 
landscapes, slow relaxation, and glassy dynamics.
Prototype examples are spin glasses~\cite{edwards_jpf_1975, mezard_spin-glass}, controlled 
by a random distribution of exchange couplings.
However, frustration has a central role also in one of the theories for the vitrification 
of liquids~\cite{kivelson_pa_1995, tarjus_jpcm_2005}, in Coulomb-frustrated 
models~\cite{emery_pc_1993, kivelson_pa_1995, nussinov_prl_1999, schmalian_prl_2000, 
schmalian_ijmpb_2001, westfahl_prb_2001, grousson_pre_2002, grousson_pre_2002b, 
westfahl_prb_2003, wu_prb_2004, tarjus_jpcm_2005} and in systems which present random 
patterns of stripes~\cite{principi_prl_2016}, relevant, for example, in Coulomb-frustrated 
charge separation~\cite{emery_pc_1993}, ferromagnetic thin films~\cite{principi_prb_2016}, 
and possibly even biology~\cite{wolf_pnas_2018}.
More precisely, we are talking about models in which frustration arises from the competition
between short-range and long-range interactions (the latter can be Coulomb, dipole-dipole, or can
have another nature).
Here in the paper we will use the term ``Coulomb-frustrated'' to refer to such models, 
just for brevity.
These models are closely connected to the phenomenon of avoided 
criticality~\cite{tarjus_jpcm_2005}.
In Coulomb-frustrated theories, complementary theoretical analyses~\cite{emery_pc_1993, 
kivelson_pa_1995, nussinov_prl_1999, schmalian_prl_2000, schmalian_ijmpb_2001, 
westfahl_prb_2001, grousson_pre_2002, grousson_pre_2002b, westfahl_prb_2003, wu_prb_2004, 
tarjus_jpcm_2005} and numerical simulations~\cite{grousson_pre_2002} have indicated the 
emergence of a glassy behavior even in absence of a quenched disorder.
Similar behavior was shown to happen in magnetic systems with dipole-dipole 
interactions~\cite{principi_prb_2016} and in Brazovskii-like models with a line of 
minima of exchange energy~\cite{brazovskii_jetp_1975, principi_prl_2016}.

A natural question is therefore: under which conditions frustration alone is sufficient
to generate a glassy state?
This question has been a subject of investigations for decades.
Many theoretical models have been analyzed, for example fully-frustrated spin
systems~\cite{villain_jpc_1977, derrida_jpf_1979, alexander_jpa_1980, yedidia_jpa_1990},
kinetically-constrained models~\cite{ritort_ap_2003}, and models of Josephson-junction
arrays~\cite{parisi_jpa_1994, chandra_prl_1995, chandra_prl_1996}.
In Ref.~\onlinecite{franz_epl_2001}, a spin model with three-spin interactions has been shown to
undergo a glass transition even in presence of a purely ferromagnetic coupling.
In this case, frustration is not present at the level of ground state, but a ``dynamical''
frustration present in finite-temperature configurations, was shown to be 
sufficient to produce a glass phase.

In the context of geometrically frustrated magnets, the possibility of a glass transition in absence
of disorder has attracted interest as an explanation to the behavior experimentally oberved in some
kagome and pyrochlore materials~\cite{chandra_jp_1993, gardner_rmp_2010, saha_prb_2021,
mitsumoto_prb_2023}.
We note, for example, that a glassy ``spin jam'' state has been predicted in a Heisenberg 
model on a lattice consisting of a triangular network of bipyramids, a structure realized 
in SrCr$_{9p}$Ga$_{12-9p}$O$_{19}$~\cite{klich_nc_2014}.
A key element of the theory of Ref.~\onlinecite{klich_nc_2014} is the prediction that, in this
structure, a quantum order-by-disorder effect can generate nonperiodic metastable states separated
by large barriers, and that the number of metastable states grows exponentially with the perimeter,
rather than exponentially with the area.

Recently, systems which combine a geometrically frustrated lattice with a long-range 
interaction have attracted attention.
In Ref.~\cite{rademaker_njp_2018}, it has been 
shown by numerical methods that an electron system with Coulomb repulsion on the 
triangular lattice undergoes an ``electron glass'' transition at quarter filling.
The mechanism proposed for the emergence of glassiness is that the long-range force lifts 
the degeneracy between the many classical ground states on the triangular lattice, 
introducing large energy barriers and a rugged landscape.
The disorder-free glass transition detected theoretically has been proposed as an 
explanation for the slow dynamics observed in an organic 
conductor~\cite{rademaker_njp_2018}.

Another theory analyzed recently which combines similarly a frustrated geometry and a 
long-range coupling is the Ising model with dipolar interaction on the kagome net. 
For this model, Ref.~\cite{hamp_prb_2018} found numerical evidence of a glass transition 
and a slow relaxation consistent with the Vogel-Fulcher law.
The model has been analyzed in Ref.~\cite{cugliandolo_prb_2020} by an analytical method, 
based on the Bethe-Peierls approximation.
The results confirmed the existence of a glass transition, although the second-order 
transition found in Ref.~\cite{cugliandolo_prb_2020} indicates a divergence of the 
relaxation time different from the Vogel-Fulcher law, and similar to that of disordered 
spin glasses.

On the experimental side, two recent works~\cite{kamber_science_2020, verlhac_np_2022} 
indicated that a disorder-free glass transition can occur even in an elemental solid: the 
rare earth neodymium (Nd).
In particular, Refs.~\cite{kamber_science_2020, verlhac_np_2022} reported evidence of 
ageing, ultraslow dynamics, and complex amorphous structures in nanoscale magnetization 
patterns imaged by spin-polarized scanning tunneling microscopy on the surface of Nd.
It was shown that the randomness of the observed patterns becomes enhanced, and not 
reduced, when the defect concentration is lowered. 
This points to a theoretical explanation in terms of a ``self-induced spin 
glass'' (``self-induced'' means an assumption that frustrations alone are sufficient for 
the spin-glass behavior in deterministic systems)~\cite{principi_prb_2016, 
principi_prl_2016}.
This scenario has been corroborated by the observation of slow relaxation in spin 
dynamics simulations, performed using exchange constants calculated from first principles.

The results found recently motivate us to investigate an exactly solvable spin model 
exhibiting a disorder-free glass transition.
In particular, we revisit a model studied by Lopatin and Ioffe~\cite{lopatin_prb_2002}, 
which has as its key ingredient a frustrated antiferromagnetic interaction on a 
large-dimensional hypercubic lattice.
In the work of Ref.~\cite{lopatin_prb_2002}, the model was introduced as a lattice 
theory for the glass transition of supercooled liquids, and presented as fundamental 
degrees of freedom a set of Ising-like binary variables representing particles ($\rho_{i} 
= 1$) and holes ($\rho_{i} = 0$).
The interaction was constructed assuming a matrix of exchange couplings of the form 
$J_{ij} = [f(\hat{t}/\sqrt{2d})]_{ij}$ with $f(x) = -u  x^{2}$ and $\hat{t}$ the 
adjacency matrix (a matrix $t_{ij}$ such that $t_{ij}=1$ if $i$ and $j$ are nearest 
neighbours and $t_{ij} = 0$ otherwise).
Ref.~\cite{lopatin_prb_2002} solved the problem for an arbitrary chemical 
potential using a replica method.
Despite the simplicitly of the model, it was shown that the system undergoes a dynamical 
and a static glass transition. 

In this work, we study a similar model in the case in which the degrees of freedom are 
classical spin vectors $\bS_{i}$.
The continuous nature of the degrees of freedom allows us to study the possibility of a 
vitrification transition from a dynamical analysis, based on a continuous Langevin 
equation.
In the main part of the work, we derive a solution of the model, exact in the limit $d \to 
\infty$, including the possibility of anisotropy.
Eventually, we apply the results to a special case: an isotropic model with $f(x) = 
J(x^{2} - 1)$, $J < 0$.
After separation of the hypercubic lattice into sublattices, this model is equivalent to 
two decoupled Heisenberg models on the fcc structure. 
In particular, the model has two nonzero couplings: a nearest neighbour interaction of 
strength $J_{1} = J/d$ and a second-nearest-neighbour interaction of magnitude $J_{2} = 
J/(2d)$.
For this model we identify a dynamical first-order glass transition, signaled by a 
jump of the Edwards-Anderson order parameter.
The transition occurs even in absence of quenched disorder: the glass phase is
self-induced.

The class of interactions $J_{ij} = [f(\hat{t}/\sqrt{2d})]_{ij}$ has a special property, 
which has not been emphasized in Ref.~\cite{lopatin_prb_2002}: for this special form 
of interaction, the Fourier transform $J(\bk)$ of the exchange coupling develops a 
$(d-1)$-dimensional surface of degenerate maxima in momentum space~\cite{balla_prb_2019, 
balla_prr_2020}.

Similar degenerate surfaces appear in the Brazovskii model and in Coulomb-frustrated 
theories~\cite{emery_pc_1993, kivelson_pa_1995, nussinov_prl_1999, schmalian_prl_2000, 
schmalian_ijmpb_2001, westfahl_prb_2001, grousson_pre_2002, grousson_pre_2002b, 
westfahl_prb_2003, wu_prb_2004, tarjus_jpcm_2005, principi_prb_2016, principi_prl_2016}.
In Coulomb-frustrated models, the emergence of a glass phase has been investigated  
intensively using broken replica symmetry, with the specific tools borrowed from the 
theory of conventional, that is, disordered, spin glasses~\cite{monasson_prl_1995}.
In particular, replica analyses based on the self-consistent screening approximation 
(SCSA)~\cite{schmalian_prl_2000} and on local mean-field 
approximations~\cite{westfahl_prb_2001, wu_prb_2004} predicted a glassy behavior of the 
random first-order type.
The presence of a glassy dynamical arrest has been corroborated by numerical simulations, 
and by an analytical study based on a mode-coupling theory and a dynamic 
SCSA~\cite{grousson_pre_2002, grousson_pre_2002b}.
In dipole-frustrated~\cite{principi_prb_2016} and for Brazovskii-like 
models~\cite{wu_prb_2004, principi_prl_2016} analogue glass transitions have been 
predicted within analyes by the replica method.

In Coulomb-frustrated models the surface of maxima occurs usually in the region of small 
momenta, at characteristic wavelengths much larger than the atomic spacing.
In addition, the interactions have an infinite range.
These factors introduce differences with respect to the model studied here, where the 
interactions are short-ranged and the degenerate surface lies in a region of large 
wavevectors, comparable with the size of the Brillouin zone.
Despite these differences, there remain similarities.
This suggests that the exactly solvable model which we study exemplifies some of the features of
the ``stripe glass'' behavior, at a dynamical mean-field level.

The methology used in our work allows to study vitrification in a physically transparent
approach, based on explicit consideration of spin dynamics at large time scale, in
analogy with the pioneering work by Edwards and Anderson on spin
glasses~\cite{edwards_jpf_1975} and to the mode coupling approach by G\"{o}tze in the
theory of structural glasses~\cite{gotze_mct}.

The range of applications of the model is even broader.
The theory analyzed here allows to investigate, in infinite dimensions, the model of a 
frustrated magnet with a degenerate surface of helical states.
Spin systems with degenerate manifolds of spiral ground states space have attracted an 
extensive interest and have been predicted to host spiral spin liquid phases (see 
Refs.~\cite{bergman_np_2007, balla_prb_2019, balla_prr_2020, yao_fp_2021,graham_prl_2023} 
for theoretical analyses and for experimental evidences of spiral spin liquid states in 
cubic spinels).

The article is organized as follows.
In Sec.~\ref{s:model}, we introduce the model analyzed in this work.
In Sec.~\ref{s:statics}, we derive an exact solution of the equilibrium statistical 
mechanics of the system in the limit $d \to \infty$.
The solution shows that renormalizations stabilize the paramagnetic solution at all 
temperatures down to $T = 0$, a situation analogue to the ``avoided 
criticality''~\cite{nussinov_prl_1999}.
In Sec.~\ref{s:dynamics}, we study the dynamics of the system, assuming a purely 
dissipative Langevin equation.
In the large-$d$ limit, we show that the problem maps to an effective single-site 
Langevin equation with a colored noise, self-consistently determined by a set of 
consistency relations.
The conditions under which the dynamics develops a glass transition are discussed in 
Sec.~\ref{s:glass_transition}.
The general results are eventually applied in the case of a isotropic Heisenberg model, 
and for a simple interaction $J_{ij} = [f(\hat{t}/\sqrt{2d})]_{ij}$ with $f(x) = 
J(x^{2}-1)$.
The results for this model are presented in Sec.~\ref{s:fcc_model}.
We conclude the article with a brief summary in Sec.~\ref{s:conclusions}.

\subsection{Notations}

Throughout the paper we will use the following notations.
The symbol $\int_{\bS_{i}}$ stands for an integral over all values of the spin $\bS_{i}$
(in spherical coordinates $\int_{\bS_{i}} = \int_{0}^{\pi} \sin \theta_{i}{\rm
d}\theta_{i} \int_{0}^{2 \pi} {\rm d}\varphi_{i}$).
$\delta_{S}(\bS - \bS')$ is a delta function in spin space, defined according to the
invariant measure (in spherical coordinates $\delta_{S}(\bS - \bS') = \delta(\theta -
\theta') \delta(\varphi - \varphi')/\sin \theta$).
$3 \times 3$ matrices carrying spin indices such as $f^{\alpha \beta}(x)$ or $\chi^{\alpha \beta}$
are written in matrix notation with a symbol $\uu{f}(x)$, $\uu{\chi}$, etc.
Matrices carrying lattice indices, such as $t_{ij}$ and $J_{ij}^{\alpha \beta}$, are 
denoted as $\hat{t}$, $\hat{J}$.
Matrix multiplications and inverses are defined assuming contraction of all internal spin 
and lattice indices ($[\hat{A} \hat{B}]^{\alpha \beta}_{ij} = \sum_{k} A^{\alpha 
\gamma}_{ik} B^{\gamma \beta}_{kj}$, $[\hat{A}^{-1}\hat{A}]_{ij}^{\alpha \beta} = \sum_{k}
A^{-1}{}^{\alpha \gamma}_{ik} A^{\gamma \beta}_{kj} = [\hat{1}]^{\alpha \beta}_{ij}
=\delta^{\alpha \beta} \delta_{ij}$ for matrices with both spin and lattice indices;
$[A B]_{ij}= \sum_{k} A_{ik} B_{kj}$ for matrices in site space only.)
We assume everywhere summation over repeated spin (greek) indices: $A^{\alpha \gamma} 
B^{\gamma \beta}=\sum_{\gamma}A^{\alpha \gamma}B^{\gamma \beta}$.

\section{Model}
\label{s:model}

Although we eventually apply the results explicitly to a particular isotropic model, in 
the main part of the work we keep the discussion more general, and analyze spin systems 
subject to a (possibly anisotropic) exchange interaction $J_{ij}^{\alpha \beta}$ and an 
on-site anisotropy $V(\bS_{i})$.
We thus assume a Hamiltonian:
\begin{equation}\label{H}
H = -\frac{1}{2} \sum_{i, j} J^{\alpha \beta}_{ij} S^{\alpha}_{i} S^{\beta}_{j} + \sum_{i} 
V(\bS_{i})~.
\end{equation}

The degrees of freedom $\bS_{i}$ are classical spins: three-dimensional vectors with 
cartesian coordinates $S^{\alpha}_{i}$, $\alpha = x, y, z$ and with the constraint of unit 
length $\bS^{2}_{i} = S^{\alpha}_{i} S^{\alpha}_{i} = 1$.
The labels $i$, $j$ run over the sites of a hypercubic lattice in $d$ dimensions.
In the analysis we take eventually the limit of large dimensionality $d \to \infty$.
The $d \to \infty$ approximation~\cite{metzner_vollhardt} has demonstrated its power for 
the consideration of lattice fermionic problems where it is the base of dynamical 
mean-field theory (DMFT), for reviews see Refs.~\cite{kotliar_rmp_2006, held_ap_2007, 
katsnelson_rmp_2008}.

The anisotropy energy $V(\bS_{i})$ in the model may be arbitrary.
We make the only assumption that it is even ($V(\bS_{i}) = V(- \bS_{i})$) in such way that $H$ is
invariant under time-reversal, that is, spin-reversal.
For the exchange interaction we consider instead a coupling of a special form: we assume that
$J_{ij}^{\alpha\beta} = [f^{\alpha \beta}(\hat{t}/\sqrt{2d})]_{ij}$, with $f^{\alpha \beta}(x)$ a $3
\times 3$ matrix of functions.
This definition is a generalization of the interaction $J_{ij} = [f(\hat{t}/\sqrt{2d})]_{ij}$,
which was introduced and analyzed in the case of Ising variables by Lopatin and
Ioffe~\cite{lopatin_prb_2002}.
In the definition, $\hat{t}$ is the adjacency matrix and powers of $\hat{t}$ are intended in the
sense of matrix multiplication.
For example, the function $f^{\alpha \beta}(x) = J_{0}^{\alpha \beta} + J_{1}^{\alpha 
\beta} x + J_{2}^{\alpha \beta} x^{2} + J_{4}^{\alpha \beta}x^{4}$ corresponds to the coupling
\begin{equation} \label{J_example}
\begin{split}
J_{ij}^{\alpha \beta} & = J_{0}^{\alpha \beta} \delta_{ij} + \frac{J_{1}^{\alpha 
\beta}}{\sqrt{2d}} t_{ij} + \frac{J_{2}^{\alpha \beta}}{2d} \sum_{k} t_{ik} t_{kj} \\
& + \frac{J_{4}^{\alpha \beta}}{4d^{2}}\sum_{k, l, m}t_{ik} t_{kl} t_{lm} t_{mj}~,
\end{split}
\end{equation}
an interaction which is nonzero between sites $i j$ having a Manhattan distance $\ell_{ij}
=\sum_{a=1}^{d} |i_{a} - j_{a}|$ at most equal to 4.
In general, the exchange coupling is constructed from linear combinations of the matrix elements
$[\hat{t}^{n}]_{ij} = \sum_{k_{1}, .., k_{n-1}} t_{ik_{1}} t_{k_{1} k_{2}} .. t_{k_{n-2}k_{n-1}}
t_{k_{n-1} j}$.
These matrix elements are equal to the number of different paths which start at $i$, end 
at $j$, and are composed by a sequence of $n$ steps along the bonds of the hypercubic 
lattice.
If $f^{\alpha \beta}(x)$ is a polynomial in $x$ the corresponding interaction is short-ranged and
the degree of the polynomial is equal to the range (measured according to the Manhattan distance).

For convenience, we assume that the on-site exchange vanishes: $J_{ii}^{\alpha \beta} =
0$.
This is convenient for the subsequent mean-field calculations and does not imply a loss of
generality, because any local interaction can be absorbed in a redefinition of $V(\bS_{i})$.
Imposing the condition $J_{ii}^{\alpha \beta} = 0$ requires to tune the constant term 
$J_{0}^{\alpha \beta}$ as a function of $J_{2}^{\alpha \beta}$, $J_{4}^{\alpha \beta}$, 
.. in such way as to cancel the contributions of the on-site matrix elements 
$[\hat{t}^{2n}/(2d)^{n}]_{ii}$~\cite{Note1}.

When the model is analyzed in an arbitrary dimension $d$ and eventually in the limit $d \to
\infty$, it is essential that the coupling constants are scaled in such way to ensure a limiting
large-$d$ model which is well-defined and non-trivial.
By taking $J_{ij}^{\alpha \beta} = [f^{\alpha \beta}(\hat{t}/\sqrt{2d})]_{ij}$ we have assumed,
following Ref.~\onlinecite{lopatin_prb_2002}, that the relevant scaling consists in assigning a
factor of order O$(d^{-1/2})$ to every power of the adjacency matrix.
The analyses in the next sections show indeed that the limit $d \to \infty$ at fixed 
$f^{\alpha \beta}(x)$ is well-defined.

The consistency of the limit, however, does not occur for arbitrary $f^{\alpha \beta}(x)$ but
requires that the function $f^{\alpha \beta}(x)$ satisfies certain conditions, which 
qualitatively correspond to the presence of a strong frustration.
In fact the scaling $\approx d^{-1/2}$ is analogue to the scaling needed in mean-field models of
spin glasses~\cite{mezard_spin-glass} and in other strongly-frustrated 
infinite-dimensional models~\cite{lopatin_prb_2002, yedidia_jpa_1990}.
If $f^{\alpha \beta}(x)$ described a non-frustrated interaction, instead, a different 
scaling would be required.
For example, in the case of a nearest-neighbour interaction, the exchange coupling would have to be
scaled as $1/d$, and not as $d^{-1/2}$, to ensure a finite $d \to \infty$
limit~\cite{yedidia_jpa_1990, georges_rmp_1996}.

Before discussing the conditions which $f^{\alpha \beta}(x)$ must satisfy in order to generate a
model consistent with the $d^{-1/2}$ scaling, it is useful to describe
some essential properties of the exchange coupling $J_{ij}^{\alpha \beta}$ which follow from the
construction $J_{ij}^{\alpha \beta} = [f^{\alpha \beta}(\hat{t}/\sqrt{2d})]_{ij}$.
As shown by a direct calculation, the Fourier transform of $J_{ij}^{\alpha \beta}$ is
\begin{equation}\label{J_transform}
J^{\alpha \beta}(\bk) = \sum_{i} {\rm e}^{-i \bk \cdot (\bx_{i} - \bx_{j})} J^{\alpha 
\beta}_{ij} = f^{\alpha \beta}(\ve_{\bk})~,
\end{equation}
with $\ve_{\bk} = \sqrt{2/d} \sum_{a=1}^{d} \cos(k_{a})$.
Thus, $J^{\alpha \beta}(\bk)$ depends on the wavevector $\bk$ only through $\ve_{\bk}$.
This implies that, for many choices of the function $f^{\alpha \beta}(x)$ the interaction can
develop degenerate surfaces of maxima in momentum space.
More precisely, consider the case in which the exchange is isotropic, $f^{\alpha \beta}(x) =
\delta^{\alpha \beta}f(x)$.
In dimension $d$, the variable $\ve_{\bk}$ is confined to the interval $-\sqrt{2d} \leq \ve_{\bk}
\leq \sqrt{2d}$.
As a result two situations can occur, depending on the form of the function $f(x)$.
If $f(x)$ reaches its maximum at one of the boundaries $\pm \sqrt{2d}$, the maxima of $J(\bk)$
occur at isolated points in momentum space (the center of the Brillouin zone $\Gamma = (k_{1}, ..,
k_{d}) = (0, .., 0)$, corresponding to $\ve_{\Gamma} = \sqrt{2d}$, or the zone corner $R = (k_{1},
.., k_{d}) = (\pi, .., \pi)$ with $\ve_{R} = - \sqrt{2d}$).
However, if $f(x)$, as a function of $x$, has maxima within the interval $-\sqrt{2d} < x <
\sqrt{2d}$ then $J(\bk)$ reaches its maximum value on a surface in momentum space (the surface
defined by the relation $\ve_{\bk} = \ve_{\rm max}$, with $\ve_{\rm max}$ the point, or points, at
which $f$ reaches its maximum.
This defines a $(d-1)$-dimensional manifold in reciprocal space).
In the anisotropic case, the same analysis remains valid since the Fourier transform has
constant values on surfaces in momentum space.
For example, the maximum eigenvalue of $J^{\alpha \beta}(\bk)$ occurs at a $(d-1)$ 
manifold in $\bk$-space, unless it occurs at the $\Gamma$ or $R$ points.

We now return to the discussion of the conditions under which the model admits a well-defined limit
for $d \to \infty$.
For large-dimensionality the interval $[-\sqrt{2d}, \sqrt{2d}]$ on which $\ve$ is defined becomes
unbounded and $\ve_{\bk}$ can assume arbitrary real values.
If the eigenvalues of $f^{\alpha \beta}(x)$ have no upper bound on the real axis $-\infty < x
<\infty$, the Fourier transform $J^{\alpha \beta}(\bk)$ can be made arbitrarily large by choosing
$\bk$ near the zone center $\Gamma$ or the zone corner $R$.
This then implies a ferromagnetic or antiferromagnetic instability at a critical temperature which
diverges for $d \to \infty$ (see Sec.~\ref{s:Weiss_MFT}).
The large-$d$ limit, in this case, is inconsistent with the assumed $d^{-1/2}$ scaling.
If instead, the eigenvalues of $f^{\alpha \beta}(x)$ are bounded from above, the maximum eigenvalue
of $J(\bk)$ occurs on a $(d-1)$-dimensional surface in momentum space and remains finite when $d
\to \infty$.
This is the case which will be considered throughout the rest of the paper, and which 
leads at large $d$ to a well-defined limit.
Throughout all the rest of this work, we assume therefore that the boundedness condition 
is satisfied: the eigenvalues of $f^{\alpha \beta}(x)$ are assumed to be bounded from 
above on the real axis.

\begin{figure}[ht]
\includegraphics[scale=0.9]{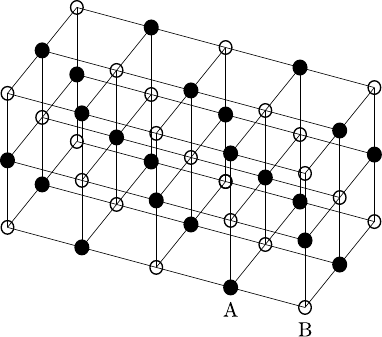}
\caption{\label{fig1} Partitioning of the cubic lattice into two interpenetrating fcc 
sublattices (A and B) for $d = 3$.
The sublattices A and B can be identified as the sets of lattice points $(x_{1}, ...., 
x_{d})$ for which $\sum_{a=1}^{d} x_{a}$ is respectively even and odd.
The separation of even and odd sites can be applied in arbitrary dimension $d$ to separate 
the hypercubic lattice into fcc sublattices~\cite{ulmke_epjb_1998}.}
\end{figure}

The boundedness condition on $f$ has a simple interpretation in terms of the exchange 
couplings in real space.
Suppose that the interaction has a range $R$, and, thus, that $f^{\alpha \beta}(\ve) =
\sum_{m=0}^{R} J_{m}^{\alpha \beta} \ve^{m}$ is a polynomial of degree $R$.
If the maximum degree $R$ is odd, it is clear that the eigenvalues cannot be bounded: depending on
the signs of $J_{R}^{\alpha \beta}$, the system presents ferromagnetic or antiferromagnetic
instabilities.
Thus $R$ must be even.
In addition, since the interaction is required to have an upper bound, the interaction of 
maximum range $J^{\alpha \beta}_{R}$ must be negative (antiferromagnetic)~\cite{Note2}.
To interpret this condition it is useful to note that the hypercubic lattice in $d$ 
dimensions is bipartite and can be partitioned into two sublattices: the sublattice A 
composed of the sites such that the sum $\sum_{a=1}^{d}x_{a}$ of the coordinates $(x_{1}, 
...., x_{d})$ is an even integer and the complementary sublattice B where the sum 
$\sum_{a=1}^{d} x_{a}$ is an odd integer (see Fig.~\ref{fig1}).
The sublattices A and B have the structure of a generalized fcc lattice in $d$ 
dimensions~\cite{ulmke_epjb_1998}.
After this partitioning, it is clear that interactions with even range couple the spins 
within the same sublattice ($A-A$ or $B-B$), while interactions with an odd range induce a 
coupling between different sublattices ($A-B$ and $B-A$).

Thus, the conditions for a well defined $d \to \infty$ limit is that that the coupling of longest
range is antiferromagnetic and takes place within the fcc sublattices, and not across the 
two different sublattices.
This shows that the well-known frustration of the fcc structure~\cite{alexander_jpa_1980,
ulmke_epjb_1998, balla_prb_2019, balla_prr_2020} plays a role in the model studied here.
We note however, that in the family of interactions which we study the frustration is not 
purely geometrical, because all models which we analyze have nonzero couplings beyond the 
nearest-neighbours.
For example, in Sec.~\ref{s:fcc_model} we study the isotropic model with $V(\bS) = 0$, 
$f^{\alpha \beta}(x) = J \delta^{\alpha \beta}(x^{2}-1)$, $J < 0$.
In this model, since $f$ is even in $x$, the interactions take place entirely within the 
fcc sublattices, and the inter-sublattice coupling vanishes.
Thus the theory describes effectively to two independent copies of a model on the fcc 
lattice.
The model, however, has two nonzero couplings on  the fcc lattice: a nearest-neighbour 
(NN) coupling of magnitude $J/d$ and a second-NN interaction of magnitude $J/(2d)$.
For this reason, the resulting problem differs significantly from the NN fcc model 
studied in Ref.~\onlinecite{alexander_jpa_1980}.
The model with interaction $f(x) = J (x^{2}-1)$, $J < 0$, rather, provides a direct 
analogue in infinite-dimensions, of the three-dimensional Heisenberg fcc antiferromagnet 
with first- and second-nearest-neighbour interactions tuned to $J_{1} = 2 J_{2}$.
This model has been analyzed in Refs.~\onlinecite{balla_prb_2019, balla_prr_2020} as a 
candidate for hosting spiral-spin-liquid behavior~\cite{bergman_np_2007, yao_fp_2021, 
graham_prl_2023}.

In general, all models analyzed here have the property of displaying surfaces of 
degeneracy in the Fourier transform of $J(\bk)$.
Thus, the models which we discuss can be viewed as a large-dimensional mean-field 
theories of spiral-spin-liquid models.
We note, however, that the special form $J_{ij} = [f(\hat{t}/\sqrt{2d})]_{ij}$ of the 
coupling which analyze is not the most general which is known to produce degenerate 
surfaces in momentum space.
Models with an interaction constructed via powers of the adjacency matrix have been 
analyzed in the context of spiral spin liquids~\cite{yao_fp_2021} but different theories, 
in which $\hat{J}$ is not reducible to a function of $\hat{t}$ have been
studied~\cite{balla_prb_2019, bergman_np_2007}.

\subsection{Analysis within the Weiss mean-field approximation}
\label{s:Weiss_MFT}

Although the exact large-$d$ results presented in the next sections lead to results which 
differ qualitatively from those of Weiss mean-field theory (MFT), a first understanding 
of the model, and in particular of the role of frustration for $d \to \infty$, can be 
obtained by analysis at the level of the MFT approximation.
To illustrate this with an example, we consider first the case of the isotropic model 
($V(\bS) = 0$) with $f^{\alpha \beta}(x) = \delta^{\alpha \beta} f(x)$, $f(x) = J(x^{2} - 
1)$ (this model will be addressed in more detail in Sec.~\ref{s:fcc_model} using the 
complete large-$d$ results).
As mentioned above, the interaction $f(x) = J(x^{2}-1)$ leads to two decoupled Heisenberg models on
two independent fcc sublattices.
A given site $i = (x_{1}, x_{2}, .., x_{d})$ is coupled with an exchange coupling of 
strength $J_{1} = J/d$ with the $2d (d-1)$ sites $j = (x_{1}\pm 1, x_{2} \pm 1, x_{3}, .. x_{d})$,
$j = (x_{1} \pm 1, x_{2}, x_{3} \pm 1, ...x_{d})$ , $j = (x_{1}, x_{2} \pm 1, x_{3} \pm 1, x_{4},
.., x_{d})$, which are NN on the fcc lattice, and with an exchange of magnitude $J/2d = J_{1}/2$ to
the $2d$ sites $j = (x_{1} \pm 2, x_{2}, ..,
x_{d})$, $(x_{1}, x_{2} \pm 2, x_{3}, ...)$, ..., which belong to the second-NN shell on 
the fcc structure.
The $-1$ in the definition $f(x) = J(x^{2} - 1)$ is needed to ensure that $J_{ii} = 0$: 
it is required to cancel the on-site matrix element $[\hat{t}^{2}/(2d)]_{ii} = 1$.
Within MFT, the susceptibility in the paramagnetic phase is given by
\begin{equation}\label{MFT_susceptibility}
\chi^{-1\alpha \beta}(\bk) = \chi_{0}^{-1\alpha \beta} - J^{\alpha \beta}(\bk)~,
\end{equation}
where $\chi_{0}^{\alpha \beta} = \delta^{\alpha \beta}/(3 k_{\rm B}T)$ is the 
susceptibility of a single spin (in the isotropic case).
The MFT predicts as a result a second-order transition at a critical temperature $T_{\rm 
MFT} = J_{\rm 
max}/(3 k_{\rm B})$ where $J_{\rm max}$ is the maximum value of $J(\bk)$.
By Eq.~\eqref{J_transform}, $J_{\rm max}$ simply equal to the maximum $f_{\rm max}$ of the 
function $f(x)$.
If the coupling is ferromagnetic ($J >0$), $f(\ve)$ is not bounded from above.
In this case the interaction is not frustrated and the critical temperature $T_{\rm MFT}$ diverges
for $d \to \infty$.
If instead $J< 0$, $J_{\rm max} = f_{\rm max} = |J|$ and the Weiss transition temperature $T_{\rm
MFT} = |J|/(3 k_{\rm B})$ remains finite.

The fact that $T_{\rm MFT}$ remains of O$(1)$ when $d \to \infty$ signals physically that the
interaction is very strongly frustrated.
In fact, the interaction in the model has a magnitude of order $J/d$ but has a coordination of $z
\approx d^{2}$ for $d$ large.
If the interactions were not strongly frustrated, we would expect a critical temperature which
diverges for $d \to \infty$ as $ z |J_{2}|d^{-1} \approx d^{2} \times d^{-1} \propto d$.
The fact that this divergence does not occur signifies that the local fields $\sum_{j} 
J_{ij}\bS_{j}$ which act on a site $i$ do not scale linearly with the dimension $d$ but 
only as their square root $\sqrt{d}$.
This behavior is analogue to that of fully connected spin-glasses~\cite{mezard_spin-glass} 
and of other mean-field frustrated models~\cite{lopatin_prb_2002, yedidia_jpa_1990}.
(See Ref.~\onlinecite{derrida_jpf_1979} for an analysis of the large-$d$ scaling of the 
local fields in the fully-frustrated hypercubic Ising model).

The same considerations apply for any $f(\ve)$ and in the anisotropic case.
If the eigenvalues of $f^{\alpha \beta}(\ve)$ are not bounded from above, the large-$d$ limit is
unstable and the MFT transition temperature diverges.
If instead the eigenvalues of $f^{\alpha \beta}(\ve)$ are bounded from above, the interaction is
strongly frustrated the finite $T_{\rm MFT}$ indicates a stable large-$d$ limit.
The finiteness of  $T_{\rm MFT}$ is again a sign of a very strong frustration.
In fact, the expansion of $J_{ij}^{\alpha \beta} = [f^{\alpha
\beta}(\hat{t}/\sqrt{2d})]_{ij}$ generates a coupling $J_{ij}^{\alpha \beta}$ of order
$d^{-\ell_{ij}/2}$ where $\ell_{ij}$ is the Manhattan distance between $i$ and $j$.
The suppression factor $d^{-\ell_{ij}/2}$ does not compensate the growth of the
coordination numbers $z_{\ell} \propto d^{\ell}$.
Thus, in absence of frustration, we would expect a divergence $T_{\rm MFT} \approx z_{R}
|J_{R}|d^{-R/2} \propto d^{R}\times d^{-R/2} \approx d^{R/2}$, where $R$ is the maximum
range of the interaction.

As a remark, we note that the MFT analysis above relied, although in an indirect way, on 
the scaling factors $d^{-1/2}$ assigned to each power of the adjacency matrix.
The role of the scaling factors have been to ensure that the condition $J_{ii}^{\alpha 
\beta} = 0$ can be imposed while keeping $f^{\alpha \beta}(x)$ fixed and finite as $d \to 
\infty$~\cite{Note1}.

\subsection{Ground states in the isotropic case}

In the isotropic case and when the function $f(x)$ is bounded from above, the interaction $J_{ij} =
[f(\hat{t}/\sqrt{2d})]_{ij}$ admits always helical ground states of the form $\bS_{i} = \mathbf{A}
\cos (\bk\cdot \bx_{i}) + \mathbf{B} \sin(\bk \cdot \bx_{i})$, with $\mathbf{A}^{2} = \mathbf{B}^{2}
= 1$, $(\mathbf{A}\cdot \mathbf{B}) = 0$ (these are exact ground states and can be derived using the
Luttinger-Tisza method)~\cite{nussinov_prl_1999, balla_prb_2019, balla_prr_2020, yao_fp_2021}.
The modulation vector $\bk$ is any wavevector belonging to the surface $\ve_{\bk} = 
\ve_{\rm max}$, with $\ve_{\rm max}$ a point at which $f(\ve)$ is maximal.
The exact ground-state energy per particle is therefore $E/N = - f(\ve_{\rm max})/2 = - 
f_{\rm max}/2$ in isotropic models.

For some interactions, the single helices are not the only type of ground states.
For example, if $f(x) = f(-x)$, so that the system breaks into two decoupled fcc 
sublattices, it is possible to construct ground states by taking independent helices, 
with arbitrary wavevectors $\bk_{A}$ and $\bk_{B}$ and arbitrary magnetization vectors 
$\mathbf{A}_{A}$, $\mathbf{B}_{A}$, $\mathbf{A}_{B}$, $\mathbf{B}_{B}$ on the two 
sublattices.

\section{Static properties}
\label{s:statics}

As in other mean-field frustrated models~\cite{mezard_spin-glass, lopatin_prb_2002}, the 
Weiss-mean field approximation is not exact for $d \to \infty$.
The exact large-$d$ solution is, rather, similar to the dynamical mean-field theory of 
correlated electron systems~\cite{georges_rmp_1996}, and has to include local self-energy 
effects.
In this section we derive an exact $d \to \infty$ solution of the statistical mechanical 
properties of the system, assuming a homogeneous state with unbroken translational and 
spin-inversion symmetries.
In contrast with the behavior predicted by the MFT approximation, the exact large-$d$ 
solution shows that the symmetric state remains locally stable at all temperatures down to 
$T = 0$.

To derive the large-$d$ solution, we use an approach based on the cavity method (in the 
context of DMFT for fermions, see Ref.~\onlinecite{georges_rmp_1996}).
In particular we use as a starting point, a methodology analogue to the cavity 
approach applied in Ref.~\cite{mezard_spin-glass} to the Sherrington-Kirkpatrick (SK) 
spin-glass model.
A key idea of this approach consists in analyzing the joint probability $P(\bS_{i}, 
\bb_{i})$ of the spin $\bS_{i}$ and the internal field $b^{\alpha}_{i} = \sum_{j} 
J^{\alpha \beta}_{ij} S^{\beta}_{j}$ acting at the same site $i$.
In equilibrium at temperature $T$~\cite{Note3}:
\begin{equation}\label{P_SB}
\begin{split}
P(\bS_{i}, \bb_{i})  & = Z^{-1} \int_{\bar{\bS}_{i_{1}}} .. 
\int_{\bar{\bS}_{i_{N}}} \big[ \delta_{S}(\bS_{i}- \bar{\bS}_{i})\delta(\bb_{i}- 
\bar{\bb}_{i})\\
& \times {\rm e}^{- \beta H(\bar{\bS}_{i_{1}}, .. \bar{\bS}_{i_{N}})}\big]~,
\end{split}
\end{equation}
where $Z$ is the partition function, $\beta=1/(k_{\rm B}T)$, $\bar{b}_{i}{}^{\alpha} = 
\sum_{j}J^{\alpha \beta}_{ij} \bar{S}_{j}{}^{\beta}$, and the integral is over all 
possible configurations of the $N$ spins.
$P(\bS_{i}, \bb_{i})$ can be interpreted, simply, as the probability that, extracting a 
random configuration at temperature $T$, the spin and the field at site $i$ have given 
values $\bS_{i}$ and $\bb_{i}$.

The reason why $P$ provides a particularly convenient framework for the analysis is that 
it describes efficiently a property of the large-$d$ limit: for $d \to \infty$ the field 
$\bb_{i}$ has strong correlations with the spin $\bS_{i}$ at the same site, but weaker
correlations with the spins $\bS_{j}$ at the other sites $j \neq i$.
Considering the joint distribution $P$ allows to isolate the strong correlation of 
$\bb_{i}$ with $\bS_{i}$, by factorizing:
\begin{equation}\label{1_cavity_factorization}
P(\bS_{i}, \bb_{i}) = Z'_{i} Z^{-1} {\rm e}^{\beta (\bb_{i} \cdot \bS_{i}) -\beta 
V(\bS_{i})} p'(\bb_{i})~.
\end{equation}

Here $Z'_{i}$ and $p'(\bb_{i})$ are, respectively, the partition function and the 
distribution of the field $\bb_{i}$ in a cavity system in which the site $i$ is removed 
(an $N-1$-body system containing all spins a part from $i$).
Explicitly, 
\begin{equation}
Z'_{i} = \int_{\bS_{i_{2}}} ... \int_{\bS_{i_{N}}} {\rm e}^{-\beta H'_{i}(\bS_{i_{1}}, 
..., \bS_{i_{N}})}~,
\end{equation}
and 
\begin{equation}\label{P_SB_p'}
\begin{split}
p'(\bb_{i}) & = Z'_{i}{}^{-1} \int_{\bS_{i_{2}}}... \int_{\bS_{i_{N}}} \big[ 
\delta(b^{\alpha}_{i}
- {\textstyle \sum}_{j}J^{\alpha \beta}_{ij} S^{\beta}_{j})\\
& \times {\rm e}^{-\beta H'_{i}(\bS_{i_{1}}, ..., \bS_{i_{N}})} \big]~,
\end{split}
\end{equation}
where
\begin{equation}
H'_{i} = -\frac{1}{2}\sum_{j\neq i, k \neq i} J^{\alpha \beta}_{jk} S^{\alpha}_{j} 
S^{\beta}_{k} + \sum_{j\neq i}V(\bS_{j})
\end{equation}
is the Hamiltonian of the cavity system.
The cavity distribution $p'(\bb_{i})$ describes the probability to find a given value of 
$b_{i}^{\alpha} = \sum_{j} J^{\alpha \beta}_{ij} S^{\beta}_{j}$ when extracting a random 
configuration $\bS_{i_{2}}$, .., $\bS_{i_{N}}$ in the cavity system, in which the effects 
of the interactions with $\bS_{i}$ are absent, and thus, is by construction uncorrelated 
to $\bS_{i}$.

Thanks to the fact that the correlation between $\bb_{i}$ and $\bS_{i}$ is subtracted in 
$p'(\bb_{i})$, the cavity distribution has simple statistical properties in large 
dimensions.
In the limit $d \to \infty$, as shown below, $p'(\bb_{i})$ is simply a Gaussian:
\begin{equation}\label{p_prime}
p'(\bb_{i}) = \frac{\exp \big[- L^{-1\alpha \beta} b^{\alpha}_{i} 
b^{\beta}_{i}/2\big]}{\sqrt{(2\pi)^{3} \det \uu{L}}}~.
\end{equation}

Here $\uu{L}$ is a $3 \times 3$ matrix defining the covariance of the cavity 
distribution and needs to be determined as a function of the temperature $T$.
Since we assumed a symmetric solution, with unbroken translation and spin-inversion
symmetries, the Gaussian distribution is centered and $\uu{L}$ does not depend 
on the lattice site $i$.

The result that the cavity distribution $p'(\bb_{i})$ is Gaussian can be derived using a
perturbative analysis which is analogue, from the diagrammatic point of view, to
DMFT~\cite{georges_rmp_1996}.
Let us discuss, in brief, the derivation.
The $\ell$-th cumulant $C'_{\ell}{}^{\alpha_{1} .. \alpha_{\ell}}$ of $p'(\bb_{i})$ is related to
the connected moments of the cavity spin distribution via $C_{\ell}'{}^{\alpha_{1} .. \alpha_{\ell}}
= \sum_{j_{1}, j_{2}, .. j_{\ell}} J_{ij_{1}}^{\alpha_{1} \beta_{1}} ... J_{i
j_{\ell}}^{\alpha_{\ell} \beta_{\ell}} \llangle S^{\beta_{2}}_{j_{2}} ...
S^{\beta_{\ell}}_{j_{\ell}} \rrangle'_{}{}^{(i)}$.
Here $\llangle ... \rrangle'_{}{}^{(i)}$ are connected averages computed in the cavity 
system (the symbol $\llangle .... \rrangle$ is used to denote connected averages, and the 
sign $\prime (i)$ reminds that the averaging is computed in a system with one cavity at 
site $i$).
The $C'_{\ell}$, in turn, can be computed by a perturbative expansion in powers of 
$J^{\alpha \beta}_{ij}$ (a high-temperature expansion with fixed $\beta V(\bS)$).
The corresponding diagrams are linked-cluster graphs (see for example Refs.~\cite{englert_pr_1963,
halvorsen_prb_2000} for discussions of the linked-cluster expansion in spin systems).
As a result, the diagrams contributing, for example, to the the cumulants $C'_{2}$ and $C'_{4}$ have
a structure of the form illustrated in
Eqs.~\eqref{C21},~\eqref{C41}.
\begin{equation}\label{C21}
\begin{split}
&  \includegraphics[scale=1]{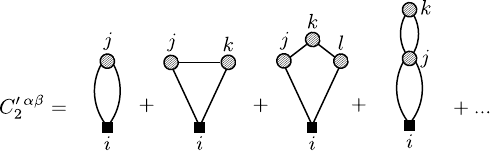}\\
& \quad = \sum_{j, k} J_{ij}^{\alpha \gamma} J_{ik}^{\delta\beta} \llangle
S^{\gamma}_{j} S^{\delta}_{k} \rrangle^{\prime(i)}\\
&\quad = \sum_{j \neq i} J^{\alpha \gamma}_{ij} \rho_{2}^{\gamma \delta} J_{ji}^{\delta \beta} +
\sum_{j, k \neq i} \beta J^{\alpha \gamma}_{ij}\rho_{2}^{\gamma \mu} J_{jk}^{\mu \nu}\rho_{2}^{\nu
\delta} J_{ki}^{\delta \beta} \\
&\quad \quad + \beta^{2} \sum_{j, k, l \neq i} J_{ij}^{\alpha \gamma} \rho_{2}^{\gamma \mu}
J_{jk}^{\mu \nu}\rho^{\nu \rho}_{2} J_{kl}^{\rho \sigma}\rho_{2}^{\sigma \delta}
J_{li}^{\delta \beta} \\
&\quad \quad + \frac{\beta^{2}}{2}\sum_{j, k \neq i} J_{ij}^{\alpha \gamma} J_{ij}^{\beta 
\delta} \rho_{4}^{\gamma \delta \mu \nu} J_{jk}^{\mu \rho} \rho_{2}^{\rho 
\sigma}J_{kj}^{\sigma \nu} + ..~,
\end{split}
\end{equation}
\begin{equation}\label{C41}
\begin{split}
&  \includegraphics[scale=1]{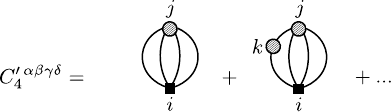}\\
& \quad = \sum_{j, k, l, m \neq i} J_{ij}^{\alpha \mu} J_{ik}^{\beta \nu} J_{il}^{\gamma \rho}
J_{im}^{\delta \sigma} \llangle S^{\mu}_{j} S^{\nu}_{k} S^{\rho}_{l}
S^{\sigma}_{m}\rrangle^{\prime (i)}\\
&\quad = \sum_{j\neq i} J_{ij}^{\alpha \mu} J_{ij}^{\beta \nu} J_{ij}^{\gamma \rho} 
J_{ij}^{\delta \sigma} \rho_{4}^{\mu \nu \rho \sigma} \\
& \quad + 4 \beta \sum_{j, k\neq i} J_{ik}^{\alpha \lambda} \rho_{2}^{\lambda 
\omega} J_{kj}^{\omega \mu} J_{ij}^{\gamma \rho} J_{ij}^{\delta \sigma} \rho_{4}^{\mu \nu 
\rho \sigma} + ..
\end{split}
\end{equation}

Here $\rho^{\alpha \beta}_{2} = \llangle S_{i}^{\alpha} S_{i}^{\beta}\rrangle^{(0)}$ and
$\rho_{4}^{\alpha \beta \gamma \delta} = \llangle S^{\alpha}_{i} S^{\beta}_{i} S^{\gamma}_{i}
S^{\delta}_{i}\rrangle^{(0)}$ are the second and fourth cumulants of the unperturbed spin
distribution. These are equal to the cumulants of the single-site distribution $\rho(\bS) = {\rm
e}^{-\beta V(\bS)}/\int_{\bar{\bS}}{\rm e}^{-\beta V(\bar{\bS})}$: $\rho_{2}^{\alpha \beta} =
\int_{\bS}\rho(\bS) S^{\alpha}S^{\beta}$,$\rho_{4}^{\alpha \beta \gamma \delta} = \int_{\bS}
\rho(\bS)S^{\alpha}S^{\beta}S^{\gamma}S^{\delta} - \rho_{2}^{\alpha \beta}\rho_{2}^{\gamma \delta} -
\rho_{2}^{\alpha \gamma} \rho_{2}^{\beta \delta}- \rho_{2}^{\alpha \delta} \rho_{2}^{\beta \gamma}$,
and are represented by vertices with 2 and 4 incoming lines in the graphs.
At higher orders the expansion involves vertices of arbitrary order (vertices with 6, 8, .. incoming
lines), representing the $m$-th order cumulants of the noninteracting distribution
$\rho_{m}^{\alpha_{1} ... \alpha_{m}} = \pa^{m}\big( \ln \int_{\bS} \rho(\bS) {\rm e}^{\bx \cdot
\bS}\big)/(\pa x^{\alpha_{1}}...\pa x^{\alpha_{m}})|_{\bx = 0}$.
The lines in the graphs, connecting different vertices, represent the interactions $J_{jk}^{\alpha
\beta}$, $J_{ij}^{\alpha \beta}$.

In general, the expansion of any $C'_{\ell}$ with $\ell$ arbitrary can be represented in terms of
graphs with the same structure of Eqs.~\eqref{C21} and~\eqref{C41}.
The graphs contributing to $C'_{\ell}$ have $\ell$ lines attached to the ``origin'' $i$,
representing the $\ell$ factors $J_{ij}^{\alpha \beta}$ in $C_{\ell}^{\prime \alpha_{1} ....
\alpha_{n}} = \sum_{j_{1}, .., j_{\ell}\neq i} J_{ij_{1}}^{\alpha_{1} \beta_{1}} .. J_{i
j_{\ell}}^{\alpha_{\ell} \beta_{\ell}} \llangle S^{\beta_{1}}_{j_{1}}...
S^{\beta_{\ell}}_{j_{\ell}}\rrangle^{\prime (i)}$, and any number of ``internal'' vertices and
lines describing the perturbative expansion of the cavity correlations.
Importantly, since $C'_{\ell}$ are moments of the cavity system, the summations over the lattice
sites $j$, $k$, $l$ range over all lattice points \emph{except} the site $i$.
The rules to associate a diagram with a perturbative term require to multiply a given term by a
numerical prefactor counting the multiplicity of the diagram.
The numerical prefactors however are irrelevant to the arguments below.

From the point of view of the large $d$ behavior, the graphs are completely analogue to the diagrams
of the cavity method in DMFT~\cite{georges_rmp_1996}.
The perturbative expansion in $J_{ij}^{\alpha \beta}$, in particular, is identical 
diagrammatically to the expansion around the atomic limit (the exchange coupling 
$J_{ij}^{\alpha \beta}$ plays the role of the electron hopping amplitude $t_{ij}$).

Applying well-known arguments of power counting in $1/d$, it can be shown that $C'_{2}$ is
of order O$(1)$ for $d \to \infty$, whereas all higher-order cumulants $C'_{\ell}$ with $\ell \geq
4$ vanish in the large-$d$ limit~\cite{georges_rmp_1996}.
In brief, for the interaction considered here, this can be shown using that the couplings
$J_{ij}^{\alpha \beta}$ are of order O$(d^{-\ell_{ij}/2})$ where $\ell_{ij}$ is the Manhattan
distance between $i$ and $j$.
As a consequence of this scaling, the diagrams contributing to $C'_{2}$ can be shown to be finite:
the summations over internal coordinates compensate the suppression O$(d^{-\ell_{ij}/2})$
coming from the scaling of the interaction.
The graphs for $C'_{4}$ or higher cumulants $C'_{\ell}$, $\ell \geq 4$, instead, are suppressed in
the large-$d$ limit, because they involve more than two lines connected to the site $i$ 
and thus, more powers of $d^{-\ell_{ij}/2}$.

A crucial subtlety is that the vanishing of the higher cumulants $C'_{\ell}$ $\ell \geq
4$ relies essentially on the cavity construction.
In fact, a diagram such as
\begin{equation*}
\includegraphics[scale=1]{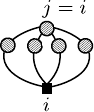}
\end{equation*}
would give a finite, O$(1)$, contribution if the summation over $j$ was allowed to run over all the
$N$ sites of the system, including the site $i$.
As is well known by standard power-counting arguments, the O$(1)$ contribution of the graph would
arise entirely from the term $j = i$, in which the internal vertex $j$ is ``collapsed'' to the
origin~\cite{georges_rmp_1996}, whereas the terms $j \neq i$ give contributions which vanish at $d
\to \infty$.
The cavity construction forces all internal summations to range over sites $j \neq i$ and thus
suppresses the O$(1)$ contribution.
As a result $C'_{4}$ picks up only contributions which vanish for $d$ large.
The same is valid for arbitrary diagrams and for any $\ell \geq 4$.

Since all cumulants $C'_{\ell}$ with $\ell$ odd vanish by spin-inversion symmetry, and 
all cumulants with $\ell \geq 4$ are negligible for $d \to \infty$, $p'(\bb_{i})$ is a 
centered Gaussian distribution, with $L^{\alpha \beta} = C'_{\ell}$.

Having derived Eq.~\eqref{p_prime}, we can use Eq.~\eqref{P_SB_p'} to reconstruct the distribution
\begin{equation}\label{P_SB1}
\begin{split}
P(\bS_{i}, \bb_{i}) & = \frac{ \exp \big[\beta (\bb_{i} \cdot \bS_{i}) - \beta V(\bS_{i}) 
- \frac{1}{2} L^{-1}{}^{\alpha \beta} b^{\alpha}_{i} b^{\beta}_{i}\big]}{Z_{1} 
\sqrt{(2\pi)^{3} \det \uu{L}}}~,
\end{split}
\end{equation}
which characterizes the statistical mechanics of the full system (without cavities).
The normalization in Eq.~\eqref{P_SB1} is
\begin{equation}\label{Z1}
Z_{1} = Z/Z'_{i} = \int_{\bS} {\rm e}^{-\beta V(\bS) + \frac{1}{2} \beta^{2} L^{\alpha 
\beta} S^{\alpha} S^{\beta}}~,
\end{equation}
and is fixed by the condition $1 = \int_{\bS} \int {\rm d}^{3} \bb~P(\bS, \bb)$.

The distribution~\eqref{P_SB1} has a form analogue to the single-site distribution $P(S_{i},
b_{i})$ of the SK model in the paramagnetic phase and for zero external fields.
In addition to the local potential $V(\bS_{i})$, there are two factors, both of order 1 for large
$d$: the fluctuations of the cavity and the term $\exp[\beta (\bb_{i} \cdot \bS_{i})]$, which
encodes the correlations between $\bS_{i}$ and $\bb_{i}$.
However in the SK model the variance of the cavity distribution is temperature-independent (equal to
$J^{2}(1-q) = J^{2}$~\cite{mezard_spin-glass}, because the overlap $q$ vanishes in the paramagnetic
state), whereas here $L^{\alpha \beta}$ has a nontrivial temperature dependence, which has to be
analyzed explicitly.

In principle, $L^{\alpha \beta}$ could be computed directly from the second moment $C_{2}'{}^{\alpha
\beta} = L^{\alpha \beta}$.
This however requires to compute correlation functions in the cavity system.
In the following, we use a different approach, based on self-consistency conditions analogue to the
self-consistency of  effective single-site problems in DMFT and other field-theoretical
approaches~\cite{de-dominicis_pr_1980, georges_rmp_1996, wu_prb_2004, lopatin_prb_2002}.

In particular, we fix $L^{\alpha \beta}$ by matching two alternative expressions for the 
site-diagonal elements $c^{\alpha \beta} = C_{ii}^{\alpha \beta}$, $a^{\alpha \beta} = 
A_{ii}^{\alpha \beta}$, $\lambda^{\alpha \beta} = \Lambda_{ii}^{\alpha \beta}$ of the full
two-point correlation functions $C_{ij}^{\alpha \beta} = \langle S^{\alpha}_{i} 
S^{\beta}_{j} \rangle$, $A^{\alpha \beta}_{ij} = \langle S^{\alpha}_{i} b^{\beta}_{j} 
\rangle$, $\Lambda^{\alpha \beta}_{ij} = \langle b^{\alpha}_{i} b^{\beta}_{j}\rangle$ 
(calculated in the complete system, without cavities).

For coincident sites $i = j$ these correlations can be calculated directly from the 
single-site distribution~\eqref{P_SB1} as:
\begin{equation}\label{c_a_lambda_def}
\begin{gathered}
C_{ii}^{\alpha \beta} = c^{\alpha \beta} = \int_{\bS} \int {\rm d}^{3} \bb~S^{\alpha} 
S^{\beta}~P(\bS, \bb)~,\\
A_{ii}^{\alpha \beta}= a^{\alpha \beta} =  \int_{\bS} \int {\rm d}^{3} \bb~S^{\alpha} 
b^{\beta}~P(\bS, \bb)~,\\
\Lambda_{ii}^{\alpha \beta} = \lambda^{\alpha \beta} =  \int_{\bS} \int {\rm d}^{3} 
\bb~b^{\alpha} b^{\beta}~P(\bS, \bb)~.\\
\end{gathered}
\end{equation}
Due to the special form of Eq.~\eqref{P_SB1}, they satisfy the "equations of motion"
\begin{equation}\label{static_eom}
\begin{gathered}
\lambda^{\alpha \beta} = L^{\alpha \beta} +  \beta^{2} L^{\alpha \gamma} c^{\gamma 
\delta} L^{\delta \beta}~,\\
a^{\alpha \beta} - \beta c^{\alpha \gamma} L^{\gamma \beta} = 0~.
\end{gathered}
\end{equation}
Eqs.~\eqref{static_eom} reflect the fact that for fixed $\bS_{i}$, the distribution of the local
field $P(\bS_{i}, \bb_{i})$ is a Gaussian with covariance $L^{\alpha \beta}$ and mean $\langle
b^{\alpha}_{i}\rangle|_{\bS_{i} {\rm fixed}} = \beta L^{\alpha \beta} S_{i}^{\beta}$.
After setting $b^{\alpha}_{i} = \beta L^{\alpha \beta} S^{\beta}_{i} + \zeta^{\alpha}_{i}$,
$\zeta^{\alpha}_{i}$ is a centered Gaussian uncorrelated from
$S_{i}^{\alpha}$ so that $\lambda^{\alpha \beta} = \langle b^{\alpha}_{i} b^{\beta}_{i}\rangle =
\beta^{2} L^{\alpha \gamma} L^{\beta \delta} \langle S^{\gamma}_{i} S^{\delta}_{i} \rangle + \langle
\zeta^{\alpha}_{i} \zeta^{\beta}_{i}\rangle = \beta^{2} L^{\alpha \gamma} c^{\gamma \delta}
L^{\delta\beta}  + L^{\alpha \beta}$, $a^{\alpha \beta} = \beta L^{\beta \gamma} \langle
S^{\alpha}_{i} S^{\gamma}_{i}\rangle + \langle S^{\alpha}_{i} \zeta^{\beta}_{i}\rangle = \beta
c^{\alpha \gamma} L^{\gamma \beta}$.
The first of Eqs.~\eqref{static_eom}, in particular, has a simple interpretation.
The covariance of the local field $\bb_{i}$ receives two contributions: a part, $L^{\alpha \beta}$,
which originates from $\zeta_{i}^{\alpha}$ and which describe the fluctuations of the cavity, and a
part due to the bias $\beta L^{\alpha \gamma} S^{\gamma}_{i}$, which is related to the Onsager
reaction field, and which fluctuates when averaged over random 
orientations of $\bS_{i}$.

The self-consistency conditions are expressed by requiring that the fluctuations computed from
Eqs.~\eqref{c_a_lambda_def} match with a separate calculation of the second-order correlations
functions.
To obtain an independent set of expressions for $C_{ij}^{\alpha \beta}$, $A_{ij}^{\alpha
\beta}$,$\Lambda_{ij}^{\alpha \beta}$ we use that at non-coincident sites $i \neq j$, these
correlations can be calculated via a generalized cavity method, with two
cavities~\cite{mezard_spin-glass}: one at $i$ and one at $j$.
As for the single cavity method, the calculation of the correlations can be approached by analyzing
the probability  $P(\bS_{i}, \bb_{i}; \bS_{j}, \bb_{j})$ to find simultaneously, given values
of the variables $\bS_{i}$, $\bb_{i}$, $\bS_{j}$, $\bb_{j}$ at the two sites $i$ and $j$.
Factorizing the Gibbs weights, this probability can be written as:
\begin{equation} \label{P_SB_2c}
\begin{split}
& P(\bS_{i}, \bb_{i}; \bS_{j}, \bb_{j}) = Z''_{ij} Z^{-1} \exp \{\beta [(\bb_{i} \cdot 
\bS_{i}) \\
& \quad + (\bb_{j} \cdot \bS_{j}) - J_{ij}^{\alpha \beta} S^{\alpha}_{i} S^{\beta}_{j} - 
V(\bS_{i}) - V(\bS_{j})]\} \\
& \quad \times p''_{ij}(b^{\alpha}_{i} - J^{\alpha \beta}_{ij} S^{\beta}_{j}, 
b^{\alpha}_{j} - J^{\alpha \beta}_{ji} S^{\beta}_{i})~.
\end{split}
\end{equation}
The terms in the first two lines describe the Gibbs-Boltzmann weight associated with the 
energy $V(\bS_{i}) + V(\bS_{j}) - (\bb_{i} \cdot \bS_{i}) - (\bb_{j} \cdot \bS_{j}) + 
J^{\alpha \beta}_{ij} S^{\alpha}_{i} S^{\beta}_{j}$ (the last term cancels the 
double-counting of the direct $i-j$ interaction).
$p_{ij}''(\mathbf{X}_{1}, \mathbf{X}_{2})$ is the distribution of the fields 
$X_{1}^{\alpha} = b^{\alpha}_{i} - J^{\alpha \beta}_{ij} S^{\beta}_{j} = \sum_{k \neq i, 
j} J_{ik}^{\alpha \beta} S_{k}^{\beta}$ and $X_{2}^{\alpha} = b^{\alpha}_{j} - J^{\alpha 
\beta}_{ji} S^{\beta}_{i} = \sum_{k \neq i, j} J^{\alpha \beta}_{jk} S^{\beta}_{k}$ 
calculated in the 2-cavity system (with the sites $i$ and $j$ removed).
The normalization factor $Z''_{ij}$ is the partition function of the 2-cavity system.

As in the case of the single-cavity distribution $p'(\bb)$, we can analyze 
$p''_{ij}(\bX_{1}, \bX_{2})$ by studying perturbatively its cumulants
\begin{equation}
\begin{split}
& C''_{\ell, m}{}^{\alpha_{1}..\alpha_{\ell} \beta_{1} \beta_{m}} = \llangle 
X_{1}^{\alpha_{1}} .. X_{1}^{\alpha_{\ell}} X_{2}^{\beta_{1}} .. 
X_{2}^{\beta_{m}}\rrangle''{}^{(ij)}\\
& = \sum_{k_{1}, .., k_{\ell}, l_{1}, .., l_{m} \neq i, j} J^{\alpha_{1} \gamma_{1}}_{i 
k_{1}} .. J^{\alpha_{\ell} \gamma_{\ell}}_{i k_{\ell}}J^{\beta_{1} \delta_{1}}_{j l_{1}} 
.. J^{\beta_{m} \delta_{m}}_{j l_{m}} \\
& \qquad \qquad \times \llangle S_{k_{1}}^{\gamma_{1}} .. S_{k_{\ell}}^{\gamma_{\ell}} 
S^{\delta_{1}}_{l_{1}} S^{\delta_{m}}_{l_{m}}\rrangle''{}^{(ij)}~.
\end{split}
\end{equation}

Here $\llangle .. \rrangle''{}^{(ij)}$ are connected averages in the system with two 
cavities at $i$ and $j$.

In the $d \to \infty$ limit we find that the cumulants $C''_{\ell, 0}{}^{\alpha_{1} ..
\alpha_{\ell}}$ and $C''_{0, \ell}{}^{\alpha_{1} .. \alpha_{\ell}}$, which involve only one of the
two sites are equal, up to negligible corrections, to the cumulants $C'_{\ell}{}^{\alpha_{1} ..
\alpha_{\ell}}$ of the system with a single-cavity.
This is due to the fact that the presence of a second cavity at a point $j$ which is
\emph{away} from the origin $i$ has only a small effects on the diagrams for the cumulants 
of $X_{1}$ (similarly the cavity at $i$ has small effects on the cumulants of $X_{2}$).

Turning to the cross-correlations between $X_{1}$ and $X_{2}$, an analysis of the mixed 
cumulants shows that the diagrams for $C''_{1, 1}{}^{\alpha \beta}$
\begin{equation*}
\includegraphics[scale=1]{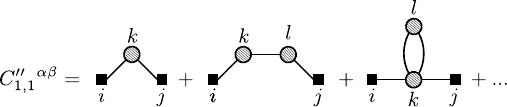}
\end{equation*}
are of order $d^{-\ell_{ij}/2}$ while the graphs for cumulants of higher order vanish in 
a faster way for large $d$.

We thus find that, at leading order for $d$ large, $p''(\bX_{1}, \bX_{2})$ is Gaussian and 
has the form:
\begin{equation}\label{p2_result}
\begin{split}
& p''_{ij}(\bX_{1}, \bX_{2}) \simeq [(2\pi)^{3} \det \uu{L}]^{-1} \exp \big\{ -\big[ \big 
(L^{-1}{}^{\alpha \beta} X_{1}^{\alpha} X_{1}^{\beta} \\
& + L^{-1}{}^{\alpha \beta} X_{2}^{\alpha} X_{2}^{\beta} \big)/2 + M_{ij}^{\alpha \beta} 
X_{1}^{\alpha} X_{2}^{\beta}\big]\big\}~.
\end{split}
\end{equation}
The off-diagonal term $M_{ij}^{\alpha \beta}$ is of order $d^{-\ell_{ij}/2}$ 
($M_{ij}^{\alpha \beta} \simeq  -L^{-1}{}^{\alpha \gamma} C''_{1, 1}{}^{\gamma \delta} 
L^{-1}{}^{\delta \beta}$).

The correlations $C_{ij}^{\alpha \beta}$, $A_{ij}^{\alpha \beta}$, and 
$\Lambda_{ij}^{\alpha \beta}$ at leading order can be calculated by plugging 
Eq.~\eqref{p2_result} into Eq.~\eqref{P_SB_2c}, by substituting $Z''_{i}/Z \simeq 
1/Z_{1}^{2}$, and by expanding to first order in $J_{ij}^{\alpha \beta}$ and 
$M_{ij}^{\alpha \beta}$ (which are both O$(d^{-\ell_{ij}/2})$).
After an explicit calculation we find the relations
\begin{equation}\label{C_A_Lambda_1st_order}
\begin{split}
C_{ij}^{\alpha \beta} & = \beta c^{\alpha \gamma} J_{ij}^{\gamma \delta} c^{\delta 
\beta} - a^{\alpha \gamma} M^{\gamma \delta}_{ij} a^{\beta \delta}~,\\
A_{ij}^{\alpha \beta} & =  c^{\alpha \gamma} J^{\gamma \beta}_{ij} + \beta c^{\alpha 
\gamma} J_{ij}^{\gamma \delta} a^{\delta 
\beta} - a^{\alpha \gamma} M_{ij}^{\gamma \delta} \lambda^{\delta \beta}~,\\
\Lambda_{ij}^{\alpha \beta} & = J_{ij}^{\alpha \gamma} a^{\gamma \beta} + a^{\gamma 
\alpha} J_{ij}^{\gamma \beta} + \beta a^{\gamma \alpha} J_{ij}^{\gamma \delta} a^{\delta 
\beta} \\
&- \lambda^{\alpha \gamma} M^{\gamma \delta}_{ij} \lambda^{\delta \beta}~,
\end{split}
\end{equation}
which are exact at leading order for $d \to \infty$.

Using Eqs.~\eqref{static_eom} and~\eqref{C_A_Lambda_1st_order}, we can eliminate 
$M_{ij}^{\alpha \beta}$ and obtain the following relations, valid for arbitrary $i$ and 
$j$ (coincident or non-coincident):
\begin{equation}\label{C_ij_Lambda_ij_self-consistency}
\begin{gathered}
C_{ij}^{\alpha \beta} = - (k_{\rm B}T)^{2} \sigma^{\alpha \beta} \delta_{ij} - k_{\rm 
B}T A^{\alpha \gamma}_{ij} \sigma^{\gamma \beta}~,\\
\Lambda^{\alpha \beta}_{ij} = - k_{\rm B}T J_{ij}^{\alpha \beta} - \beta 
\sigma^{-1}{}^{\alpha \gamma} A_{ij}^{\gamma \beta}~.
\end{gathered}
\end{equation}
Here $\sigma^{\alpha \beta} = \lambda^{-1}{}^{\alpha \beta} - L^{-1}{}^{\alpha \beta}$.

Eqs.~\eqref{C_ij_Lambda_ij_self-consistency} now do not make any reference to the cavity system;
they express relations between the "true" correlation functions $C$, $A$, and $\Lambda$.
The calculation can be completed using that the correlations have to satisfy by definition
$A_{ij}^{\alpha \beta} = \sum_{k} C_{ik}^{\alpha \gamma} J_{kj}^{\gamma \beta}$, 
$\Lambda_{ij}^{\alpha \beta} = \sum_{k} J_{ik}^{\alpha \gamma} A_{kj}^{\gamma \beta}$.

As a result we obtain:
\begin{equation}\label{C_A_Lambda_solution}
\begin{gathered}
\hat{C} = - (\beta^{2}\hat{\sigma}^{-1} + \beta \hat{J})^{-1}~,\\
\hat{A} = \hat{C}\hat{J}~,\\
\hat{\Lambda} = \hat{J} \hat{C} \hat{J} = -k_{\rm B}T \hat{J} + (\hat{\sigma} + \beta 
\hat{J}^{-1})^{-1}~.
\end{gathered}
\end{equation}

Here the matrix $\hat{\sigma}$ is site-diagonal, and is equal to $\sigma^{\alpha \beta}_{ij} =
\delta_{ij}\sigma^{\alpha \beta}$.
It plays the role of the local self-energy of DMFT~\cite{georges_rmp_1996}.

The self-consistency conditions can finally be imposed by requiring that the onsite elements
$C^{\alpha \beta}_{ii}$, $A^{\alpha \beta}_{ii}$, and $\Lambda^{\alpha \beta}_{ii}$
computed from Eqs.~\eqref{C_A_Lambda_solution} coincide with the correlations $c^{\alpha 
\beta}$, $a^{\alpha \beta}$, $\lambda^{\alpha \beta}$ of the single-site problem, 
computed from Eqs.~\eqref{c_a_lambda_def} and the distribution~\eqref{P_SB1}.
Using Eqs.~\eqref{static_eom} it can be checked that the three matching conditions 
$C^{\alpha \beta}_{ii} = c^{\alpha \beta}$, $A^{\alpha \beta}_{ii} = a^{\alpha \beta}$, 
and $\Lambda^{\alpha \beta}_{ii} = \lambda^{\alpha \beta}$ are equivalent, so $L^{\alpha 
\beta}$ can be determined by imposing any one of them.

This completes the solution of the problem for $d \to \infty$, at least in the case of a
symmetric solution, with unbroken symmetries.

Since the two-site correlation $C_{ij}^{\alpha \beta}$ is connected to the static 
susceptibility $\chi_{ij}^{\alpha \beta}$ by the thermodynamic relation $\chi_{ij}^{\alpha 
\beta} = \beta C_{ij}^{\alpha \beta}$, the first of Eqs.~\eqref{C_A_Lambda_solution} can 
be written as $\hat{\chi}^{-1} = - \beta \hat{\sigma}^{-1} - \hat{J}$, 
and, after a Fourier transform
\begin{equation}
\chi^{-1}{}^{\alpha \beta}(\bk) = - \beta \sigma^{-1}{}^{\alpha \beta} - J^{\alpha 
\beta}(\bk)~.
\end{equation}

The result is common in systems with large coordination numbers: the susceptibility has 
the same structure as the Weiss mean-field susceptibility $\chi^{-1}{}^{\alpha \beta} = 
\chi_{0}^{-1}{}^{\alpha \beta} - J^{\alpha \beta}(\bk)$ (Eq.~\eqref{MFT_susceptibility}),
but $\chi_{0}^{-1}{}^{\alpha \beta}$ is now replaced by a renormalized term 
$-\beta\sigma^{-1}{}^{\alpha \beta}$, local in real space, and determined by 
self-consistency conditions.
The term $-\beta\sigma^{-1}{}^{\alpha \beta}$ plays the role of the "locator 
matrix"~\cite{lopatin_prb_2002}, here in the special case of a state with zero magnetization.

\subsection{Internal energy and free energy}

The effective single-site problem and the self-consistency equations give access to the 
thermodynamic properties of the system.
In fact, since the energy can be recast as $H = \sum_{i} V(\bS_{i}) + \frac{1}{2} 
\sum_{i} (\bb_{i} \cdot \bS_{i})$, the thermodynamic internal energy per site can be 
computed as:
\begin{equation}
\frac{E(\beta)}{N} = \langle V(\bS) \rangle_{1} - \frac{1}{2} a^{\alpha \alpha}~.
\end{equation}
Here $\langle V(\bS)\rangle_{1} = Z_{1}^{-1} \int_{\bS} V(\bS) \exp\big[- \beta V(\bS) + 
\beta^{2}L^{\alpha \beta} S^{\alpha} S^{\beta}/2\big]$ is the average anisotropy energy 
computed with the distribution~\eqref{P_SB1} and $a^{\alpha \alpha}$ is the trace of $a^{\alpha
\beta} = \langle S^{\alpha} b^{\beta}\rangle$.
Integrating over temperatures and using the self-consistency conditions we find that the 
free energy is given by an expression with a standard "${\rm tr} \ln$" 
form~\cite{lopatin_prb_2002}:
\begin{equation}\label{free-energy}
\begin{split}
& \frac{F(\beta)}{N} = - T S_{\infty} +k_{\rm B}T \int_{0}^{\beta} {\rm 
d}\beta'~E(\beta')\\
& = - k_{\rm B}T \ln Z_{1}(\beta, L^{\alpha \beta}) - \frac{k_{\rm B}T}{2} {\rm tr} \ln 
[-\beta^{2}( \uu{\sigma}^{-1} + \uu{L})]\\
& \quad + \frac{k_{\rm B}T}{2} {\rm tr} \ln [- (\beta^{2} \hat{\sigma}^{-1} + \beta 
\hat{J})]~.
\end{split}
\end{equation}

Here $Z_{1}(\beta, L^{\alpha \beta}) = \int_{\bS} \exp\big[- \beta V(\bS) + 
\beta^{2}L^{\alpha \beta} S^{\alpha} S^{\beta}/2\big]$ is the single-site partition 
function [Eq.~\eqref{Z1}] and the integration constant $S_{\infty}$ is equal to the 
entropy per site in the infinite-temperature limit.
In the classical model analyzed here $S_{\infty}$ is an arbitrary constant with no 
physical meaning (only entropy differences are meaningful).
In the second line of Eq.~\eqref{free-energy}, we have chosen $S_{\infty} = k_{\rm B} \ln 
(4 \pi)$.

We also note that the self-consistency equations can be obtained from a variational 
principle on $F$.
If we regard the free energy as a function $F(\beta, L^{\alpha \beta}, \sigma^{\alpha 
\beta})$ of three independent variables $\beta$, $L^{\alpha \beta}$, and $\sigma^{\alpha 
\beta}$, defined by Eq.~\eqref{free-energy}, then the self-consistency conditions are 
equivalent to the requirements that $F$ is stationary with respect to variations of 
$L^{\alpha \beta}$ and $\sigma^{\alpha \beta}$.
In fact the relations $\pa F/\pa L^{\alpha \beta}|_{\beta, \sigma} = 0$ and $\pa F/\pa 
\sigma^{\alpha \beta}|_{\beta, L} = 0$ give $c^{\alpha \beta} = C_{ii}^{\alpha \beta}$ 
and 
\begin{equation}\label{c_sigma}
\beta^{2} c^{\alpha \beta} = - (\uu{\sigma}^{-1} + \uu{L})^{-1}{}^{\alpha \beta}~,
\end{equation}
a relation which, using Eqs.~\eqref{static_eom}, can be shown to be equivalent to 
the condition $\sigma^{\alpha \beta} = \lambda^{-1}{}^{\alpha \beta} - L^{-1}{}^{\alpha 
\beta}$.

\subsection{Local stability of the symmetric solution}
\label{s:stability}

In the derivation, the system has been assumed to present unbroken spin-inversion and 
translation symmetries.
In other words, the analysis described a system which, starting from a high-temperature 
paramagnetic phase, is continously cooled down.
This symmetric solution can be consistent only if the susceptibility matrix $\chi_{ij}^{\alpha
\beta}$ is positive-definite, in such way that the system is locally stable against a spontaneous
modulation.
In this section, we show that this local stability condition is satisfied at any temperature, down
to $T = 0$: the disordered phase is never destabilized, at arbitrarily low $T$.

To analyze stability, consider first the isotropic case.
In this case, $c^{\alpha \beta} = \delta^{\alpha \beta}/3$, $L^{\alpha \beta} = L
\delta^{\alpha \beta}$, $\sigma^{\alpha \beta} = \sigma \delta^{\alpha \beta}$, $\lambda^{\alpha
\beta} = \lambda \delta^{\alpha \beta}$ and several of the expressions
simplify.
In particular, expressing $C_{ii}^{\alpha \beta}$ as a Fourier integral and using the
infinite-dimensional density of states~\cite{georges_rmp_1996, lopatin_prb_2002}
\begin{equation}
\nu(\ve) = \frac{1}{(2\pi)^{d}} \int_{-\pi}^{\pi}{\rm d}k_{1}.. \int_{-\pi}^{\pi}{\rm 
d}k_{d} \delta(\ve - \ve_{\bk}) \substack{\to \\ d \to \infty} \frac{{\rm 
e}^{-\ve^{2}/2}}{\sqrt{2\pi}}~,
\end{equation}
the self-consistency condition $c^{\alpha \beta}=C_{ii}^{\alpha \beta}$ reduces to the
scalar equation
\begin{equation}\label{I_beta_sigma}
\frac{\beta}{3} = -\int_{-\infty}^{\infty} {\rm d}\ve~\frac{\nu(\ve)}{\beta/\sigma + 
f(\ve)} = I(\beta/\sigma)~.
\end{equation}

As discussed in Sec.~\ref{s:model}, $f(\ve)$ is assumed to be bounded from above, with a 
maximum value ${\rm max}_{\ve}[f(\ve)] = f_{\rm max}$.
The integral $I(\beta/\sigma)$ therefore is well defined for $\beta/\sigma < -f_{\rm 
max}$.
Since $I(\beta/\sigma)$ tends to $0$ for $\beta/\sigma \to -\infty$ and to 
$+\infty$ for $\beta/\sigma$ to $-f_{\rm max}$, there is always a value of $\beta/\sigma$ 
in the interval $-\infty < \beta/\sigma < -f_{\rm max}$ which solves 
Eq.~\eqref{I_beta_sigma}, at any temperature $T$.
Thus, the self-consistency relations admit a solution at any $T$.
In addition, the solution corresponds always to a locally-stable state because the susceptibility
$\chi^{-1}(\bk) = -\beta/\sigma - J(\bk) = -\beta/\sigma - f(\ve_{\bk}) \geq f_{\rm max} -
f(\ve_{\bk})$ is automatically positive for all $\bk$.

We see therefore that the second-order transition predicted by the Weiss theory
(Sec.~\ref{s:Weiss_MFT}) disappears in the exact large-$d$ solution.

The reason why the instability is suppressed at arbitrarily low $T$ can be traced to the 
fact that $J(\bk) = f(\ve_{\bk})$ has its maximum value $f_{\rm max}$ on an entire surface 
in momentum space, and not on isolated $\bk$-points.
Due to this geometrical property, there is a large number of modes near the degenerate 
surface $\ve_{\bk} = \ve_{\rm max}$ which become simultaneously soft as $\beta/\sigma$ 
approaches $-f_{\rm max}$.
The fluctuation of these modes make the integral $I(\beta/\sigma)$ diverge for $\beta/\sigma \to
-f_{\rm max}$, ensuring that Eq.~\eqref{I_beta_sigma} has always a solution.

The mechanism at play is closely analogue to that occurring in the Brazovskii model, and 
in other models with Coulomb- or dipole-frustrated 
interactions~\cite{schmalian_ijmpb_2001, wu_prb_2004, principi_prb_2016, 
principi_prl_2016, nussinov_prl_1999}, where  the modes near a degenerate surface remove a
second-order instability.

Before continuing the discussion, let us analyze the properties of the solution in more detail.
The condition $J_{ii} = 0$, implies that $\int_{-\infty}^{\infty} {\rm 
d}\ve~\nu(\ve)f(\ve) = J_{ii} = 0$.
This, in particular, shows that $f_{\rm max}$ is always positive and $\beta/\sigma < 
-f_{\rm max}$ is always negative.
Using $\int_{-\infty}^{\infty}{\rm d}\ve~\nu(\ve) = 1$ and analyzing Eq.~\eqref{I_beta_sigma} it can
then be shown that $\beta/\sigma < {\rm min}(\{-f_{\rm max}, -3/\beta\})$.
At high temperatures $\beta/\sigma \approx -3/\beta + {\rm O}(\beta^{-3})$.
The effective single-atom susceptibility $\tilde{\chi} = -\sigma/\beta$ is approximately 
equal to its Weiss mean-field value $\chi_{0} = \beta/3$.
At low temperatures, however, $\tilde{\chi}$ becomes much smaller than $\chi_{0}$ and 
eventually saturates to a finite value $\tilde{\chi} = -\sigma/\beta  \approx 1/f_{\rm 
max}$ when $T \to 0$.
The temperature-dependence of $\tilde{\chi}$ is such that $\tilde{\chi}^{-1}$ is always 
larger than $J(\bk)$ at all wavelengths, so that the system is locally stable at all $T$.

In the anisotropic case, the analysis is more complex.
However, we expect that from the point of view of the local stability of the disordered solution,
the result is the same.
When all eigenvalues of $f^{\alpha \beta}(x)$ are bounded from above, we expect that the
paramagnetic state remains locally stable at all $T$.

This local stability analysis rules out that the system may order at low temperatures via a
continuous transition.
It is not excluded, however, that the model may undergo a first-order transition into an ordered
state.
In this case, the symmetric solution which we presented corresponds to the one of lowest
free-energy only for temperatures higher than the temperature $T_{\rm f.o.}$ of the transition,
while for $T < T_{\rm f.o.}$ it has to be interpreted as a metastable supercooled phase ($T_{\rm
f.o.}$ denotes the temperature of the first-order transition).

A study of possible first-order transitions requires an analysis of the broken-symmetry solutions
of the large-$d$ model, which is beyond the scope of this work~\cite{Note4}.
We give, however, some remarks on this question.
In the context of continuous field theories with degenerate surfaces of soft modes, the Brazovskii
model with scalar degrees of freedom has been predicted to present a fluctuation-induced first-order
transition to a modulated phase~\cite{brazovskii_jetp_1975, tarjus_jpcm_2005}.
Coulomb-frustrated models with vector degrees of freedom and O$(n)$ symmetry have been argued,
instead, to behave differently.
In particular, the analyses in Refs.~\onlinecite{nussinov_prl_1999, tarjus_jpcm_2005} 
indicate that for $n \geq 3$ Coulomb-frustrated models present no ordered phases when 
there is an exact degeneracy of low-lying modes at a momentum-space surface (although a 
lifting of the degeneracy due to lattice effects beyond the continuum limit induce a
finite-temperature transition).

In the models studied here, the mechanism by which the symmetric state is stabilized at
arbitrarily low $T$ is analogue to that occurring in Coulomb-frustrated models, since it originates
from the anomalous density of states near a surface in momentum space.
It should be noted, however, that the lattice spin systems which we study differ from continuous
field theories because of a significantly different shape of the degenerate surface.
In Coulomb-frustrated models, degenerate surfaces of soft modes arise due to competing interactions
on different length scales.
In the limit of small frustration, the surface of soft modes occurs in a region of wavelengths much
larger than the lattice spacing.
In the lattice systems of interest in this work, instead, frustration occurs at the 
microscopic scale and the surface of soft modes occurs at large wavevectors, comparable to 
the size of the Brillouin zone. The degenerate surface of soft modes, as a result, has a 
highly non-spherical shape, determined by the condition $\ve_{\bk} = \ve_{\rm max}$.

Due to the non-spherical shape, it is possible that a mechanism of \emph{order by
disorder}~\cite{bergman_np_2007} selects at low temperatures a modulated phase.
Order by disorder has indeed been found in the $J_{1} = 2J_{2}$ fcc antiferromagnet in three
dimensions: in this model Ref.~\onlinecite{balla_prb_2019} predicted the entropic selection at low
temperatures of an ordered helical state, with a modulation vector oriented along a high-symmetry
direction~\cite{balla_prb_2019}.
It is likely therefore that, also in the large-$d$ models studied here, an ordered state is selected
entropically as the most stable low-temperature phase.

Since however we showed that the disordered phase is locally stable in the $d \to \infty$ model, we
can consistently study its properties at all $T$ (in the region $T < T_{\rm f.o.}$ the analysis
then describes a metastable, supercooled phase).

In the following sections we will focus on the disordered solution and, using a dynamical analysis,
we will explore the possibility that, at a certain temperature, it may undergo a dynamical
glass transition, and a consequent breaking of ergodicity.
If the glass transition competes with a first-order transition and $T_{\rm f.o.}> T_{\rm g}$ we
assume that the system can be supercooled down to $T_{\rm g}$.

In other words, the disordered solution represents the state of the system in a spin-liquid
phase~\cite{bergman_np_2007}, and by the dynamical analysis in the next sections we study whether
this liquid state can freeze into a glass.
A first-order transition into a modulated phase, is instead analogue to a crystallization.
The assumption which we make is that the crystallization, if present, can be avoided by
supercooling.

\subsection{Distribution of the internal field and low-temperature limit of the internal energy in
in isotropic models}
\label{s:internal_field}

Before analyzing dynamical properties, we discuss some additional properties of the static
solution focusing on the isotropic case.
In isotropic models, the cavity distribution $p'(\bb)$ is, at any temperature, a
rotationally-invariant Gaussian of width $L$.
The variance of the cavity field distribution $L$ is related to the self-energy $\sigma$ by the
relation  $L = -3/\beta^{2} - 1/\sigma$ (which follows from Eqs.~\eqref{static_eom} using that in
the rotationally-invariant case $c^{\alpha \beta} = \delta^{\alpha \beta}/3$ at all $T$).
At high temperatures, $L$ is controlled by the leading orders of perturbation theory and 
is approximately $L \approx {\cal J}^{2}_{2}/3$, with ${\cal J}^{2}_{2} = 
\int_{-\infty}^{\infty}{\rm d}\ve~\nu(\ve) f^{2}(\ve)=  \sum_{k} (J_{ik})^{2}$.
When the temperature is lowered, $L$ decreases.
Eventually in the limit $T \to 0$, $L$ vanishes as $L \approx -1/\sigma \approx k_{\rm B} 
T f_{\rm max}$.

Consider now the complete distribution of the field $p(\bb) = \int_{\bS} P(\bS,
\bb)$~\cite{mezard_spin-glass}.
At high temperatures, the reaction field is weak and the distribution $p(\bb)$ is approximately
equal to the cavity distribution $p'(\bb)$.
In particular, the variance $\lambda$ of $p(\bb)$ is approximately the same as the 
variance $L$ of $p'(\bb)$.

When the temperature is lowered, the correlation between $\bb_{i}$ and $\bS_{i}$ becomes 
more important and the distribution $p(\bb_{i})$ becomes significantly different from 
$p'(\bb_{i})$.
Eventually in the limit of low temperatures, the distribution of the field $\bb_{i}$ is 
completely dragged by the spin $\bS_{i}$.
In fact, using $L \approx k_{\rm B}T f_{\rm max}$ we find:
\begin{equation}\label{P_SB_T=0}
\begin{split}
P(\bS, \bb) & \approx \frac{\exp \{\beta [(\bb \cdot \bS) - b^{2}/(2 f_{\rm max}) - 
f_{\rm max}/2]\}}{4 \pi \sqrt{(2\pi)^{3} \det \uu{L}}}\\
& \substack{\to\\\beta \to \infty} \frac{1}{4\pi} \delta(\bb - f_{\rm max} \bS)~.
\end{split}
\end{equation}

In all relevant configurations, $\bb_{i}$ is equal to $f_{\rm max} \bS_{i}$ up to a small 
fluctuation (with root mean square $\sqrt{L} \approx \sqrt{k_{\rm B}T f_{\rm max}}$).
The probability $p(\bb) = \int_{\bS} P(\bS, \bb) $ is then concentrated within a narrow 
spherical surface of radius $f_{\rm max}$.
The variance of the distribution $\lambda = 1/\sigma + \beta^{2}/(3 \sigma^{2})$ remains 
finite and approaches $\lambda \approx f_{\rm max}^{2}/3$ for $T \to 0$.

Eq.~\eqref{P_SB_T=0} implies, in addition, that the internal energy per site for $T \to 0$ tends to
$E_{0} = - N a^{\alpha \alpha}/2 = - N f_{\rm max}/2$ in isotropic models.
$E_{0}$ is exactly equal to the energy of the spiral ground states, which is the exact ground state
energy of the system.
This suggests that the configurations contributing to the $T \to 0$ limit of the symmetric state
are built predominantly from a superposition of Fourier modes with $\bk$ lying near the degenerate
surface.
However, we expect that the correlations in this disordered phase remain short-ranged, since the
correlation function $C_{ij}^{\alpha \beta}$ evolves smoothly when the system is cooled down
starting from high temperatures.

We note as a remark that $E_{0}$ describes the $T \to 0$ limit of the \emph{equilibrium} internal
energy.
This energy may be unreachable if the system remains trapped in a glassy state at a vitrification
temperature $T_{\rm g}$.
The state in which the system freezes at $T_{\rm g}$ may have an energy larger than $E_{0}$ when it
is eventually cooled down to $T \to 0$.
In this case, the relevant thermodynamic properties should be analyzed taking into account the
trapping into a metastable state (see Ref.~\onlinecite{lopatin_prb_2002} for a replica theory of
the thermodynamics within a glass phase).

\section{Dynamical properties}
\label{s:dynamics}

The analysis in Sec.~\ref{s:statics} described the properties of the disordered phase from the
point of view of equilibrium statistical mechanics.
In this section, we extend the analysis and study the equilibrium dynamics of the system.
The results of this dynamical analysis will be used in Sec.~\ref{s:glass_transition} to analyze the
conditions under which the disordered phase can become non-ergodic, and develop a glassy behavior.

To study the system from a dynamical point of view, we assume a purely dissipative Langevin
equation:
\begin{equation}\label{langevin}
\begin{split}
\dot{\bS}_{i} & = - \bS_{i} \times (\bS_{i} \times (\bN_{i} + \bnu_{i}))\\
& = \bN_{i} + \bnu_{i} - \bS_{i} (\bS_{i} \cdot (\bN_{i} + \bnu_{i}))~.
\end{split}
\end{equation}
Here $\bN_{i}$ is the vector
\begin{equation}\label{N_i}
\bN_{i}  = -\frac{\pa H}{\pa \bS_{i}} = \bb_{i} + \bF_{i}~,
\end{equation}
$b^{\alpha}_{i} = \sum_{j} J_{ij}^{\alpha \beta} S^{\beta}_{j}$ is the instantaneous internal
field, and $F^{\alpha}(\bS_{i}) = - \pa V(\bS_{i})/\pa S^{\alpha}_{i}$ is a contribution from the
on-site anisotropy.
$\bnu_{i}(t)$ is a random torque, considered as a white Gaussian noise with zero mean and 
\begin{equation}\label{bare_noise_variance}
\langle \nu^{\alpha}_{i}(t) \nu^{\beta}_{j}(t')\rangle = 2 k_{\rm B}T\delta^{\alpha 
\beta} 
\delta_{ij} \delta(t-t')~.
\end{equation}
For simplicity we have adopted a rescaled set of units, absorbing the gyromagnetic ratio 
and the Gilbert damping constant in a redefinition of the time scale.

This purely dissipative dynamics is a particular case, in the limit of strong damping, of 
the stochastic Landau-Lifshitz-Gilbert (LLG) equation~\cite{garcia-palacios_prb_1998, 
eriksson2017atomistic}.
We restrict the analysis to the overdamped case to simplify the presentation, but the 
derivations could be adapted to include a precession term.
We expect that the emergence of vitrification depends on the structure of the energy 
landscape, and not on the specific nature of the dynamical equations.
Thus, it seems natural to assume that a more detailed, precessional dynamics, would give 
the same conditions for the emergence of glassiness.

In the analysis, we focus on equilibrium dynamics: we consider correlation functions
averaged over the Gaussian distribution of the noises $\bnu_{i}$ \emph{and} over the initial
conditions $\bS_{i_{1}0}$, ..,$\bS_{i_{N}0}$ at time $t=0$, assigning to the initial conditions a
weight given by the equilibrium Gibbs distribution $Z^{-1}\exp [-\beta H(\bS_{i_{1}0}, ...,
\bS_{i_{N}0})]$.

As in the case of the static properties, the $d \to \infty$ limit allows to reduce the dynamics of
the $N$-body problem to an effective single-site problem and a set of self-consistency equations.
We present detailed derivations of the dynamical properties in
Secs.~\ref{s:effective_1_site},~\ref{s:properties_single_site},~\ref{s:self_consistency_equations},
and in appendix~\ref{a:dynamical_two_point_functions}, but for compactness we summarize the main
results throughout this and the next page.

In the limit $d \to \infty$ we find that the dynamics of a single spin can be replaced by a
non-Markovian Langevin equation
\begin{equation}\label{1_site_langevin}
\dot{\bS}_{i} = - \bS_{i} \times (\bS_{i} \times (\bb_{i}(t) + \bF_{i} + \bnu_{i}(t)))~,
\end{equation}
subject to a time-dependent field $\bb_{i}(t)$ given by:
\begin{equation}\label{bi_t}
\begin{split}
b_{i}(t) & = \zeta_{i}^{\alpha}(t) + \beta l^{\alpha \beta}(t) S^{\beta}_{i0} \\
& + \int_{0}^{t} {\rm d}t'~K^{\alpha \beta}(t-t') S^{\beta}_{i}(t')~.
\end{split}
\end{equation}

In Eq.~\eqref{bi_t}, $\bzeta_{i}(t)$ is a colored Gaussian noise with zero mean and correlation
$\langle \zeta^{\alpha}_{i}(t) \zeta^{\beta}_{i}(t')\rangle = l^{\alpha \beta}(t-t')$.
The kernel $K^{\alpha \beta}(t-t')$ is related to the spectrum of the noise by the
fluctuation-dissipation relation:
\begin{equation} \label{FDT_Kl}
K^{\alpha \beta}(t-t') = -\beta \Theta(t-t') \frac{{\rm d}}{{\rm d}t} l^{\alpha 
\beta}(t-t')~.
\end{equation}
Finally the term $\beta l^{\alpha \beta}(t) S^{\beta}_{i0}$ in Eq.~\eqref{bi_t} is controlled by
the same function $l^{\alpha \beta}(t)$ which defines the correlation of $\zeta^{\alpha}_{i}(t)$,
and depends on $\bS_{i0}$, which is the initial condition of the spin $\bS_{i}$ at time $t =
0$.

Eqs.~\eqref{1_site_langevin},~\eqref{bi_t} define a single-site problem which encodes all
time-dependent correlation functions of the spin $\bS_{i}$ and the field $\bb_{i}$ at a single site
$i$.
In other words, instead, of solving the $N$-body dynamics, the local correlations $\langle
S^{\alpha_{1}}_{i}(t_{1}) ... S_{i}^{\alpha_{n}}(t_{n}) b_{i}^{\beta_{1}}(t'_{1}) ...
b^{\beta_{\ell}}_{i}(t'_{\ell})\rangle$ can be calculated equivalently by solving the single-site
dynamics, and by averaging the solutions of the 1-site Langevin equations over the realizations of
$\bnu_{i}(t)$, $\bzeta_{i}(t)$, and over the initial conditions of $\bS_{i0}$.
In this procedure, the initial conditions $\bS_{i0}$ must be averaged with the probability
distribution
\begin{equation}\label{P1}
\begin{split}
P_{1{\rm eq}}(\bS_{i0}) & = Z^{-1} \int_{\bS_{i_{2}}} ..\int_{\bS_{i_{N}}} {\rm 
e}^{-\beta 
H(\bS_{i0}, \bS_{i_{2}}, .., \bS_{i_{N}})}\\
& = \int {\rm d}^{3}\bb~P(\bS_{i0}, \bb)\\
& = Z^{-1}_{1} {\rm e}^{-\beta V(\bS_{i0}) + \frac{1}{2}\beta^{2} L^{\alpha \beta} 
S_{i0}^{\alpha} S^{\beta}_{i0}}
\end{split}
\end{equation}
which is the Gibbs probability of the spin $\bS_{i0}$.

The non-Markovian single-site problem is entirely specified by the correlation function $l^{\alpha
\beta}(t-t')$.
At equal times $t = t'$, $l^{\alpha \beta}(t=t') = l^{\alpha \beta}(0)$ can be shown to be equal
to the matrix $L^{\alpha \beta}$ found in the static analysis of Sec.~\ref{s:statics}.
For arbitrary times, $l^{\alpha \beta}(t-t')$ can be fixed by a set of self-consistency conditions
satisfied by the correlations $C^{\alpha \beta}_{ij}(t-t') = \langle S^{\alpha}_{i}(t)
S^{\beta}_{j}(t')\rangle$, $A^{\alpha \beta}_{ij}(t-t') = \langle S^{\alpha}_{i}(t)
b^{\beta}_{j}(t')\rangle$, $\Lambda^{\alpha \beta}_{ij}(t-t') = \langle b^{\alpha}_{i}(t)
b^{\beta}_{j}(t')\rangle$ and their site-diagonal elements $c^{\alpha \beta}(t-t') = C_{ii}^{\alpha
\beta}(t-t')$, $a^{\alpha \beta}(t-t') = A_{ii}^{\alpha \beta}(t-t')$, $\lambda^{\alpha \beta}(t-t')
= \Lambda_{ii}^{\alpha \beta}(t-t')$.

In particular, we find that the self-consistency relations can be reduced to a compact form when
expressed in terms of the time derivatives $\dot{C}_{+ij}^{\alpha \beta}(t-t') = \Theta(t-t') {\rm
d}C^{\alpha \beta}_{ij}(t-t')/{\rm d}t$, $\dot{A}_{+ij}^{\alpha \beta}(t-t') = \Theta(t-t') {\rm d}
A_{ij}^{\alpha \beta}(t-t')/{\rm d}t$, $\dot{\Lambda}_{+ij}^{\alpha \beta}(t-t') = \Theta(t-t') {\rm
d}\Lambda^{\alpha \beta}_{ij}(t-t')/{\rm d}t$, restricted to the retarded region $t > t'$.
$\dot{C}_{+}$ is simply related to the response function $G^{\alpha\beta}_{ij}(t-t') = 
\langle \delta S^{\alpha}_{i}(t)/\delta \nu_{j}^{\beta}(t')\rangle$ by the 
fluctuation-dissipation theorem (FDT) $G_{ij}^{\alpha \beta}(t-t') = - \beta 
\dot{C}_{+ij}^{\alpha \beta}(t-t')$ (see appendix~\ref{a:FP_and_FDT}).

In the limit $d \to \infty$, we find that $\dot{C}_{+}$, $\dot{A}_{+}$, 
$\dot{\Lambda}_{+}$ satisfy the relations:
\begin{equation}\label{C_dot_plus}
\begin{split}
&  \dot{C}_{+ij}^{\alpha \beta}(t-t') - \delta_{ij} \dot{c}^{\alpha \beta}_{+}(t-t') =  
\\
& \beta \int_{t'}^{t} {\rm d}t'' \int_{t'}^{t''} {\rm d}t''' \big[\dot{C}_{+ij}^{\alpha 
\gamma}(t-t'') \\
& \qquad \times K^{\gamma \delta}(t''-t''') \dot{c}_{+}^{\delta \beta}(t'''-t')\Big] \\
& -\beta \int_{t'}^{t} {\rm d}t''~\dot{A}^{\alpha \gamma}_{+ij}(t-t'') \dot{c}_{+}^{\gamma 
\beta}(t''-t')~,
\end{split}
\end{equation}
\begin{equation}\label{A_dot_plus}
\begin{split}
& \dot{A}^{\alpha \beta}_{+ij}(t-t')  = \dot{c}_{+}^{\alpha \gamma}(t-t') J^{\gamma 
\beta}_{ij} \\
& - \beta \int_{t'}^{t} {\rm d}t'' \dot{c}_{+}^{\alpha \gamma}(t-t'') 
\Big[\dot{\Lambda}_{+ij}^{\gamma \beta}(t''-t') \\
& - \int_{t'}^{t''}{\rm d}t'''~K^{\gamma \delta}(t''-t''')\dot{A}_{+ij}^{\delta 
\beta}(t'''-t')\Big]~,
\end{split}
\end{equation}
or, in frequency space:
\begin{equation}\label{Cdot_Adot_Fourier}
\begin{split}
\dot{C}_{+}^{\alpha \beta}(\bk, \omega) & - \dot{c}_{+}^{\alpha \beta}(\omega) = \beta 
\dot{C}_{+}^{\alpha \gamma}(\bk, \omega) K^{\gamma \delta}(\omega) \dot{c}_{+}^{\delta 
\beta}(\omega) \\
& - \beta \dot{A}_{+}^{\alpha \gamma}(\bk, \omega) \dot{c}_{+}^{\gamma \beta}(\omega)~,\\
\dot{A}_{+}^{\alpha \beta}(\bk, \omega) & = \dot{c}_{+}^{\alpha \gamma}(\omega) J^{\gamma 
\beta}(\bk) - \beta \dot{c}_{+}^{\alpha \gamma}(\omega)\dot{\Lambda}_{+}^{\gamma 
\beta}(\bk, \omega)\\
& + \beta \dot{c}^{\alpha \gamma}_{+}(\omega) K^{\gamma \delta}(\omega) 
\dot{A}_{+}^{\delta \beta}(\bk, \omega)~.
\end{split}
\end{equation}

Here $\dot{c}_{+}^{\alpha \beta}(t-t') = \Theta(t-t') {\rm d}c^{\alpha \beta}(t-t')/{\rm 
d}t$ is the time-derivative of the single-site correlation $\langle S^{\alpha}(t) 
S^{\beta}(t')\rangle$, related to the single-site response function $g^{\alpha 
\beta}(t-t') = \langle \delta S_{i}^{\alpha}(t)/\delta \nu^{\beta}_{i}(t') \rangle$ by the FDT
$g^{\alpha \beta}(t-t') = - \beta \dot{c}_{+}^{\alpha \beta}(t-t')$.
$\dot{c}_{+}^{\alpha \beta}(\omega) = \int_{-\infty}^{\infty} {\rm d}t~{\rm e}^{i \omega 
t} \dot{c}_{+}^{\alpha \beta}(t-t')$ is the Fourier transform of $\dot{c}_{+}$.

Together with the relations $\dot{A}_{+}^{\alpha \beta}(\bk, \omega) = \dot{C}_{+}^{\alpha 
\gamma}(\bk, \omega) J^{\gamma \beta}(\bk)$, $\dot{\Lambda}_{+}^{\alpha \beta}(\omega, 
\bk) = J^{\alpha \gamma}(\bk, \omega) \dot{A}^{\gamma \beta}_{+}(\bk, \omega)$, 
Eqs.~\eqref{Cdot_Adot_Fourier} fix the correlations as functionals of $K$ and 
$\dot{c}_{+}$.

The solutions are
\begin{equation}\label{Cp_Ap_Lp}
\begin{gathered}
\dot{C}_{+}^{\alpha \beta}(\bk, \omega)  = \big[(\uu{\dot{c}_{+}}(\omega))^{-1} - \beta 
\uu{K}(\omega)  + \beta \uu{J}(\bk)\big]^{-1}{}^{\alpha \beta} ~,\\
\dot{A}_{+}^{\alpha \beta}(\bk, \omega)  = \dot{C}_{+}^{\alpha \gamma}(\bk, \omega) 
J^{\gamma \beta}(\bk) ~,\\
\dot{\Lambda}_{+}^{\alpha \beta}(\bk, \omega)  = J^{\alpha \gamma}(\bk) 
\dot{C}_{+}^{\gamma \delta}(\bk, \omega) J^{\delta \beta}(\bk)~.
\end{gathered}
\end{equation}

This result has a DMFT-like form, as expected due to the limit of large dimensionality: the
self-energy depends on the frequency $\omega$ but not on the momentum $\bk$~\cite{georges_rmp_1996}.

The self-consistency conditions can be imposed by requiring that the site-diagonal correlations
$\dot{C}_{+ii}^{\alpha \beta}(t-t')$, $\dot{A}_{+ii}^{\alpha \beta}(t-t')$,
$\dot{\Lambda}_{+ii}^{\alpha \beta}(t-t')$ computed by Fourier transformation from
Eqs.~\eqref{Cp_Ap_Lp} coincide with the corresponding quantities $\dot{c}_{+}^{\alpha \beta}(t-t')$,
$\dot{a}^{\alpha \beta}_{+}(t-t') = \Theta(t-t'){\rm d}a^{\alpha \beta}(t-t')/{\rm dt} =\Theta(t-t')
{\rm d}\langle S^{\alpha}_{i}(t) b^{\beta}_{i}(t')\rangle /{\rm d}t$, $\dot{\lambda}_{+}^{\alpha
\beta}(t-t') = \Theta(t-t') {\rm d}\lambda^{\alpha \beta}(t-t')/{\rm d}t = \Theta(t-t') {\rm
d}\langle b^{\alpha}_{i}(t) b^{\beta}_{i}(t')\rangle /{\rm d}t$ computed from the single-site
Langevin equation.

These conditions can be expressed in frequency space as:
\begin{equation}\label{self-consistency}
\begin{split}
\uu{\dot{c}_{+}}(\omega) & = \int_{-\pi}^{\pi} \frac{{\rm d}^{d}k}{(2\pi)^{d}} 
\big[(\uu{\dot{c}_{+}}(\omega))^{-1} - \beta \uu{K}(\omega)  + \beta 
\uu{J}(\bk)\big]^{-1}~,\\
\uu{\dot{a}_{+}}(\omega) & = \int_{-\pi}^{\pi} \frac{{\rm d}^{d}k}{(2\pi)^{d}} \big\{
\big[(\uu{\dot{c}_{+}}(\omega))^{-1} - \beta \uu{K}(\omega) \\
& \qquad \qquad \qquad + \beta \uu{J}(\bk)\big]^{-1} \uu{J}(\bk)\big\}~,\\
\uu{\dot{\lambda}_{+}}(\omega) & = \int_{-\pi}^{\pi} \frac{{\rm d}^{d}k}{(2\pi)^{d}} 
\big\{
\uu{J}(\bk)\big[(\uu{\dot{c}_{+}}(\omega))^{-1} - \beta \uu{K}(\omega) \\
& \qquad \qquad \qquad + \beta \uu{J}(\bk)\big]^{-1} \uu{J}(\bk)\big\}~.
\end{split}
\end{equation}

The three matching conditions for $\dot{c}_{+}$, $\dot{a}_{+}$, $\dot{\lambda}_{+}$ are equivalent
to each other, because Eqs.~\eqref{C_dot_plus},~\eqref{A_dot_plus}, and thus
Eqs.~\eqref{self-consistency}, are consistent by construction with the relations
\begin{equation}\label{dynamic_eom}
\begin{split}
& \dot{a}_{+}^{\alpha \beta}(\omega) = \dot{c}_{+}^{\alpha \gamma}(\omega) K^{\gamma 
\beta}(\omega)~,\\
& \dot{\lambda}_{+}^{\alpha \beta}(\omega)  = - \frac{1}{\beta} K^{\alpha \beta}(\omega) 
+ K^{\alpha \gamma}(\omega) \dot{c}_{+}^{\gamma \delta}(\omega) K^{\delta \beta}(\omega)~,
\end{split}
\end{equation}
which are automatically satisfied by the effective single-site Langevin equation (see
Sec.~\ref{s:properties_single_site}).
Thus, we can choose equivalently to impose any one of the three relations~\eqref{self-consistency}
to fix the self-consistency.

As a remark, note that the self-consistency equations fix the time derivatives and not directly the
correlation functions.
However, since the equal-time correlations must be equal to the static correlations, computed in
Sec.~\ref{s:statics}, we can deduce any time-dependent average by integrating over time (using the
static averages of Sec.~\ref{s:statics} as a boundary condition at $t = t'$).

\subsection{Derivation: effective single-site problem}
\label{s:effective_1_site}

To derive the dynamical results we used, as in Sec.~\ref{s:statics}, a combination of perturbation
theory and the cavity method~\cite{mezard_spin-glass, georges_rmp_1996}.
The starting point is a methodology analogue to the cavity approach used in 
Ref.~\onlinecite{mezard_spin-glass} for the dynamics of the SK model.
A fundamental idea of this cavity approach is that a single spin $\bS_{i}(t) = 
\bS_{i_{1}}(t)$ has only a weak effect on the trajectories of the other $N-1$ spins  
$\bS_{i_{2}}(t)$, ..., $\bS_{i_{N}}(t)$.
For each set of initial conditions $\bS_{i 0}$, $\bS_{i_{2}0}$, ..., $\bS_{i_{N}0}$ and 
for each realization of the noise $\bnu_{i}(t)$, $\bnu_{i_{2}}(t)$ ..., 
$\bnu_{i_{N}}(t)$, this allows to expand the trajectories of the $N-1$ spins $i_{2}$, 
..., $i_{N}$~\cite{mezard_spin-glass} 
\begin{equation}\label{response_series}
\begin{split}
 S_{j}^{\alpha}(t) &= S_{j}^{\alpha}{}'{}^{(i)}(t) + \sum_{k} \int_{0}^{t} {\rm d}t'~
G_{jk}^{(1) \alpha, \mu}{}'{}^{(i)}(t, t') J_{ki}^{\mu \beta} S_{i}^{\beta}(t')\\
& + \frac{1}{2} \sum_{k, l} \int_{0}^{t} {\rm d}t' \int_{0}^{t}{\rm d}t''~G^{(2)}_{j, k 
l}{}^{\alpha, \mu \nu}{}'{}^{(i)}(t; t', t'') \\
& \times J_{ki}^{\mu \beta} J_{li}^{\nu \gamma} S^{\beta}_{i}(t') S_{i}^{\nu}(t'') + ...~,
\end{split}
\end{equation}
and the magnetic field $b^{\alpha}_{i} = \sum_{j} J^{\alpha \beta}_{ij} S^{\beta}_{j}(t)$ acting on
site $\bS_{i}$,
\begin{equation}\label{b_i_response_theory}
\begin{split}
& b_{i}^{\alpha}(t) = \eta_{i}^{\alpha}(t) + \int_{0}^{t}{\rm d}t'~K^{(1) \alpha, 
\beta}(t, t') S_{i}^{\beta}(t') \\
& +\frac{1}{2} \int_{0}^{t} {\rm d}t' \int_{0}^{t} {\rm d}t''~K^{(2)\alpha, \beta 
\gamma}(t; t', t'') S_{i}^{\beta}(t') S_{i}^{\gamma}(t'') \\
& + ...
\end{split}
\end{equation}
as power series in the interaction between the spin $\bS_{i}$ and the cavity.

The zero order term in Eq.~\eqref{response_series}, denoted as $S_{j}{}^{\alpha \prime (i)}(t)$, is
a solution of the Langevin equations of the cavity system (Eqs.~\eqref{langevin} in absence of the
site $i$, that is, $\dot{\bS}^{\prime (i)}_{j} = -\bS_{j}^{\prime (i)} \times (\bS_{j}^{\prime (i)}
\times (\bN_{j}^{\prime (i)} + \bnu_{j}))$ with $\bN_{j}^{\prime (i)} = \bF(\bS_{j}^{\prime (i)}) +
\bb_{j}^{\prime (i)}$ and $b_{j}^{\alpha \prime (i)} = \sum_{k \neq i} J^{\alpha \beta}_{jk}
S^{\beta \prime (i)}_{k}$).
$G^{(n)}$ are the $n$-th order response functions
\begin{equation}
G_{j, k_{1} .. k_{n}}^{(n)\alpha, \mu_{1} ..\mu_{n}\prime (i)}(t; t_{1}, .., t_{n}) = \frac{\delta
S_{j}^{\alpha\prime(i)}(t)}{\delta \nu^{\mu_{1}}_{k_{1}}(t_{1}) .. \delta
\nu^{\mu_{n}}_{k_{n}}(t_{n})}~,
\end{equation}
calculated along the trajectory $S'_{j}{}^{\alpha}(t)$ ($G^{(n)}$ are trajectory-dependent).

In Eq.~\eqref{b_i_response_theory}, $\eta_{i}^{\alpha}(t) = \sum_{j} J_{ij}^{\alpha \beta}
S_{j}{}^{\beta \prime (i)}(t)$ is the magnetic field at zero order.
The kernels $K^{(n)}(t)$ describe the $n$-th order corrections to the instantaneous field due to
the interactions of the cavity with $\bS_{i}$, and are related to the response functions via:
\begin{equation}
\begin{split}
&K^{(n)\alpha, \mu_{1} .. \mu_{n}}(t; t_{1}, .., t_{n}) = \sum_{j, k_{1}, .., k_{n}}J_{ij}^{\alpha
\beta} \\
& \qquad \times G^{(n)\beta, \nu_{1} ..\nu_{n}\prime (i)}_{j, k_{1} .. k_{n}}(t; t_{1}, ..,
t_{n}) J_{k_{1}i}^{\nu_{1} \mu_{1}} .. J_{k_{n}i}^{\nu_{n} \mu_{n}}~.
\end{split}
\end{equation}

By construction $S_{j}{}^{\alpha \prime (i)}(t)$, $G^{(n)}$, $\bbeta_{i}(t)$ and 
$K^{(n)}$ do not depend explicitly on $\bS_{i0}$ and $\bnu_{i}$, but only on the noises  
$\bnu_{i_{2}}(t)$, ..., $\bnu_{i_{N}}(t)$ and the initial conditions $\bS_{i_{2}0}, ..., 
\bS_{i_{N}0}$ of the cavity.
For varying initial conditions and realizations of the noises, they become random 
variables.

In order to derive Eq.~\eqref{bi_t}, it is necessary to show that the complex 
expression~\eqref{b_i_response_theory} for the instantaneous field simplifies for $d \to 
\infty$.
In particular we need to derive four properties:
1) The nonlinear response terms have a negligible effect at large $d$, so that the series 
in Eq.~\eqref{b_i_response_theory} can be truncated keeping only the term $\eta_{i}^{\alpha}(t)$ and
the linear-response part $\int_{0}^{t}{\rm d}t' K^{(1)\alpha \beta}(t, t') S^{\beta}_{i}(t')$.
2) The linear response kernel $K^{(1)}(t, t')$ is effectively a deterministic quantity, 
with a negligible variance.
For any trajectory at temperature $T$, $K^{(1)\alpha \beta}(t, t')$ is then equal to its
thermal average $K^{\alpha \beta}(t-t') = \langle K^{(1)\alpha \beta}(t, t')\rangle$, up 
to negligible fluctuations.
3) The random field $\eta_{i}(t)$ has for $d \to \infty$ a Gaussian distribution, with 
mean $\beta l^{\alpha \beta}(t) S^{\beta}_{i0}$ and correlation $l^{\alpha \beta}(t-t')$.
4) The kernel $K^{\alpha \beta}(t-t')$ is related to $l^{\alpha \beta}(t-t')$ by the FDT.

These results are similar to those valid in the SK model~\cite{mezard_spin-glass}, and can be
derived by studying in a perturbative expansion the cumulants of $\bbeta_{i}(t)$, and the $K^{(n)}$.
Let us consider first the cumulants of $\bbeta_{i}(t)$.
In equilibrium, the initial conditions $\bS_{i_{2}0}$, ..., $\bS_{i_{N}0}$ are drawn with 
a distribution which, for each $\bS_{i0}$, is given by:
\begin{equation}\label{p_known_si0}
p(\bS_{i_{2}0}, .., \bS_{i_{N}0}|\bS_{i0}) = \frac{Z^{-1}}{P_{1{\rm eq}}(\bS_{i0})} {\rm 
e}^{-\beta H(\bS_{i0}, \bS_{i_{2}0}, ..., \bS_{i_{N}0})}~.
\end{equation}
Thus we can study the fluctuations of $\bbeta$ by analyzing the cumulants
\begin{equation}
C^{\prime \alpha_{1}, .., \alpha_{n}}_{\ell|\bS_{i0}} = \llangle 
\eta^{\alpha_{1}}_{i}(t_{1}) .. \eta^{\alpha_{\ell}}_{i}(t_{\ell}) \rrangle_{\bnu_{2}, .., 
\bnu_{N}; \bS_{i_{2}0}, .., \bS_{i_{N}0}|\bS_{i0}}~.
\end{equation}
Here $\llangle ... \rrangle_{\bnu_{2}, .., \bnu_{N}; \bS_{i_{2}0}, .., \bS_{i_{N}0}|\bS_{i0}}$ are
connected averages over $\bnu_{2}$, .., $\bnu_{N}$, and the initial conditions $\bS_{i_{2}0}$, ..,
$\bS_{i_{N}0}$, weighted with the distribution~\eqref{p_known_si0}.

For large $d$, $p(\bS_{i_{2}0}, .., \bS_{i_{N}0}|\bS_{i0})$ is close to the cavity distribution
$Z_{i}'{}^{-1}\exp[-\beta H'_{i}(\bS_{i_{2}0}, .., \bS_{i_{N}0})]$ (the difference between
$p(\bS_{i_{2}0}, .., \bS_{i_{N}0}|\bS_{i0})$ and the cavity distribution comes from the interaction
$-\sum_{j} J^{\alpha \beta}_{ij} S_{i0}^{\alpha} S_{j0}^{\beta}$ which is small for large $d$).

Thus we can at first ignore the corrections due to $\bS_{i0}$ and calculate the cumulants
\begin{equation}\label{C_prime_dynamical}
\begin{split}
& C^{\prime \alpha_{1}, .., \alpha_{n}}_{\ell}  = \llangle \eta^{\alpha_{1}}_{i}(t_{1}) 
.. 
\eta^{\alpha_{\ell}}_{i}(t_{\ell}) \rrangle^{\prime (i)}_{\bnu_{2}, .., \bnu_{N}; 
\bS_{i_{2}0}, .., \bS_{i_{N}0}}\\
& \quad = \sum_{j_{1}, .., j_{\ell}} J^{\alpha_{1} \beta_{1}}_{ij_{1}} .. 
J^{\alpha_{\ell} \beta_{\ell}}_{i j_{\ell}}\llangle S^{\beta_{1} \prime 
(i)}_{j_{1}}(t_{1})\\ 
& \quad \qquad \qquad \times ... \times S^{\beta_{\ell}\prime 
(i)}_{j_{\ell}}(t_{\ell})\rrangle^{\prime (i)}_{\bnu_{2}, .., \bnu_{N}; \bS_{i_{2}0}, .., 
\bS_{i_{N}0}}~,
\end{split}
\end{equation}
assuming that the averages over initial conditions are weighted simply with the Gibbs 
distribution $Z_{i}'{}^{-1}\exp[-\beta H'_{i}(\bS_{i_{2}0}, .., \bS_{i_{N}0})]$ of the 
cavity.
After this simplifications, the cumulants in Eq.~\eqref{C_prime_dynamical} reduce to averages
calculated in the equilibrium dynamics of the cavity system.

The $C'_{\ell}$ can be studied perturbatively by expanding at the same time the solutions 
of the cavity Langevin equations and the distribution of initial conditions $Z'_{i}{}^{-1} \exp
(-\beta H'_{i})$ as a series in the interactions $J_{jk}^{\alpha \beta}$ acting within the cavity
system.
To this end, we note that the expansion of the solutions $\bS^{\prime (i)}(t)$, for any fixed
realization of the initial conditions $\bS_{i_{2}0}$, ..., $\bS_{i_{N}0}$ and of the noises
$\bnu_{i_{2}}(t)$, ... $\bnu_{i_{N}}(t)$, can be represented in terms of tree diagrams of the form:
\begin{equation}\label{Sprime}
\centering 
\begin{split}
& \includegraphics[scale=1]{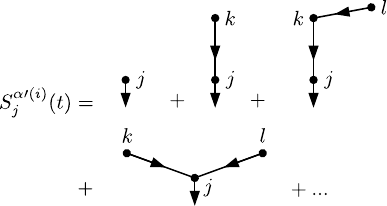}\\
& = S_{j}^{\alpha (0)}(t) + \sum_{k} \int_{0}^{t} {\rm d}t'~ {\cal G}_{j}^{(1)\alpha 
\beta}(t, t') J_{jk}^{\beta \gamma}S_{k}^{\gamma(0)}(t')\\
& \qquad + \sum_{k, l} \int_{0}^{t} {\rm d}t'~\int_{0}^{t'} {\rm d}t''~ {\cal 
G}_{j}^{(1)\alpha \beta}(t, t') J_{jk}^{\beta \gamma}\\
& \qquad \qquad \times {\cal G}_{k}^{(1) \gamma \delta}(t', t'') J_{kl}^{\delta \mu} 
S_{l}^{\mu(0)}(t'') \\
& \qquad  + \sum_{k, l} \int_{0}^{t} {\rm d}t'~\int_{0}^{t'}{\rm d}t''~{\cal 
G}_{j}^{(2)\alpha \beta \gamma}(t; t', t'') \\
& \qquad \qquad \times J_{jk}^{\beta \delta} S_{k}^{(0)\delta}(t')  J_{jl}^{\gamma \mu} 
S_{l}^{(0) \mu}(t'') + ...
\end{split}
\end{equation}

The zero-order term in Eq.~\eqref{Sprime} is simply the solution $S_{j}^{\alpha(0)}(t)$ of the
non-interacting Langevin equation $\dot{\bS}_{j} = - \bS_{j} \times (\bS_{j} \times (\bF_{j} +
\bnu_{j}(t)))$, in absence of exchange interactions.
The diagrammatic corrections can be visualized as a stream of processes acting one after the other
in time.
The first correction, represented in the second graph of~\eqref{Sprime}, describes the first-order
perturbation of the trajectory $\bS_{j}$ due to the field of the other spins $\bS_{k}$, and
involves the linear response function ${\cal G}_{j}^{(1)\alpha \beta}(t, t') = \delta
S_{j}^{(0)\alpha}(t)/\delta \nu_{j}^{\beta}(t')$ of the spin $\bS_{j}$, computed in the
non-interacting problem.
The third graph describes two perturbative processes acting in chain: the field due to
$\bS_{l}^{(0)}(t'')$ polarizes the spin $\bS_{k}(t')$ at a later time $t'$ and the correction
induced on $\bS_{k}(t')$ in turn polarizes $\bS_{j}(t)$ at $t$.
The last graph in Eq.~\eqref{Sprime} describes the second-order non-linear response of 
$\bS_{j}$ to the field induced on it by $\bS_{k}^{(0)}$ and $\bS_{l}^{(0)}$.

In general, terms of higher order involve an arbitrary number of interaction lines and non-linear
response functions of arbitrary order.
In the diagrammatic representation, we have used the following graphical conventions.
The solution $\bS_{j}^{(0)}(t)$ of the non-interacting Langevin equations of site $j$ is represented
by a dot with $1$ outgoing line, located at the lattice site $j$.
The non-interacting response functions ${\cal G}_{j}^{(n)\alpha, \beta_{1} .. \beta_{n}}(t;
t_{1}, .. t_{n})$ of $\bS_{j}$ are represented by dots with $n$ incoming lines and $1$ outgoing
line, located at $j$.
Finally, the interactions $J_{jk}^{\alpha \beta}$ are represented by lines connecting different
sites.

Every interaction $J_{jk}^{\alpha \beta}$ describes the effect of $\bS_{k}$ as a source of
perturbation to the trajectory $\bS_{j}$ or vice versa.
This implies that all lines in the graph connect one outgoing leg in a vertex to one ingoing leg of
a different vertex.
As a result, the arrows distinguishing ingoing and outgoing directions can be drawn 
directly on the midpoints of the interaction lines, as in the example~\eqref{Sprime}.

Every term represented in a graph has to be summed over all internal sites, and integrated over
all times.
Since all response functions ${\cal G}^{(n)}$ are causal, the outward line at each vertex has a
time $t$ which is larger than the times $t_{1}$, .., $t_{n}$ of the incoming lines.
The orientation of the arrows thus describe the direction of growing times.

The expansion~\eqref{Sprime} can be used to describe the perturbative solution of $\bS'_{j}(t)$ for
any given noise realization and for any fixed initial condition.
To determine cumulants of $\bbeta_{i}(t)$ the solution must be averaged over the $\bnu_{j}(t)$ and
the $\bS_{j0}$.
In order to calculate the cumulant $\llangle
\eta^{\alpha_{1}}_{i}(t_{1})..\eta^{\alpha_{\ell}}_{i}(t_{\ell})\rrangle$ perturbatively we need to
draw $\ell$ tree diagrams of the type~\eqref{Sprime}, representing the solutions $S_{j_{1}}^{\prime
\beta_{1} (i)}(t_{1})$, ..., $S_{j_{n}}^{\prime \beta_{\ell} (i)}(t_{\ell})$, add $\ell$ lines
representing the factors $J_{ij_{1}}^{\alpha_{1}\beta_{1}}$, ... $J_{ij_{\ell}}^{\alpha_{n}
\beta_{\ell}}$ in Eq.~\eqref{C_prime_dynamical}, and average the resulting product of graphs.

The perturbative terms needed in this expansion involve averages of products of $\bS^{(0)}_{j}(t)$,
 of non-interacting response functions ${\cal G}^{(n)}$, and of initial conditions $\bS_{j0}$,
calculated in the nonperturbed problem (with $J_{jk}^{\alpha \beta} = 0$).
As in the linked-cluster expansion, the perturbative terms are conveniently handled by
separating these averages, which we denote as $\langle \bS_{j_{1}}^{(0)} ...\bS_{j_{n}}^{(0)} ..
{\cal G}_{k_{1}}^{(n_{1})} .. {\cal G}_{k_{m}}^{(n_{m})} ... \bS_{l_{1}0}
.. \bS_{l_{p}0}\rangle^{(0)}$, as sums of connected averages (cumulants).

This leads to an expansion analogue to the LCE in which the vertices are connected averages of the
type $\llangle \bS_{j}^{(0)} ... \bS_{j}^{(0)} .. {\cal G}_{j}^{(n_{1})} .. {\cal G}_{j}^{(n_{m})}
... \bS_{j0} ..\bS_{j0}\rrangle^{(0)}$, computed at zero order (with $J_{ij}^{\alpha \beta} = 0$).
These vertices are fully local because at zero order all spins are non-interacting, and all
connected averages which are not site-diagonal vanish.
In the following the vertices will be represented graphically in the form:
\begin{equation}\label{vertices}
\includegraphics[scale=1]{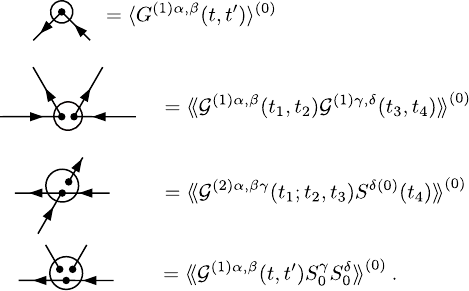}
\end{equation}

Here the dots with incoming and outgoing arrows have the same meaning as before: they represent the
solutions of the non-interacting Langevin equations and the $n$-th order non-interacting response
functions ${\cal G}_{j}^{(n)}$.
Dots connected to lines without arrows, instead, are introduced to represent factors of
$S^{\alpha}_{j0}$, where $\bS_{j0}$ is the initial condition of $\bS_{j}$ at time $t = 0$.
The circle stands for a connected average of all quantities inside it, over $\bnu_{j}(t)$, and
$\bS_{j0}$, calculated using the zero-order Langevin dynamics and the zero-order ensemble
$P(\bS_{j0}) = \rho(\bS_{j0}) = \exp[-\beta V(\bS_{j0})]/\int_{\bS} \exp [-\beta V(\bS)]$.
Since the system is homogeneous, the value of the vertices does not depend on the site $j$.
in addition, since $P(\bS_{j0}) = \rho(\bS_{j0}) $ is the equilibrium probability for the
non-interacting Langevin equation, the vertices are time-translation invariant.

The perturbative expansion involves in general vertices of the type~\eqref{vertices}, with
arbitrarily many ``dots``, representing the connected average of arbitrary products of $S^{(0)}(t)$,
${\cal G}^{(n)}$ and $\bS_{i0}$.

Using this graphical representation, the first few terms of the cumulant $C_{2}^{\prime \alpha
\beta}(t-t') = \llangle \eta^{\alpha}_{i}(t) \eta^{\beta}_{i}(t') \rrangle^{\prime
(i)}_{\bnu_{i_{2}}, ..., \bnu_{i_{N}}; \bS_{i_{2}0}, .., \bS_{i_{N}0}}$ can be represented as:
\begin{equation}\label{l_diagrams}
\begin{split}
& \includegraphics[scale=1]{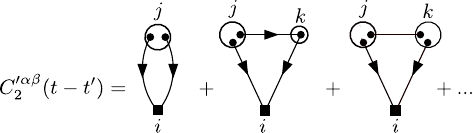}\\
& = \sum_{j}J_{ij}^{\alpha \gamma} J_{ij}^{\beta \delta} \langle 
S_{j}^{\beta(0)}(t) S_{k}^{\gamma(0)}(t')\rangle^{(0)}\\
& + \sum_{j, k} J_{ij}^{\alpha \gamma} J_{jk}^{\delta \mu} J_{ik}^{\beta \nu} 
\int_{0}^{t'} {\rm d}t'' \Big[\langle S_{j}^{(0) \gamma}(t) S_{j}^{(0)\delta}(t'') 
\rangle^{(0)} \\
& \qquad \times \langle {\cal G}^{(1)\mu \nu}_{k}(t', t'')\rangle^{(0)}\Big]\\
& + \sum_{j, k} \beta J_{ij}^{\alpha \gamma} J_{jk}^{\delta \mu} J_{ik}^{\beta \nu} 
\langle S^{(0) \gamma}_{j}(t) S_{j0}^{\delta}\rangle^{(0)}\\
& \qquad \times \langle S^{(0) \gamma}_{k}(t) S_{k0}^{\delta }\rangle^{(0)} + ...
\end{split} 
\end{equation}

Here the two ``external'' lines connected to the origin $i$ correspond to the factors
$J_{ij}^{\alpha \beta}$, $J_{ik}^{\alpha \beta}$ in the definition of the local field
$\eta^{\alpha}_{i}(t) = \sum_{j} J_{ij}^{\alpha \beta} S_{j}^{\beta \prime (i)}(t)$.
The inner part of the graphs describes the average of the product of $S_{j}^{\beta \prime (i)}(t)$
and $S_{k}^{\beta \prime (i)}(t')$.
The third graph contains a term coming from the first-order expansion of the distribution
$Z'_{i}{}^{-1} {\rm e}^{-\beta H'_{i}(\bS_{i_{2}0}, .., \bS_{i_{N}0})}$ of the initial conditions.
The contribution of the corresponding interaction is represented by a "static line" (illustrated in
the diagram as a line without an arrow).

At higher order the diagrams involve arbitrary numbers of lines with arrows (representing the
perturbative solution of the Langevin equations) and without arrows (representing the expansion of
the Gibbs distribution of the cavity).
The expansion of cumulants $C'_{\ell}$ with arbitrary $\ell$ is given by diagrams with the same
structure, but with $\ell$ lines connected to the origin $i$.

We can now discuss the scaling of diagrams in the large-$d$ limit.
Although the vertices of the dynamic problem are more complex than those of the static calculation,
the power counting for $d \to \infty$ is the same.
As a result, it turns out that the only finite cumulant for $d \to \infty$ is  $C_{2}^{\prime \alpha
\beta}(t-t')$, because it is represented by graphs in which two lines are attached to $i$.
All higher cumulants of $\bbeta(t)$, instead, are represented by graphs in which more than two
lines are attached to the origin $i$.
Since by the cavity construction the internal sites are summed over all lattice sites except $i$,
this implies, as in Sec.~\ref{s:statics}, that all cumulants beyond the second are suppressed for
$d \to \infty$.
This shows that $\bbeta_{i}(t)$ has a Gaussian distribution in large dimension.
We identify the cumulant $C_{2}^{\prime \alpha \beta}(t-t')$ with the correlation function
$l^{\alpha \beta}(t-t')$ describing the fluctuation of the noise in the effective single-site
dynamics.

The same power counting analysis which has been used for correlations of $\bbeta$ can be applied to
any cumulant involving both $\bbeta$ and the kernels $K^{(n)}$.
The result is the same: the only terms of order 1 are those for which the corresponding perturbative
graphs have two lines, and no more, connected to $i$.
Using this, we see that the average of the linear kernel $K^{(1)\alpha \beta}(t, t')$ is of order 1,
because it has an expansion given by diagrams
\begin{equation}\label{K_diagrams}
\begin{split}
& \includegraphics[scale=1]{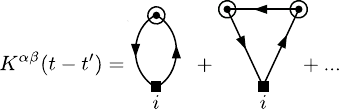}\\
& = \sum_{j} J^{\alpha \gamma}_{ij} J_{ij}^{\beta \delta} \langle {\cal G}^{(1)\gamma 
\delta}(t, t') \rangle^{(0)}\\
& + \sum_{j, k} \int_{t'}^{t} {\rm d}t''~\big(J^{\alpha \gamma}_{ij} J_{jk}^{\delta 
\mu} J_{ki}^{\nu \beta} \\
& \qquad \times \langle {\cal G}_{j}^{(1)\gamma \delta}(t, t'') \rangle^{(0)}\langle 
{\cal G}_{k}^{(1)\mu \nu}(t, t'') \rangle^{(0)}\big) + ...
\end{split}
\end{equation}

All higher response functions ($K^{(n)}$, $n \geq 2$) are, instead, negligible for $d \to \infty$,
because their averages and cumulants are represented by graphs in which more than two lines are
attached to $i$.
In particular, the variance of $K^{(1)}$ is given by a graph with four lines connected to the origin
$i$, and is negligible.
Thus $K^{(1)}$ is deterministic: it has negligible fluctuations in the limit $d \to \infty$.

In addition, we note that since the fluctuation $l^{\alpha \beta}(t-t') = C_{2}^{\prime
\alpha \beta}(t-t')$ and the kernel $K^{\alpha \beta}(t-t')$ are related to the correlation and the
linear response functions of the cavity system in equilibrium.
Thus it can be shown that they are related by the FDT, Eq.~\eqref{FDT_Kl}~(see
appendix~\ref{a:FP_and_FDT}).

To complete the calculation we have to discuss one last point: the fact that the correct averages
over initial conditions should be computed with the weight~\eqref{p_known_si0} and not with the
cavity distribution $Z_{i}^{\prime-1} \exp (-\beta H'_{i})$.
The difference between the two distributions leads to negligible corrections to all cumulants, a
part from one: the mean $\langle \bbeta(t)\rangle_{\nu_{2}, .., \nu_{N}, \bS_{i_{2}}, ..,
\bS_{i_{N}}|\bS_{i0}}$.
In fact, the average of $\bbeta(t)$ computed with the cavity distribution is zero by inversion
symmetry.
The mean $\langle \bbeta(t)\rangle_{\nu_{2}, .., \nu_{N}, \bS_{i_{2}}, .., \bS_{i_{N}}|\bS_{i0}}$
at fixed $\bS_{i0}$, instead, receives a O$(1)$ contribution.
For $d \to \infty$ this contribution can be calculated expanding the
distribution~\eqref{p_known_si0} to first order in the coupling $\beta \sum_{j}J^{\alpha \beta}_{ij}
S^{\alpha}_{i0} S^{\beta}_{j0}$.
As a result we find:
\begin{equation}\label{eta_mean}
\begin{split}
&\langle \eta^{\alpha}_{i}(t) \rangle_{\bnu_{i_{2}}, .. \bnu_{i_{N}}, \bS_{i_{2}0}, ..., 
\bS_{i_{N}0}|\bS_{i0}} \\
& \substack{\to \\ d\to \infty} \sum_{j, k} \beta J^{\alpha \gamma}_{ij} J^{\beta 
\delta}_{ik} \llangle S^{\gamma \prime (i)}_{j}(t) S_{k0}^{\delta}\rrangle^{\prime 
(i)}_{\bnu_{i_{2}}.. \bnu_{i_{N}},
\bS_{i_{2}0}..\bS_{i_{N}0}}\\
& = \beta l^{\alpha \beta}(t) S^{\beta}_{i0}~.
\end{split}
\end{equation}

Diagrammatically this term is represented by:
\begin{equation}
\centering 
\includegraphics[scale=1]{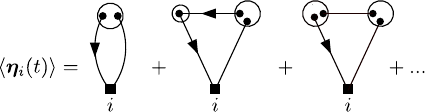}~,
\end{equation}
which, again, are graphs in which only two lines are connected to the origin.
One of the two lines is now a static line, representing the term $\beta J^{\alpha \beta}_{ij}
S^{\alpha}_{i0}S^{\beta}_{j0}$ in the distribution~\eqref{p_known_si0}.

In all other terms, the difference between the distribution~\eqref{p_known_si0} and the 
cavity distribution is negligible.
In particular, the variance $l^{\alpha \beta}(t-t')$ and the kernel $K^{\alpha 
\beta}(t-t')$ calculated above remain exact for $d \to \infty$.
This can be seen systematically expanding 
\begin{equation}
\begin{split}
& \llangle \eta^{\alpha_{1}} .. \eta^{\alpha_{n}} K^{(n_{1})} ...K^{(n_{m})} 
\rrangle_{\bnu_{i_{2}}, ..., \bnu_{i_{N}}, \bS_{i_{2}0}, 
..., \bS_{i_{N}0}|\bS_{i0}}\\
& = \sum_{m=0}^{\infty} \frac{\beta^{m}}{m!} \sum_{j_{1}, .., j_{m}} J_{i 
j_{1}}^{\beta_{1} \gamma_{1}} ... J_{i j_{m}}^{\beta_{m} \gamma_{m}} S_{i0}^{\beta_{1}} .. 
S_{i0}^{\beta_{m}} \llangle S_{j_{1}0}^{\gamma_{1}} .. S_{j_{m}0}^{\gamma_{m}}\\
& \qquad \times \eta^{\alpha_{1}} .. \eta^{\alpha_{n}} K^{(n_{1})} ...K^{(n_{m})} 
\rrangle^{\prime (i)}_{\bnu_{i_{2}} .. \bnu_{i_{N}}, \bS_{i_{2}0}, \bS_{i_{N}0}}
\end{split}
\end{equation}
as a series of connected averages with insertions of $\bS_{j0}$ calculated in the cavity 
ensemble.
Every additional interaction with $\bS_{i0}$ carries an additional external static line 
attached to $i$, and suppresses the graph.

This concludes the derivation.
Introducing $\zeta_{i}^{\alpha}(t) = \eta_{i}^{\alpha}(t) - \beta l^{\alpha \beta}(t)$, 
and using Eq.~\eqref{b_i_response_theory} we find that the instantaneous magnetic field 
is for large $d$: $b^{\alpha}_{i}(t) = \zeta_{i}^{\alpha}(t) + \beta l^{\alpha \beta}(t) 
S_{i0}^{\beta} + \int_{0}^{t} {\rm d}t~K^{\alpha \beta}(t-t') S_{i}^{\beta}(t')$, with  
$\bdelta$ a zero-mean Gaussian noise having $\langle \zeta^{\alpha}_{i}(t) 
\zeta^{\beta}_{i}(t')\rangle = l^{\alpha \beta}(t-t')$.

\subsection{General properties of the single-site dynamical problem}
\label{s:properties_single_site}

The non-Markovian Langevin equations~\eqref{1_site_langevin},~\eqref{bi_t} derive from the
equilibrium dynamics of the system and thus satisfy time-translation invariance (TTI) and the
fluctuation-dissipation theorem.
As a result, the single-site response function $g^{\alpha \beta}(t-t') = \langle \delta
S^{\alpha}_{i}(t)/\delta \nu^{\beta}_{i}(t') \rangle $ is related to the single-site correlation
$c^{\alpha \beta}(t-t') = \langle S_{i}^{\alpha}(t) S_{i}^{\beta}(t') \rangle$ via the FDT
$g^{\alpha \beta}(t-t') = - \beta \Theta(t-t') {\rm d}c^{\alpha \beta}(t-t')/{\rm d}t  =
\dot{c}^{\alpha \beta}_{+}(t-t')$.
The TTI is not immediately manifest in the single-site equations because the term $\beta 
l^{\alpha \beta}(t) S^{\beta}_{i0}$ and the integral $\int_{0}^{t}{\rm d}t'~K^{\alpha 
\beta}(t-t') S^{\beta}_{i}(t')$ seem to depend explicitly on the choice of the origin of 
time $t_{0} = 0$.
However, the two non-invariant terms (the truncation of the integral at the lower limit 
and the field $\beta l^{\alpha \beta}(t) S^{\beta}_{i0}$) compensate each other, leading 
to results which do not depend on the origin of time.
This can be verified, for example, by mapping the Langevin dynamics to an equivalent 
Markovian problem, in which the colored noise is mimicked by a bath of harmonic 
oscillators~\cite{cugliandolo_jpsj_2000, maimbourg_prl_2016}.
Integrating out the bath leads to an equation of motion in which $\bb_{i}(t)$ has the 
form~\eqref{bi_t}.

In addition to TTI and the FDT, another essential property of the single-site Langevin problem is
that it is consistent with the static results of Sec.~\ref{s:statics}.
All equal-time correlations of $\bS_{i}(t)$ and $\bb_{i}(t)$ coincide with the static correlations
described by the distribution~\eqref{P_SB1}, when the matrix $L^{\alpha \beta}$ is identified with
the instantaneous correlation $l^{\alpha \beta}(0)$.
This can be seen conveniently by studying the correlations at time $t = 0$, where the
field~\eqref{bi_t} reduces to $b^{\alpha}_{i}(0) = \zeta_{i}^{\alpha}(0) + \beta l^{\alpha \beta}(0)
S^{\beta}_{i0}$.
$\zeta_{i}^{\alpha}(0)$ is a Gaussian field with $\langle \zeta_{i}^{\alpha}(0)
\zeta_{i}^{\beta}(0)\rangle = l^{\alpha \beta}(0)$.
Identifying $l^{\alpha \beta}(0)= L^{\alpha \beta}$ and using that $\bS_{i0}$ is weighted by the
distribution~\eqref{P1}, it is simple to verify that the joint distribution of $\bS_{i0}$ and
$\bb_{i}(0)$ is exactly the static distribution $P(\bS_{i}, \bb_{i})$ in Eq.~\eqref{P_SB1}.
TTI implies that $P(\bS_{i}(t), \bb_{i}(t))$ remains the same at all later times.

We finally discuss two equations of motion, which are useful in the subsequent derivations.
Using the property of Gaussian averages~\cite{zinn-justin_qft} $\langle S^{\alpha}_{i}(t)
\zeta_{i}^{\beta}(t') \rangle = \int_{0}^{\infty} {\rm d}t''~l^{\beta \gamma}(t'-t'') \langle \delta
S^{\alpha}_{i}(t)/\delta \zeta^{\gamma}_{i}(t'')\rangle $ and the fluctuation-dissipation relations
$g^{\alpha \beta}(t-t') = -\beta \dot{c}_{+}^{\alpha \beta}(t-t')$, $K^{\alpha \beta}(t-t') =
-\beta \Theta(t-t') {\rm d}l^{\alpha \beta}(t-t')/{\rm d}t$ we find for $t \geq t'$:
\begin{equation}\label{a_c_dynamics}
\begin{split}
a^{\alpha \beta}(t-t') & = \langle S^{\alpha}_{i}(t) b^{\beta}_{i}(t') \rangle = \beta  c^{\alpha
\gamma}(t-t')l^{\gamma \beta}(0)  \\
& + \int_{t'}^{t} {\rm d}t'' g^{\alpha \gamma}(t-t'') l^{\gamma \beta}(t''-t')~,
\end{split}
\end{equation}
\begin{equation}\label{lambda_c_dynamics}
\begin{split}
& \lambda^{\alpha \beta}(t-t') = \langle b^{\alpha}_{i}(t) b^{\beta}_{i}(t') \rangle = l^{\alpha
\beta}(t-t') \\
& + \beta^{2}l^{\alpha \gamma}(t-t') c^{\gamma \delta}(0) 
l^{\delta \beta}(0) \\
& + \beta \int_{t'}^{t} {\rm d}t''~K^{\alpha \gamma}(t-t'') c^{\gamma \delta}(t''-t') 
l^{\delta \beta}(0)\\
& + \int_{t'}^{t} {\rm d}t'' \int_{t'}^{t''}{\rm d}t'''~\big[K^{\alpha \gamma}(t-t'') \\
& \qquad \times g^{\gamma \delta}(t''-t''') l^{\delta \beta}(t'''-t')\big] ~.
\end{split}
\end{equation}

The $\lambda^{\alpha\beta}(t-t')$ correlation in the region $t < t'$ follows from
Eq.~\eqref{lambda_c_dynamics} by symmetry.

For equal times, these relations reduce to the static equations~\eqref{static_eom}.
Taking a time derivative and using the fluctuation-dissipation relations we find after a 
Fourier transform Eqs.~\eqref{dynamic_eom}.
We note also that Eqs.~\eqref{a_c_dynamics},~\eqref{lambda_c_dynamics} are fully consistent with the
time-translation invariance of the colored single site dynamics.

\subsection{Self-consistency equations}
\label{s:self_consistency_equations}

To conclude, we discuss the derivation of Eqs.~\eqref{C_dot_plus},~\eqref{A_dot_plus}, which are
needed to fix the self-consistency of the single-site problem.
When $i = j$, Eqs.~\eqref{C_dot_plus} and~\eqref{A_dot_plus} can be shown to be equivalent to
Eqs.~\eqref{dynamic_eom}, and, thus, are satisfied automatically.

For non-coincident sites $i \neq j$, the correlations can be studied by a two-cavity method, with
two cavities at $i$ and $j$.
In this framework, the correlations are analyzed by studying an effective 2-body Langevin
equation, in which the effects of the remaining $N - 2$ spins are described via a fluctuating bath.

At leading order for $d \to \infty$, the effective 2-spin problem is equivalent to two independent
copies of the single-site Langevin equations~(Eqs.~\eqref{1_site_langevin} and~\eqref{bi_t}).
Nontrivial correlations between the two sites arise from the leading corrections, which are of
order O$(d^{-\ell_{ij}/2})$, and which couple the motion of the two spins.
These corrections are due to: 1) the direct contribution of $\bS_{i}$ to the field $\bb_{j}$ (and,
vice versa of $\bS_{j}$ to $\bb_{i}$) and 2) indirect contributions, mediated by the bath.

A diagrammatic analysis analogue to that discussed in Sec.~\ref{s:effective_1_site} shows that the
2-site equations can be written in the form:
\begin{equation}\label{2_site_langevin}
\begin{split}
& \dot{\bS}_{i} = - \bS_{i} \times (\bS_{i} \times (\bF_{i} + \bb_{i}(t) + \bnu_{i}(t)))~,\\
& \dot{\bS}_{j} =  -\bS_{j} \times (\bS_{j} \times (\bF_{j} + \bb_{j}(t) + \bnu_{j}(t)))~,
\end{split}
\end{equation}
where the components of the fields $b^{\alpha}_{i}(t)$ and $b^{\alpha}_{j}(t)$ are:
\begin{equation}\label{b_ij}
\begin{split}
& b^{\alpha}_{i}(t) = \zeta_{i}^{\alpha}(t) +  \beta l^{\alpha \beta}(t) S^{\beta}_{i0} +
\int_{0}^{t}{\rm d}t'~K^{\alpha \beta}(t-t') S^{\beta}_{i}(t')\\
& + J_{ij}^{\alpha \beta} S^{\beta}_{j}(t) + \beta r_{ij}^{\alpha \beta}(t)
S^{\beta}_{j0}  + \int_{0}^{t}{\rm d}t'~
R_{ij}^{\alpha \beta}(t-t') S^{\beta}_{j}(t')~,\\
& b^{\alpha}_{j}(t) = \zeta_{j}^{\alpha}(t) + \beta l^{\alpha \beta}(t) S^{\beta}_{j0} +
\int_{0}^{t}{\rm d}t'~ K^{\alpha \beta}(t-t') S^{\beta}_{j}(t') \\
& + J_{ji}^{\alpha \beta} S^{\beta}_{i}(t) + \beta r^{\alpha \beta}_{ji}(t) S^{\beta}_{i0}
+ \int_{0}^{t}{\rm d}t'~ R_{ji}^{\alpha \beta}(t-t') S^{\beta}_{i}(t')~,
\end{split}
\end{equation}

Here, $\bzeta_{i}(t)$ and $\bzeta_{j}(t')$ are Gaussian noises with zero mean.
The site-diagonal correlations $\langle \zeta^{\alpha}_{i}(t) \zeta^{\beta}_{i}(t')\rangle = \langle
\zeta^{\alpha}_{j}(t) \zeta^{\beta}_{j}(t')\rangle$ are of order 1 and are equal, up to negligible
corrections to the correlation $l^{\alpha \beta}(t-t')$ defining the single-site Langevin equation.
In addition, the random fields have a correlation $\langle \zeta^{\alpha}_{i}(t)
\zeta^{\beta}_{j}(t')\rangle = r^{\alpha \beta}_{ij}(t-t')$.

The kernels, similarly, have site-diagonal parts $K^{\alpha \beta}(t-t')$ which are equal for $d$
large to the kernels $K^{\alpha \beta}(t-t')$ entering the single-site equation~\eqref{bi_t}.
In addition, there are off-diagonal response terms $R_{ij}^{\alpha \beta}(t-t')$,
$R_{ji}^{\alpha \beta}(t-t')$.

Diagrammatically, the off-diagonal corrections encoded in Eq.~\eqref{b_ij} correspond to graphs of
the form:
\begin{equation*}
\includegraphics[scale=1]{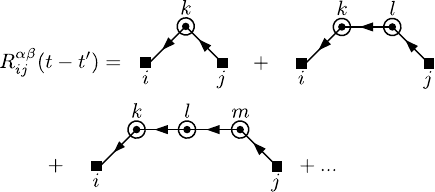}
\end{equation*}
\begin{equation*}
\includegraphics[scale=1]{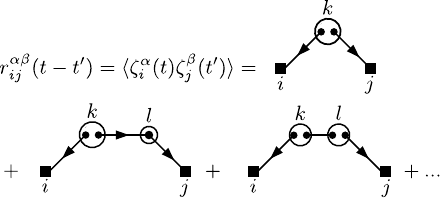}
\end{equation*}
\begin{equation*}
\includegraphics[scale=1]{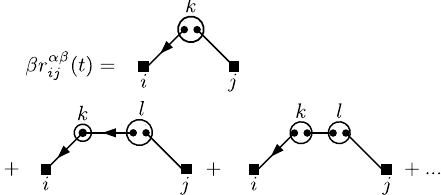}
\end{equation*}

The diagrams imply that correlation and response functions $r_{ij}$, $R_{ij}$ are of order
$d^{-\ell_{ij}/2}$, with $\ell_{ij}$ the Manhattan distance between $i$ and $j$.
At this order, no other diagram contributes, and thus Eqs.~\eqref{b_ij} are exact.

The intersite terms are related via $r_{ij}^{\alpha \beta}(t-t') = \sum_{k, l \neq i, j}
J^{\alpha \gamma}_{ik} J^{\beta \delta}_{jl} \llangle S_{k}^{\gamma \prime \prime 
(ij)}(t) S_{l}^{\delta \prime \prime (ij)}(t') \rrangle^{\prime \prime (ij)}$, 
$R_{ij}^{\alpha \beta}(t-t') = \sum_{k, l \neq i, j} J^{\alpha \gamma}_{ik} J^{\beta 
\delta}_{jl} \llangle G^{(1) \gamma, \delta \prime \prime (ij)}(t, t')_{kl} 
\rrangle^{\prime \prime (ij)}$, $R_{ji}^{\beta \alpha}(t-t') = \sum_{k, l \neq i, j} 
J^{\alpha \gamma}_{ik} J^{\beta \delta}_{jl} \llangle G^{(1)\delta, \gamma \prime \prime 
(ij)}_{lk}(t, t') \rrangle^{\prime \prime (ij)}$ to the equilibrium correlations $\llangle 
S_{k}^{\gamma \prime \prime (ij)}(t) S_{l}^{\delta \prime \prime (ij)}(t') 
\rrangle^{\prime \prime (ij)}$ and to the response functions $\llangle G^{(1) \gamma, 
\delta \prime \prime (ij)}(t, t')_{kl} \rrangle^{\prime \prime (ij)}$ of the 2-cavity 
system.
Thus they satisfy the fluctuation-dissipation relations:
\begin{equation}
\begin{split}
R_{ij}^{\alpha \beta}(t-t') & = - \beta \Theta(t-t') \frac{{\rm d}}{{\rm 
d}t}r_{ij}^{\alpha \beta}(t-t')~,\\
R_{ji}^{\beta\alpha}(t-t') & = - \beta \Theta(t-t') \frac{{\rm d}}{{\rm d}t} 
r_{ji}^{\beta \alpha}(t-t')~.
\end{split}
\end{equation}

The 2-site equations~\eqref{2_site_langevin},~\eqref{b_ij} allows to calculate the correlations
$C_{ij}^{\alpha \beta}(t-t')$, $A^{\alpha \beta}_{ij}(t-t')$, $\Lambda^{\alpha \beta}_{ij}(t-t')$.
The explicit calculations are presented in appendix~\ref{a:dynamical_two_point_functions}, and use
that for $d$ large, the equations for the two spins can be solved at first order in $r_{ij}$,
$R_{ij}$, and $J_{ij}$.
As a result we find:
\begin{equation}\label{C_ij_plus}
\begin{split}
\dot{C}_{+ij}^{\alpha \beta}(\omega) & = - \beta \dot{c}_{+}^{\alpha \gamma}(\omega) 
\big(J_{ij}^{\gamma \delta} + R_{ij}^{\gamma \delta}(\omega)\big)\dot{c}_{+}^{\delta 
\beta}(\omega)~,
\end{split}
\end{equation}
\begin{equation} \label{A_ij_plus_Lambda_ij_plus}
\begin{split}
\dot{A}_{+ij}^{\alpha \beta}(\omega) & = \dot{c}^{\alpha \gamma}_{+}(\omega) (J^{\gamma 
\beta}_{ij} + R_{ij}^{\gamma \beta}(\omega)) + \dot{C}_{+ij}^{\alpha \gamma}(\omega) 
K^{\gamma \beta}(\omega)~,\\
\dot{\Lambda}^{\alpha \beta}_{+ij}(\omega) & = -\frac{1}{\beta} R_{ij}^{\alpha 
\beta}(\omega) + K^{\alpha \gamma}(\omega) \dot{C}^{\gamma \delta}_{+ij}(\omega) 
K^{\delta \beta}(\omega)\\
& + (J_{ij}^{\alpha \gamma} + R_{ij}^{\alpha \gamma}(\omega)) 
\dot{c}_{+}^{\gamma \delta}(\omega) K^{\delta \beta}(\omega)\\
& + K^{\alpha \gamma}(\omega) \dot{c}_{+}^{\gamma \delta}(\omega) (J_{ij}^{\delta \beta} 
+ R_{ij}^{\delta \beta}(\omega))~.
\end{split}
\end{equation}

Eq.~\eqref{C_ij_plus} has a very simple interpretation: the response function at non-coincident
sites $i \neq j$ receives two contributions of the same order of magnitude ($\approx
d^{-\ell_{ij}/2}$): one from the direct interaction $J_{ij}^{\alpha \beta}$ and one mediated by the
cavity.

Combining Eqs.~\eqref{C_ij_plus},~\eqref{A_ij_plus_Lambda_ij_plus}, and using 
Eqs.~\eqref{dynamic_eom} we find relations which do not depend on $R_{ij}$, and which are 
valid both for $i = j$ and for $i \neq j$:
\begin{equation}
\begin{split}
& \dot{C}_{+ij}^{\alpha \beta}(\omega) - \delta_{ij} \dot{c}^{\alpha \beta}_{+}(\omega) = 
\beta \dot{C}_{+ij}^{\alpha \gamma}(\omega) K^{\gamma \delta}(\omega) \dot{c}_{+}^{\delta 
\beta}(\omega) \\
& \qquad \qquad - \beta \dot{A}^{\alpha \gamma}_{+ij}(\omega) \dot{c}^{\gamma 
\beta}_{+}(\omega) 
~,\\
& \dot{A}^{\alpha \beta}_{+ij}(\omega) = \dot{c}_{+}^{\alpha \gamma} J_{ij}^{\gamma 
\beta} - \beta \dot{c}_{+}^{\alpha \gamma}(\omega)\dot{\Lambda}_{+ij}^{\gamma 
\beta}(\omega) \\
& \qquad \qquad + \beta \dot{c}_{+}^{\alpha \gamma}(\omega) K^{\gamma \delta}(\omega) 
\dot{A}_{+ij}^{\delta \beta}(\omega)~.
\end{split}
\end{equation}

These are equivalent to Eqs.~\eqref{Cdot_Adot_Fourier}, and in real time, to 
Eqs.~\eqref{C_dot_plus},~\eqref{A_dot_plus}.

\section{Glass transition}
\label{s:glass_transition}

The dynamical analysis allows to distinguish under which conditions the paramagnetic 
phase has a liquid- or a glass-like behavior. 
The liquid and the glass phase can be discerned by the large-time behavior 
of the time-dependent correlations: $\lim_{|t-t'| \to \infty} l^{\alpha \beta}(t-t') = 
L_{2}^{\alpha \beta}$, $\lim_{|t-t'| \to \infty} c^{\alpha \beta}(t-t') = c_{2}^{\alpha 
\beta}$,  $\lim_{|t-t'| \to \infty} C_{ij}^{\alpha \beta}(t-t') = C_{2ij}^{\alpha 
\beta}$, $\lim_{|t-t'| \to \infty}A_{ij}^{\alpha \beta}(t-t') = A_{2ij}^{\alpha 
\beta}$....
In the ergodic phase all correlations decay to zero at large times.
In the glass phase, instead, $L_{2}, c_{2}$, $\lambda_{2}$, $C_{2ij}$ are different from 
zero.
In particular, $c_{2}^{\alpha \beta}$, the large-time limit of the spin correlation, can 
be identified with the Edwards-Anderson order parameter: $c_{2}^{\alpha \beta} = q^{\alpha 
\beta}_{\rm EA}$.

To locate the occurence of a glass transition, it is not necessary to solve the dynamics 
in detail.
In analogy with the theory of supercooled liquids in $d \to 
\infty$~\cite{maimbourg_prl_2016}, the order parameters $L_{2}$, $c_{2}$, $a_{2}$, 
$\lambda_{2}$ are fixed by closed equations which make reference only to the static 
averages and to the long-time limit of the correlations, and not to their transient 
behavior.

To derive these equations, we need to discuss the $|t-t'| \to \infty$ limit of both the 
single site problem and the self-consistency relations.

We begin by discussing the single site problem.
When ergodicity is broken and $L_{2}^{\alpha\beta} \neq 0$, the field $\bb_{i}(t)$ 
develops a static component, which remains constant for an infinitely long time. 
The single site colored Langevin equation in presence of this quenched component can be 
analyzed by methods analogue to the approach used in Ref.~\cite{maimbourg_prl_2016} for 
the vitrification of a supercooled liquid in infinite dimensions.
In presence of a quenched component, $\bzeta_{i}(t)$ 
can be separated as the sum $\bzeta_{i}(t) = \bzeta_{1i}(t) + \bzeta_{2i}$ 
of two independent parts $\bzeta_{1i}$ and $\bzeta_{2i}$, both having Gaussian 
distributions with zero mean.
$\bzeta_{1i}(t)$ is a dynamic noise, with a correlation $\langle \zeta^{\alpha}_{1i}(t) 
\zeta_{1i}^{\beta}(t')\rangle = l^{\alpha \beta}(t-t') - L_{2}^{\alpha \beta} = 
l_{1}^{\alpha \beta}(t-t')$ which decreases to zero at large times.
$\bzeta_{2i}$ is instead a static noise, with $\langle \zeta_{2i}^{\alpha} 
\zeta_{2i}^{\beta}\rangle = L_{2}^{\alpha \beta}$ describing the time-persistent part 
of the random field $\bzeta_{i}$.
The dynamic and static parts of the noise are mutually uncorrelated: $\langle 
\zeta_{1i}^{\alpha}(t) \zeta_{2i}^{\beta}\rangle = 0$.

The field $\bb_{i}(t)$ which enters in the single-site Langevin equations breaks 
similarly into static and dynamic parts:
\begin{equation}\label{b2i}
\begin{split}
b^{\alpha}_{i}(t) & = b_{2i}^{\alpha} + \zeta_{1i}^{\alpha}(t) + \beta l_{1}^{\alpha 
\beta}(t) S^{\beta}_{i0} \\
& + \int_{0}^{t} {\rm d}t'~K_{1}^{\alpha \beta}(t-t') S^{\beta}_{i}(t')~,
\end{split}
\end{equation}
with $b_{2i}^{\alpha} = \zeta_{2i}^{\alpha} + \beta L_{2}^{\alpha \beta} S^{\beta}_{i0}$ 
and $K_{1}^{\alpha \beta}(t-t') = - \beta \Theta(t-t') {\rm d}l_{1}^{\alpha 
\beta}(t-t')/{\rm d}t = K^{\alpha \beta}(t-t')$ (the constant $L_{2}^{\alpha \beta}$ does 
not contribute to the time derivative defining the memory kernel $K^{\alpha \beta}$).

We can assume that the dynamical noises $\bnu_{i}$ and $\bzeta_{1}$ drive the system into 
an equilibrium state dependent on the static field mean field $\bb_{2i}$.
In other words, we assume that after running the Langevin dynamics from the initial 
time $t = 0$ to a large time, the system relaxes to a Gibbs probability, subject to the 
static field $\bb_{2i}$.

Analyzing the colored Langevin equations with the methodology of 
Refs.~\cite{cugliandolo_jpsj_2000, maimbourg_prl_2016}, we find that the asymptotic 
large-time distribution is:
\begin{equation}\label{P_b2}
P_{1}(\bS_{i}|\bb_{2i}) = \frac{{\rm e}^{-\beta V(\bS_{i}) + \beta (\bb_{2i}\cdot \bS_{i}) 
+ \frac{1}{2}\beta^{2} L_{1}^{\alpha \beta} S^{\alpha}_{i} 
S^{\beta}_{i}}}{\int_{\bar{\bS}} {\rm e}^{-\beta V(\bar{\bS}) + \beta (\bb_{2i} \cdot 
\bar{\bS}) + \frac{1}{2} \beta^{2} L_{1}^{\gamma \delta} 
\bar{S}^{\gamma}\bar{S}^{\delta}}}~,
\end{equation}
with $L_{1}^{\alpha \beta} = l_{1}^{\alpha \beta}(0) = L^{\alpha \beta} - L_{2}^{\alpha 
\beta}$.
This distribution can be interpreted as the probability of a single spin within an 
individual metastable state.
The quenched component $\bb_{2i}$ represents a mean field, present within the metastable 
state, and acting on the spin $\bS_{i}$.
When the Gibbs distribution breaks into a superposition of many nonergodic sectors, the 
field $\bb_{2i}$ is itself a random variable.
The probability to extract a value of $\bb_{2i}$ can be calculated using the Gaussian 
distribution of $\bzeta_{2i}$ and the equilibrium distribution $P_{1{\rm eq}}(\bS_{i0})$ 
of the initial conditions $\bS_{i0}$.
The result is
\begin{equation}\label{P_slow}
\begin{split}
& P_{\rm slow}(\bb_{2i}) = \int_{\bS_{0}} \int {\rm d}^{3}\bzeta_{2}~\Big\{ 
\delta(b_{2i}^{\alpha} - \beta L_{2}^{\alpha \beta} S^{\beta}_{0} - \zeta_{2}^{\alpha})\\
&  \times \frac{\exp \big(-L_{2}^{-1\alpha \beta} \zeta_{2}^{\alpha} 
\zeta_{2}^{\beta}/2\big)}{\sqrt{(2\pi)^{3} \det \uu{L_{2}}}} \times \frac{1}{Z_{1}} \exp 
\big[-\beta V(\bS_{0}) \\
& \qquad + \beta^{2} L^{\alpha \beta} S_{0}^{\alpha} S_{0}^{\beta}/2\big]\Big\}\\
&=\frac{1}{Z_{1} \sqrt{(2\pi)^{3} \det \uu{L_{2}}}} \int_{\bS_{0}} \exp 
\big[-\beta V(\bS_{0}) \\
& + \beta (\bb_{2i} \cdot \bS_{0}) - L_{2}^{-1 \alpha \beta} b_{2i}^{\alpha} 
b_{2i}^{\beta}/2 + \beta^{2} L_{1}^{\alpha \beta} S_{0}^{\alpha} S_{0}^{\beta}/2\big]~,
\end{split}
\end{equation}
and can be interpreted as the probability to find a given value of the mean field, selecting a
metastable state at random in the Gibbs ensemble.

Since $P_{1{\rm eq}}(\bS_{i}) = \int {\rm d}^{3}\bb_{2i}~P_{\rm slow}(\bb_{2i}) 
P_{1}(\bS_{i}|\bb_{2i})$, the equal-time averages are consistent with the static 
analysis.
The separation of the average into a quenched and a vibrational part, however, allows 
also to calculate the correlations at large time separations.
In particular, the Edwards-Anderson order parameter can be calculated as:
\begin{equation}\label{q_EA}
\begin{split}
& q^{\alpha \beta}_{\rm EA} = \lim_{|t-t'| \to \infty} \langle S_{i}^{\alpha}(t) 
S_{i}^{\beta}(t') \rangle\\
& \qquad = \int {\rm d}^{3}\bb_{2i}~P_{\rm slow}(\bb_{2i})~\langle 
S^{\alpha}_{i}\rangle|_{\bb_{2i}}\langle S^{\beta}_{i}\rangle|_{\bb_{2i}}\\
& \qquad = \overline{\langle S^{\alpha}_{i}\rangle|_{\bb_{2i}}\langle 
S^{\beta}_{i}\rangle|_{\bb_{2i}}}~.
\end{split}
\end{equation}
Here $\langle \bullet \rangle|_{\bb_{2i}} = \int_{\bS} ~ \bullet ~ P(\bS|\bb_{2i})$ 
is an average calculated in the ensemble~\eqref{P_b2}, with $\bb_{2i}$ fixed, and the 
overline symbol $\overline{{}\bullet{}} = \int {\rm d}^{3}\bb_{2i} \bullet P_{\rm 
slow}(\bb_{2i})$ is an average over $\bb_{2i}$, weighted with the distribution $P_{\rm 
slow}(\bb_{2i})$.

More generally, arbitrary averages $\langle \Phi_{1}(\bS(t_{1})) \Phi_{2}(\bS(t_{2})) ... 
\Phi_{n}(\bS (t_{n}))\rangle$ in the limit in which all time differences are large can 
be calculated as:
\begin{equation}\label{large-time-average}
\overline{\langle \Phi_{1}(\bS(t_{1}))\rangle|_{\bb_{2i}} ... \langle 
\Phi_{n}(\bS(t_{n}))\rangle|_{\bb_{2i}}}~.
\end{equation}

Eq.~\eqref{large-time-average} assumes that within a single state the dynamics loses 
correletion at large time, so that for a fixed $\bb_{2i}$, $\langle \Phi_{1}(\bS(t_{1})) 
.. \Phi_{n}(\bS(t_{n})) \rangle|_{\bb_{2i}} \approx \langle 
\Phi_{1}(\bS(t_{1}))\rangle|_{\bb_{2i}} \times ... \times \langle \Phi_{n}(\bS(t_{n})) 
\rangle|_{\bb_{2i}} $ when the time separations are large.
A frozen correlation, which persists for large times emerges after averaging over the 
static field $\bb_{2i}$.

Eq.~\eqref{q_EA} provides an equation linking the Edwards-Anderson order parameter to
$L^{\alpha \beta}$, $L_{2}^{\alpha \beta}$, and $L_{1}^{\alpha \beta}$.
To fix the order parameters, we also need to discuss the large-time limit of the 
self-consistency equations.
To this end, it is convenient to integrate Eqs.~\eqref{C_dot_plus} over time.
In the limit $t-t' \to + \infty$ we find:
\begin{equation}\label{self-consistency_1}
\begin{split}
& \int_{t'}^{t} {\rm d}t''~\dot{C}_{+ij}^{\alpha \beta}(t''-t') 
= - C_{1ij}^{\alpha \beta} \\
& = -\delta_{ij}c_{1}^{\alpha \beta}  - \beta \int_{t'}^{t} {\rm 
d}t_{1}~ \{[A_{ij}^{\alpha \gamma}(t-t_{1}) - A_{ij}^{\alpha \gamma}(0)]\\
& \times \dot{c}_{+}^{\gamma \beta}(t_{1}-t')\} + \beta \int_{t'}^{t} {\rm d}t_{1} 
\int_{t'}^{t_{1}} {\rm d}t_{2} \big[C_{ij}^{\alpha \gamma}(t-t_{1}) \\
& - C_{ij}^{\alpha \gamma}(0)\big] K^{\gamma \delta}(t_{1} - t_{2}) \dot{c}_{+}^{\delta 
\beta}(t_{2} - t')~,
\end{split}
\end{equation}
where $C_{1ij}^{\alpha \beta} = C_{ij}^{\alpha \beta}(0) - C_{2ij}^{\alpha \beta}$ and 
$c_{1}^{\alpha \beta} = c^{\alpha \beta}(0) - c_{2}^{\alpha \beta}$.
Since $\dot{c}_{+}^{\alpha \beta}(t_{1} -t)$ and $K^{\gamma \delta}(t_{1} - t_{2})$ are 
defined by time derivatives of correlations, they do not contain quenched parts.
Thus $\dot{c}_{+}^{\alpha \beta}(t_{1}-t')$ and $\int_{t'}^{t_{1}} {\rm d}t_{2} 
K^{\gamma \delta}(t_{1} - t_{2}) \dot{c}_{+}^{\delta \beta}(t_{2} - t')$ vanish for 
$|t_{1} - t'| \to \infty$.
As a result the integrals are dominated by the region in which $t_{1}$ is near $t'$ and 
far from $t$.
This allows to replace in the integrals $A_{ij}^{\alpha \gamma}(t-t_{1}) = 
\lim_{|t-t_{1}|\to \infty} A^{\alpha \gamma}(t-t_{1}) =  A_{2ij}^{\alpha \beta}$, 
$C_{ij}^{\alpha \gamma}(t-t_{1}) = \lim_{|t - t_{1}|\to \infty} = C_{2ij}^{\alpha 
\gamma}$.
Introducing $C_{1ij}^{\alpha \beta} = C_{ij}^{\alpha \beta}(0) - C_{2ij}^{\alpha 
\beta}$,$A_{1ij}^{\alpha \beta} = A_{ij}^{\alpha \beta}(0) - A_{2ij}^{\alpha \beta}$ and 
using the FDT we find that Eq.~\eqref{self-consistency_1} behaves for $|t-t'|$ large as:
\begin{equation}\label{self-consistency_2}
\begin{split}
& \beta^{2} C_{1ij}^{\alpha \gamma} \int_{t'}^{t}{\rm d}t_{2}~\dot{c}_{+}^{\delta 
\beta}(t_{2} - t') \big[l^{\gamma \delta}(t - t_{2}) - l^{\gamma \delta}(0)\big]\\
& = \beta^{2}C_{1ij}^{\alpha \gamma} L_{1}^{\gamma \delta} c_{1}^{\delta \beta} = 
-C_{1ij}^{\alpha \beta}  + \delta_{ij} c_{1}^{\alpha \beta} + \beta A_{1ij}^{\alpha 
\gamma}c_{1}^{\gamma \beta}~.
\end{split}
\end{equation}
In Eq.~\eqref{self-consistency_2} we have used again that the integration is dominated by 
the region in which $t_{2}$ is far from $t$.

Analyzing in a similar way Eqs.~\eqref{A_dot_plus},~\eqref{a_c_dynamics}, 
and~\eqref{lambda_c_dynamics}, and using that the equal-time correlations satisfy the 
static relations~\eqref{static_eom} we find:
\begin{equation}\label{self-consistency_3}
\begin{gathered}
A_{1ij}^{\alpha \beta}  +  \beta^{2} c_{1}^{\alpha \gamma} L_{1}^{\gamma \delta} 
A_{1}^{\delta \beta} =  c_{1}^{\alpha \gamma} J_{ij}^{\gamma \beta} + \beta c_{1}^{\alpha 
\gamma} \Lambda_{1ij}^{\gamma \beta}~,\\
a_{1}^{\alpha \beta} = \beta c_{1}^{\alpha \gamma} L_{1}^{\gamma \beta}~,\\
\lambda_{1}^{\alpha \beta} = L_{1}^{\alpha \beta} + \beta^{2} L_{1}^{\alpha \gamma} 
c_{1}^{\gamma \delta} L_{1}^{\delta \beta}~,
\end{gathered}
\end{equation}
where $\Lambda_{1ij}^{\alpha \beta} = \Lambda_{ij}^{\alpha \beta}(0) - 
\Lambda_{2ij}^{\alpha \beta}$, $a^{\alpha \beta}_{1} = a^{\alpha \beta}(0) - a_{2}^{\alpha 
\beta}$, $\lambda^{\alpha \beta}_{1} = \lambda^{\alpha \beta}(0) - \lambda_{2}^{\alpha 
\beta}(t-t')$, $\Lambda_{2ij}^{\alpha \beta} = \lim_{|t-t'| \to \infty} \Lambda^{\alpha 
\beta}_{ij}(t-t')$, $a_{2}^{\alpha \beta} = \lim_{|t-t'| \to \infty} \langle 
S_{i}^{\alpha}(t) b^{\beta}_{i}(t')\rangle$, $\lambda_{2}^{\alpha \beta}(t-t') = 
\lim_{|t-t'| \to \infty} \langle b^{\alpha}_{i}(t) 
b^{\beta}_{i}(t')\rangle$.
Introducing the "self-energy" $\sigma_{1ij}^{\alpha \beta} = \delta_{ij} 
(\lambda_{1}^{-1\alpha \beta} - L_{1}^{-1\alpha \beta})$ the solutions can be written as:
\begin{equation}\label{C_A_Lambda_vibrational}
\begin{gathered}
\hat{C}_{1} = - (\beta^{2} \hat{\sigma}_{1}^{-1} + \beta \hat{J})^{-1}\\
\hat{A}_{1} = \hat{C}_{1}\hat{J}\\
\hat{\Lambda}_{1} = \hat{J} \hat{C}_{1} \hat{J} =  -k_{\rm B}T \hat{J} + 
(\hat{\sigma}_{1} + \beta \hat{J})^{-1}~.
\end{gathered}
\end{equation}

These equations are identical in form to the relations~\eqref{C_A_Lambda_solution}.
They show that the vibrational parts $\hat{C}_{1}$, $\hat{A}_{1}$, $\hat{\Lambda}_{1}$ 
satisfy the same relations of the full correlations $\hat{C}$, $\hat{A}$, 
$\hat{\Lambda}$, but with a different self-energy.

Combining Eqs.~\eqref{q_EA} and ~\eqref{C_A_Lambda_vibrational} we can write the 
self-consistency equations for the order parameters as:
\begin{equation}\label{glass_self_consistency}
\begin{split}
& c_{1}^{\alpha \beta}  = c^{\alpha \beta} - q_{\rm EA}^{\alpha \beta}\\
& = -\int_{-\pi}^{\pi} \frac{{\rm d}^{d}k}{(2\pi)^{d}} 
\big[\big(\beta^{2}\uu{\sigma_{1}}^{-1} + \beta \uu{J}(\bk)\big)^{-1}\big]^{\alpha 
\beta}\\
& = - \int_{-\infty}^{\infty} {\rm d}\ve~\nu(\ve) \Big[(\beta^{2} 
\uu{\sigma_{1}}^{-1} + \beta \uu{f}(\ve))^{-1}\Big]^{\alpha \beta}~.
\end{split}
\end{equation}

In the paramagnetic phase, the solution is trivial: $q_{\rm EA}^{\alpha \beta} = 0$, 
$c_{1}^{\alpha \beta} = c^{\alpha \beta}$, $\sigma_{1}^{\alpha \beta} = \sigma^{\alpha 
\beta}$, $\lambda_{1}^{\alpha \beta} = \lambda^{\alpha \beta}$, ... 
In the glass phase, a solution with $q_{\rm EA} \neq 0$ appears.
In this case the static correlations $C_{ij}^{\alpha \beta}$ differ from $C_{1ij}^{\alpha 
\beta}$.
The statistical mechanical fluctuations calculated in Sec.~\ref{s:statics} do not have a 
completely vibrational origin but, rather, a part of the fluctuations becomes 
configurational.

\section{Application: isotropic Heisenberg model on the infinite-dimensional fcc lattice}
\label{s:fcc_model}

In this section we apply the results to a specific case: a isotropic antiferromagnet with $V(\bS) =
0$, $f^{\alpha \beta}(x) = f(x) \delta^{\alpha \beta} =  J (x^{2} -1) \delta^{\alpha \beta}$, and $J
< 0$.
As discussed in Sec.~\ref{s:model}, this interaction couples only the spins which 
reside in the same fcc sublattice (A or B) of the simple hypercubic crystal and, as a result, the
model can be viewed as describing two independent copies of a spin system on the fcc
lattice~\cite{ulmke_epjb_1998}.
The interaction $f(x) = J(x^{2} - 1)$, in particular, is equivalent to a model on the fcc lattice
with two nonzero antiferromagnetic couplings: a nearest-neighbour interaction $J_{1} = J/d$ and a
second-nearest neighbour interaction $J_{2} = J_{1}/2 = J/(2d)$.

Since the maximum of $f(x)$ occurs at $x = 0$, the system admits a degenerate manifold of helical
ground states, with modulation vectors belonging to the $(d-1)$-dimensional surface defined by the
condition $\ve_{\bk} = 0$.

\begin{figure}[h]
\centering 
\includegraphics[scale=1]{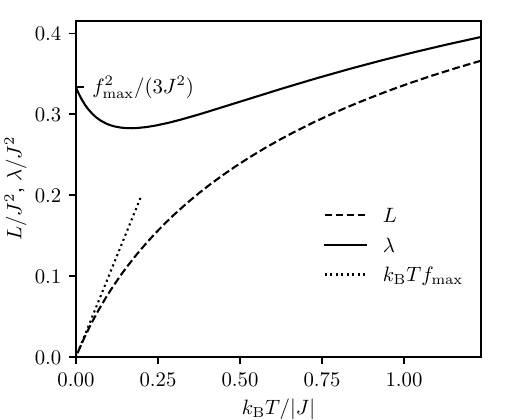}
\includegraphics[scale=1]{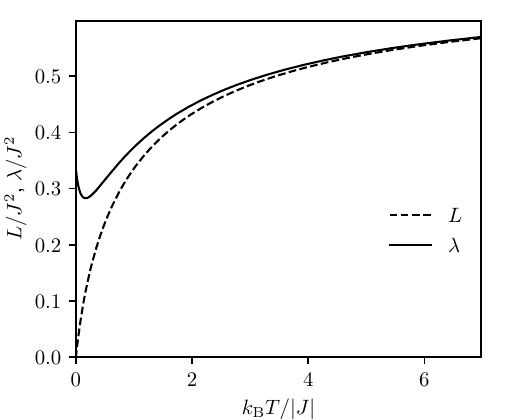}
\caption{\label{fig2} Temperature dependence of the variance of the cavity distribution 
$L$ (dashed line) and of the full equilibrium distribution $\lambda$ (solid line), in the 
isotropic model with $V(\bS) = 0$, $f^{\alpha \beta}(x) = J (x^{2} - 1)\delta^{\alpha 
\beta}$.
The top panel shows the low-temperature region, the bottom panel a wider range of temperatures.
At low $T$ $L(T) \approx k_{\rm B}T f_{\rm max} = k_{\rm B}T |J|$. (The dotted line is a guides to
the eye highlighting the asymptotic behavior at $T \to 0$).
The variance $\lambda$, instead, approaches a finite limit $f^{2}_{\rm max}/(3 J^{2})$.
The temperature dependence $\lambda(T)$ is non-monotonic: $\lambda(T)$ has a minimum at $T \simeq
0.167 |J|$.
At higher temperatures the curves $L(T)$ and $\lambda(T)$ approach each other, and 
eventually saturate to $L(T) \simeq \lambda(T) \simeq 2 J^{2}/3$ in the limit $T \to 
\infty$.
}
\end{figure}

\begin{figure}[h]
 \centering
 \includegraphics[scale=1]{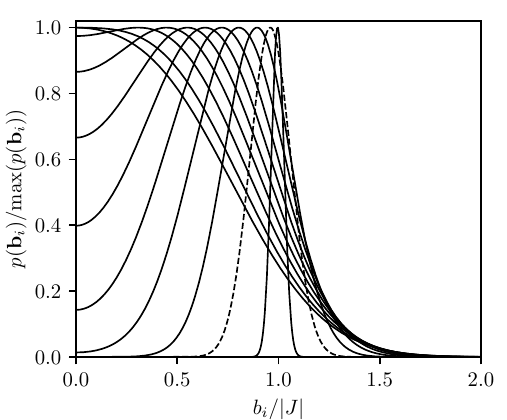}
 \caption{\label{fig3} Equilibrium distribution of the field $\bb_{i}$ for 10 different 
temperatures, equally spaced between $10^{-3}|J|/k_{\rm B}$ and $0.25|J|/k_{\rm B}$ (solid lines).
The 10 values of the temperature are, with 3-digit precision, $k_{\rm B}T/|J|$ = 0.001, 0.029,
0.056, 0.084, 0.112, 0.139, 0.167, 0.195, 0.222, 0.25.
The dashed line shows the equilibrium distribution of the field $\bb_{i}$ at the glass temperature
$T_{\rm g} \simeq 0.0103 |J|/k_{\rm B}$.
For the higher temperatures in the plot, the distribution differs from a Gaussian, but has
a bell-like shape with a maximum at $b_{i} = 0$.
At lower temperatures, the maximum shifts away from $b_{i} = 0$.
For $T \to 0$, the distribution becomes concentrated on a narrow shell near the spherical 
surface $|\bb_{i}| = |J| = f_{\rm max}$.}
\end{figure}

The static properties of the model, computed from the equilibrium Gibbs distribution as discussed
in Sec.~\ref{s:statics}, are shown in Figs.~\ref{fig2},~\ref{fig3}.
In particular, Fig.~\ref{fig2} shows the temperature dependence of $L$ and $\lambda$, 
calculated numerically using Eq.~\eqref{I_beta_sigma}, and the relations $L = -3/\beta^{2} 
- 1/\sigma$, $\lambda = 1/\sigma + \beta^{2}/(3 \sigma^{2})$.
In the limit of small temperatures, the variance of the cavity distribution $L = \langle
\bb_{i}^{2}\rangle^{\prime (i)}/3$ decreases with $T$ and vanishes for $T \to 0$ as $L(T)
\approx k_{\rm B}T f_{\rm max} = k_{\rm B}T |J|$.
The full variance of the equilibrium distribution $\lambda = \langle 
\bb_{i}^{2}\rangle/3$, instead, tends to $f^{2}_{\rm max}/3 = |J|^{2}/3$ for $T \to 
0$.
In the opposite limit of high temperatures, $L \approx \lambda \approx {\cal 
J}_{2}^{2}/3 = \sum_{j} J_{ij}^{2}/3  = 2 |J|^{2}/3$.

The numerical solution shows a further interesting feature: the dependence of $\lambda$ on 
temperature is non-monotonic.
$\lambda$ reaches a minimum at a temperature $T \simeq 0.167 |J|$ and, for lower 
temperatures, starts to \emph{grow} when $T$ is decreased.

The behavior ${\rm d}\lambda/{\rm d}T < 0$ occurs in a region of temperatures for which 
the equilibrium distribution of the field  $p(\bb_{i}) = \int_{\bS} P(\bS_{i}, \bb_{i}) = 
\sinh(\beta |\bb_{i}|)\exp[-(\beta^{2} L + L^{-1} \bb_{i}^{2})/2]/[(2\pi L)^{3/2} \beta 
|\bb_{i}|]$ is significantly different from a Gaussian function.
The shape of the distribution $p(\bb_{i})$ is shown in Fig.~\ref{fig3} for various 
temperatures.

To investigate the occurrence of a glass transition in the model, we study the self-consistent
equations for the order parameter (Eqs.~\eqref{q_EA} and~\eqref{glass_self_consistency}), which in
the isotropic case reduce to:
\begin{equation}\label{order_parameter_fcc}
\begin{split}
&c_{1} = \frac{1}{3} - q_{\rm EA} = \frac{1}{\beta} I(\beta/\sigma_{1})~,\\
&\beta^{2} c_{1}  = -1/(\sigma_{1}^{-1} + L_{1})~,\\
&q_{\rm EA} = \frac{1}{3}~\overline{\langle S^{\alpha} \rangle \langle 
S^{\alpha}\rangle} = \frac{1}{3} \overline{\big({\cal L}(\beta |\bb_{2}|)\big)^{2}}\\
& \quad ~~ = \frac{1}{3} (1 - I_{1}(\beta^{2} L_{2}))~,\\
& L = L_{1} + L_{2}~.
\end{split}
\end{equation}

In these relations $I(x) = - \int_{-\infty}^{\infty} {\rm d}\ve \nu(\ve)/[x + f(\ve)]$ is the
integral introduced in Eq.~\eqref{I_beta_sigma}, ${\cal L}(x) = 1/\tanh(x) - 1/x$ is the Langevin
function, and
\begin{equation}
I_{1}(x) = \int_{-\infty}^{\infty} {\rm d}y~\frac{y[1- ({\cal L}(y))^{2}]}{(2\pi 
x^{3})^{1/2}} \exp \bigg(-\frac{(y-x)^{2}}{2x}\bigg)~.
\end{equation}

\begin{figure}[h]
\centering 
\includegraphics[scale=1]{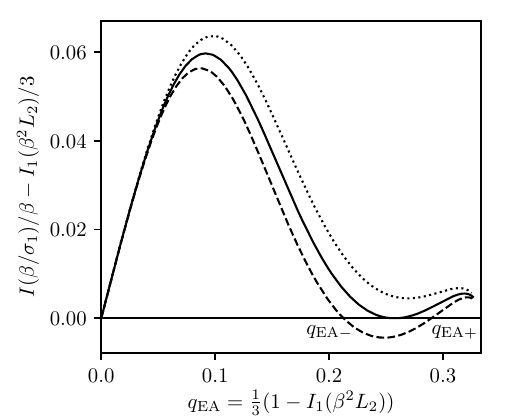}
\caption{\label{fig4} Numerical analysis of Eqs.~\eqref{order_parameter_fcc}.
The curves show the function $I(\beta/\sigma_{1})/\beta - I_{1}(\beta^{2} L_{2})/3$
calculated substituting $\beta/\sigma_{1} = - 1/(\beta c_{1}) - \beta L_{1} = -3/(\beta 
I_{1}(\beta^{2}L_{2})) + \beta L_{2} - \beta L$.
The $x$ axis is parametrized by the order parameter $q_{\rm EA}(\beta^{2} L_{2}) = (1 - 
I_{1}(\beta^{2} L_{2}))/3$.
The dotted, solid, and dashed curves represent the curves calculated at temperatures 
respectively equal to $T = 0.012 |J|/k_{\rm B}$, $0.0103 |J|/k_{\rm B}$, and $0.009 
|J|/k_{\rm B}$.
The solutions of the self-consistency problem are defined by the points at which the 
curves cross 0.
For high-temperature the curve intersects 0 only at the trivial point $q_{\rm EA} = 0$ 
(dotted line).
A nontrivial solution appears at $T_{\rm g} \simeq 0.0103 |J|/k_{\rm B}$.
Below $T_{\rm g}$, the equations have two solutions $q_{{\rm EA} \pm}$.
The physical solution is $q_{{\rm EA}+}$, the one with the largest order parameter.
}
\end{figure}

The results of a numerical study of these equations is presented in Fig.~\ref{fig4}.
At high temperatures, the only solution is the trivial one.
However, at a temperature $T_{\rm g} \simeq 0.0103 |J|/k_{\rm B}$ a nontrivial solution 
appears.
We interpret $T_{\rm g}$ as the temperature of a dynamical glass transition, at which the system
vitrifies.
It is interesting to note that $T_{\rm g}$ is an order of magnitude smaller than the 
temperature scales which characterize the behavior of the static solution.
For example $T_{\rm g}$ is much smaller than the temperature $T \simeq 0.167 |J|/k_{\rm B}$ at
which the derivative ${\rm d}\lambda/{\rm d}T$ changes sign.

At $T_{\rm g}$, the Edwards-Anderson order parameter $q_{\rm EA}$ jumps discontinuously 
from 0 to $q_{\rm EA}(T_{\rm g}) \simeq 0.2575$.
This value corresponds to an average local magnetization at $T_{\rm g}$ equal to $|\mathbf{m}|=
[\overline{\langle S^{\alpha}\rangle \langle S^{\alpha} \rangle}]^{1/2} = \sqrt{3 q_{\rm EA}(T_{\rm
g})} \simeq 0.879$.
We interpret this as the average magnitude of the local moments in an amorphous state in which the
system vitrifies at $T_{g}$.

\begin{figure}[h]
 \centering 
 \includegraphics[scale=1]{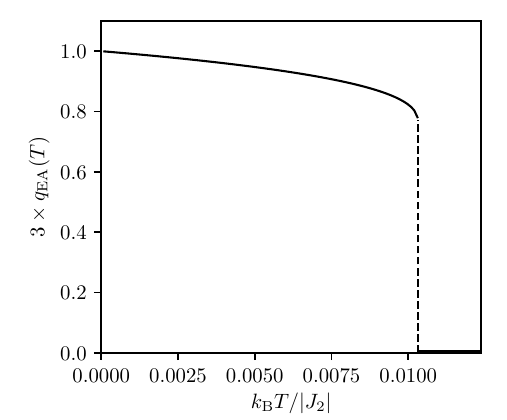}
 \caption{\label{fig5} Equilibrium average of the Edwards-Anderson order parameter 
$q_{\rm EA}(T)$ obtained from a numerical solution of Eqs.~\eqref{order_parameter_fcc}.
The order parameter represented in the figure describes the large-time limit $3 q_{\rm 
EA}(T) = \lim_{|t-t'|\to \infty} \langle S^{\alpha}(t) S^{\alpha}(t') \rangle$, with an 
averaging $\langle ... \rangle$ weighted with the equilibrium Gibbs distribution at 
temperature $T$.
This equilibrium $q_{\rm EA}$ may differ in general from the average order parameter 
which the system presents after a cooling from the paramagnetic phase $T > T_{\rm g}$, 
because the spins remain trapped in a single metastable state at $T_{\rm g}$ and fall out 
of equilibrium at lower temperatures.}
\end{figure}

Below $T_{\rm g}$ Eqs.~\eqref{order_parameter_fcc} present two solutions $q_{{\rm EA} 
\pm}(T)$, which bifurcate from the transition point.
The branch with smaller overlap, which we denote as $q_{{\rm EA}-}(T)$, is an unphysical solution
because it leads to an order parameter which decreases when $T$ is lowered.
The temperature dependence of the solution $q_{\rm EA}(T) = q_{{\rm EA}+}(T)$, with larger overlap,
is illustrated in Fig.~\ref{fig5}.
Starting from the transition point, $q_{\rm EA}(T_{\rm g})$ grows at small $T$ and 
eventually $3 q_{{\rm EA}+}$ reaches 1 for $T \to 0$.

The transition is of a dynamical first order type, similar to the transition which occurs 
in supercooled liquids in $d \to \infty$~\cite{maimbourg_prl_2016} and which was 
predicted in stripe systems~\cite{schmalian_prl_2000, schmalian_ijmpb_2001, 
westfahl_prb_2001, grousson_pre_2002, grousson_pre_2002b, westfahl_prb_2003, wu_prb_2004}.
At $T_{\rm g}$, the equilibrium statistical properties remain smooth, without any signature of a
phase transition, but the system undergoes a dynamical arrest.
This contrasts, for example, with the second-order transitions found in random spin 
glasses~\cite{mezard_spin-glass}.

Below $T_{\rm g}$ the system remains trapped into a single metastable state and falls out of
equilibrium.
After cooling from the high temperature liquid regime $T > T_{\rm g}$ to the glassy region 
$T < T_{\rm g}$, we can expect that the physical thermal properties are not controlled by 
the equilibrium distribution at $T$, but by the distribution of metastable states at the 
temperature at $T_{\rm g}$, at which the system fell out of equilibrium for the first 
time.
The replica theory is useful to discuss this regime~\cite{lopatin_prb_2002}.

The analogy with the theory of stripe glasses and with the Ising model analyzed in
Ref.~\onlinecite{lopatin_prb_2002} suggests that the system may present an
ideal glass transition at a temperature $T_{\rm s} < T_{\rm g}$, signaled by the 
vanishing of the configurational entropy.
We leave open the question whether this transition is present in the Heisenberg model studied here.

In conclusion, we note that the model analyzed in this section is a direct extension in 
infinite dimensions of the fcc $J_{1}-J_{2}$ Heisenberg antiferromagnet with $J_{1} = 2 
J_{2}$.
In three dimensions this antiferromagnetic fcc model has been studied by Balla \emph{et
al.}~\cite{balla_prb_2019, balla_prr_2020}.
The analysis of Ref.~\cite{balla_prb_2019} showed that the 3D $J_{1}-J_{2}$ fcc model ultimately
orders in the limit $T \to 0$, due to an order-by-disorder mechanism.
Our analysis in infinite dimensions predicts that the disordered phase is locally stable, 
so that ordering must occur via a first-order transition. The dynamics within the 
disordered solution, in addition undergoes a dynamical glass transition at a finite $T$.

\section{Conclusions}
\label{s:conclusions}

In this article, we have analyzed a class of infinite-dimensional frustrated spin models,
characterized by the property that their interaction $J(\bk)$ presents degenerate surfaces in
momentum space.
Using the cavity method, we derived in the limit of infinite dimensionality $d \to \infty$ the exact
solution of the statistical mechanical properties within the disordered, paramagnetic phase.
By a study of the equilibrium dynamics of the system, we derived a set of consistency equations for
the glass order parameters.
The theory was applied explicitly to the case of a isotropic model, equivalent after bipartition of
the hypercubic lattice to a Heisenberg model on the fcc lattice.
For this model we identified the temperature of dynamical vitrification.
The transition is of a dynamical first order type and is similar to the vitrification 
predicted in the theory of stripe glassiness.
Although the exact results were obtained in the limit of infinite dimensions, the power demonstrated
by DMFT methods in describing correlated electron
systems~\cite{kotliar_rmp_2006, held_ap_2007, katsnelson_rmp_2008} indicates the relevance of the
results also for the realistic case of a system in three dimensions.

In comparison to the previous works hypothesizing self-induced spin-glass states in 
deterministic frustrated spin models~\cite{principi_prb_2016, principi_prl_2016} our new 
approach seems to have two advantages.
First, it has an explicit small parameter, albeit formal, and second it does not use the 
replica trick, whose applicability for deterministic systems is not obvious.
Instead, we study the large-time behavior of correlation functions, in the spirit of the 
original approach by Edwards and Anderson in the theory of spin glasses, and of the  
G\"{o}tze mode-coupling theory in the context of structural glasses.
Importantly, both approaches lead to the same conclusion that frustration only can be 
sufficient for vitrification.
This qualitatively supports the interpretation of the experimental data of 
Refs.~\cite{kamber_science_2020, verlhac_np_2022} as an observation of a self-induced 
spin glass state in elemental neodymium at low temperatures.
Another interesting technical question concerning the relation between dynamical and 
replica approaches will be considered elsewhere.

\acknowledgments

This work was supported by the European Research Council (ERC) under the European Union's 
Horizon 2020 research and innovation program, grant agreement 854843-FASTCORR and by 
the Dutch Research Council (NWO) via the Spinoza Prize.
We are thankful to Alex Khajetoorians, Lorena Niggli, and Tom Westerhout for stimulating 
discussions.

\appendix 

\section{Fokker-Planck equation and fluctuation-dissipation theorem}
\label{a:FP_and_FDT}

The Fokker-Planck (FP) equation associated with the Brownian motion of magnetic moments 
has been discussed in several works (see for example Refs.~\cite{brown_pr_1963, 
kubo_ptps_1970, garcia-palacios_prb_1998, aron_jsm_2014}).

Here we give a brief derivation of the FP equation, and discuss the FDT and the Hilbert 
space representation of the dynamics.
To derive the equations, we parametrize the spins by two coordinates $(\theta^{1}_{i}, 
\theta^{2}_{i})$.
A natural choice are the spherical coordinates $\theta^{1}_{i} = \theta_{i}$, 
$\theta_{i}^{2} = \varphi_{i}$, $\bS(\theta^{1}, \theta^{2}) = (\sin \theta^{1} \cos 
\theta^{2}, \sin \theta^{1} \sin \theta^{2}, \cos \theta^{1})$, but an arbitrary 
pametrization can be used.
In the following we denote as $\theta^{\mu}_{i}$, $\mu = 1, 2$ the coordinates.
To avoid confusion, letters from the beginning of the Greek alphabet $\alpha$, $\beta$, 
$\gamma$, $\delta$ are used to denote the cartesian components of the spins, as in the 
main text; letters from the final part of the alphabet $\mu$, $\nu$, $\rho$, ... are 
used, instead, to denote the two-dimensional parametrization $\theta^{\mu}$.

In terms of the $\theta^{\mu}_{i}$, the Langevin equation can be written as:
\begin{equation}\label{langevin_spherical}
\dot{\theta}^{\mu}_{i} = g^{\mu \nu}(\btheta_{i}) (\bu_{\nu}(\btheta_{i})\cdot 
\bE_{i})~,
\end{equation}
where $\bE_{i} = -\bS_{i} \times (\bS_{i} \times (\bN_{i} + \bnu_{i}))$ is the right-hand 
side of the equation of motion~\eqref{langevin}, 
$\bu_{\mu}(\btheta) = \pa \bS/\pa \theta^{\mu}$ are the tangent vectors, and $g^{\mu 
\nu}(\btheta)$ is the inverse of the metric tensor $g_{\mu \nu}(\btheta) = 
(\bu_{\mu}(\btheta) \cdot \bu_{\nu}(\btheta))$.

Using the theory of Langevin processes on manifolds~\cite{zinn-justin_qft} and introducing 
the reparametrization-invariant distribution
\begin{equation}\label{rho_invariant}
\rho(\btheta_{1}, ...., \btheta_{N}) = \frac{\lt \langle  \prod_{i=1}^{N} 
\delta(\btheta_{i} - \btheta_{i}(t))\rt \rangle_{\bnu}}{\prod_{j=1}^{N} 
\sqrt{g(\btheta_{j})}}
\end{equation}
we find after some computations:
\begin{equation}\label{FP}
\begin{split}
& \frac{\pa \rho}{\pa t}  + \sum_{i=1}^{N} \frac{1}{\sqrt{g(\btheta_{i})}} \frac{\pa}{\pa 
\theta_{i}^{\mu}} \bigg\{\sqrt{g(\btheta_{i})}  g^{\mu \nu}(\btheta_{i}) \\
& \times \bigg[(\bu_{\nu}(\btheta_{i}) \cdot\bN_{i})  \rho - k_{\rm B}T \frac{\pa 
\rho}{\pa \theta_{i}^{\nu}}\bigg] \bigg\} = 0~.
\end{split}
\end{equation}

In Eq.~\eqref{rho_invariant} the factor $1/\prod_{j=1}^{N}\sqrt{g(\btheta_{j})}$, 
$g(\btheta_{i}) = \det g_{\mu \nu}(\btheta_{i})$ is introduced in order to make the 
probability invariant.
With this normalization $\rho$ is the invariant probability distribution.
It gives the probability to find the system in an infinitesimal element ${\rm d}\Omega$ 
of the configuration space:  ${\rm d}P = \rho(\btheta_{1}, .., 
\btheta_{N}) {\rm d}\Omega =\rho(\btheta_{1}, .., \btheta_{N}) \prod_{i} 
\sqrt{g(\btheta_{i})} {\rm d}^{2}\theta_{i}$.

In spherical coordinates $g(\btheta_{i})= \sin^{2} \theta_{i}$ and the infinitesimal 
element is the usual ${\rm d}\Omega = \prod_{i=1}^{N} \sin\theta_{i} {\rm d}\theta_{i} 
{\rm d}\varphi_{i}$.
Eq.~\eqref{FP} becomes in these coordinates:
\begin{equation}
\begin{split}
& \frac{\pa \rho}{\pa t} + \sum_{i=1}^{N} \bigg\{\frac{1}{\sin \theta_{i}} 
\frac{\pa}{\pa \theta_{i}} \bigg[\sin \theta_{i} \bigg((N_{i}^{x}\cos \theta_{i} \cos 
\varphi_{i}\\
& + N_{i}^{y} \cos \theta_{i} \sin \varphi_{i} - N_{i}^{z} \sin \theta_{i}) \rho - 
k_{\rm B}T \frac{\pa \rho}{\pa \theta_{i}}\bigg)\bigg]\\
& + \frac{1}{\sin^{2}\theta_{i}} \frac{\pa}{\pa 
\varphi_{i}}\bigg[(-N_{i}^{x}\sin \varphi_{i} + N_{i}^{y} \cos 
\varphi_{i})\sin \theta_{i} \rho\\
& \qquad -k_{\rm B}T \frac{\pa \rho}{\pa \varphi_{i}}\bigg]\bigg\}= 0~.
\end{split}
\end{equation}

As a remark, we note that the methodolody which leads to Eq.~\eqref{FP} assumes the 
Stratonovich stochastic calculus.
Thus in Eq.~\eqref{langevin} we implicitly assume a Langevin equation~\eqref{langevin} 
defined according to the Stratonovich prescription (see Ref.~\cite{aron_jsm_2014} for 
different prescriptions).

The time evolution can be written more compactly introducing the angular momentum 
operators~\cite{kubo_ptps_1970}, which in a general frame of coordinates can be written 
as:
\begin{equation}\label{Li}
\begin{split}
\bm{\mathscr{L}}^{\alpha}_{i} \bullet & = -i \epsilon^{\alpha \beta \gamma} g^{\mu 
\nu}(\btheta_{i}) (S^{\beta}_{i} u^{\gamma}_{\mu}(\btheta_{i})) \frac{\pa}{\pa 
\theta_{i}^{\nu}} \bullet \\
& = - \frac{i \epsilon^{\alpha \beta \gamma}}{\sqrt{g(\btheta_{i})}} \frac{\pa}{\pa 
\theta_{i}^{\nu}} \sqrt{g(\btheta_{i})} g^{\mu \nu}(\btheta_{i}) S^{\beta}_{i} 
u_{\mu}^{\gamma}(\btheta_{i}) \bullet ~.
\end{split}
\end{equation}

These are generators of spin rotations and satisfy the commutation relations 
$[\mathscr{L}^{\alpha}_{i}, S^{\beta}_{j}] = i \delta_{ij} \epsilon^{\alpha \beta 
\gamma} S^{\gamma}_{i}$, $[\mathscr{L}^{\alpha}_{i}, \mathscr{L}^{\beta}_{j}] = i 
\delta_{ij} \epsilon^{\alpha \beta \gamma} \mathscr{L}_{i}^{\gamma}$.
Note that $[S^{\alpha}_{i}, S^{\beta}_{j}] = 0$ because the spins here are simply 
classical variables.

In terms of the $\mathscr{L}_{i}^{\alpha}$, the FP equation~\eqref{FP} can be recast 
as:
\begin{equation}\label{FP_operator}
\begin{split}
\frac{\pa \rho}{\pa t} & + \mathscr{H} \rho = 0~,
\end{split}
\end{equation}
with
\begin{equation}\label{FP_Hamiltonian}
\begin{split}
\mathscr{H}& = \sum_{i=1}^{N} \mathscr{L}_{i}^{\alpha} (i\epsilon^{\alpha \beta \gamma} 
S^{\beta}_{i} N^{\gamma}_{i} + k_{\rm B}T \mathscr{L}^{\alpha}_{i})~.
\end{split}
\end{equation}
and the components of the angular momentum operators read as usual
\begin{equation}
\begin{split}
\mathscr{L}_{i}^{x} & = i \sin \varphi_{i} \frac{\pa}{\pa \theta_{i}} + \frac{i \cos 
\varphi_{i}}{\tan \theta_{i}} \frac{\pa}{\pa \varphi_{i}}~, \\
\mathscr{L}_{i}^{y} & = -i \cos \varphi_{i} \frac{\pa}{\pa \theta_{i}} + \frac{i \sin 
\varphi_{i}}{\tan \theta_{i}} \frac{\pa}{\pa \varphi_{i}}~, \\
\mathscr{L}_{i}^{z} & = -i \frac{\pa}{\pa \varphi_{i}}~.
\end{split}
\end{equation}

Using that $\mathscr{L}_{i}^{\alpha} H = i \epsilon^{\alpha \beta \gamma}S^{\beta}_{i} 
N_{i}^{\gamma}$, it is simple to show that the operator $i \epsilon^{\alpha \beta 
\gamma}S^{\beta}_{i} N^{\gamma}_{i} + k_{\rm B}T \mathscr{L}_{i}^{\alpha}$ annihilates 
the Gibbs distribution $\rho_{\rm G} = Z^{-1} {\rm e}^{-\beta H}$. Thus, $\rho_{\rm G}$ 
is stationary in time: $\pa \rho_{\rm G}/\pa t = - \mathscr{H} \rho_{\rm G} = 0$.

Time-dependent correlation functions can be studied by setting up a Hilbert-space 
representation~\cite{zinn-justin_qft, kurchan_jpf_1992}.
Here, in order to maintain a notation which is manifestly reparametrization invariant, we 
find it convenient to define the scalar product of two functions $|\Phi_{1} \rangle = 
\Phi_{1}(\btheta_{1}, ..., \btheta_{N})$, $| \Phi_{2} \rangle =\Phi_{2}(\btheta_{1}, ..., 
\btheta_{N})$ as $\langle \Phi_{1}|\Phi_{2}\rangle = \int \prod_{j} {\rm d}^{2}\theta_{j} 
\sqrt{g(\btheta_{j})} \Phi^{*}_{1}(\btheta_{1}, .., \btheta_{N}) \Phi_{2}(\btheta_{1}, 
..., \btheta_{N})$, including  in the definition of the product the measure $\sqrt{g}$.
(In spherical coordinates $\sqrt{g(\theta, \varphi)} = \sin \theta$ and the integrals are 
the usual $\int {\rm d}^{2}\theta_{j} \sqrt{g(\btheta_{j})} = 
\int_{0}^{\pi} {\rm d}\theta_{j} \int_{0}^{2\pi} {\rm d}\varphi_{j} \sin \theta_{j}$).

If the distribution of the initial conditions at time $t = 0$ is $|\rho\rangle$, the 
average of a product of spin variables at later times can be represented as:
\begin{equation}
\begin{split}
\langle S_{i_{1}}^{\alpha_{1}}(t_{1}) & ... S^{\alpha_{l}}_{i_{l}}(t_{l})\rangle  = 
\langle 1| S_{i_{1}}^{\alpha_{1}} {\rm e}^{-\mathscr{H} (t_{1}-t_{2})} \times \\
& ... \times {\rm e}^{-\mathscr{H}(t_{l-1} - t_{l})}S_{i_{l}}^{\alpha_{l}} {\rm 
e}^{-\mathscr{H}t_{l}}|\rho\rangle~,
\end{split}
\end{equation}
where it is assumed that the times are arranged in the order as $t_{1} > t_{2} > ... > 
t_{l} > 0$.
$|1\rangle$ is simply a function equal to 1, so that the scalar product $\langle 1|\Phi 
\rangle$ is the integral of $\Phi(\btheta_{1}, .., \btheta_{N})$ over the configuration 
space, computed with the invariant measure $\prod_{j} {\rm d}^{2} \theta_{j} 
\sqrt{g(\btheta_{j})}$.
In particular, the condition that $|\rho\rangle$ is normalized is expressed by $\langle 
1|\rho\rangle = 1$.

It can be verified that $\langle 1|\mathscr{L}^{\alpha}_{i} = 0$.
In fact, for any function $|\Phi \rangle$ Eq.~\eqref{Li} implies:
\begin{equation}
\begin{split}
\langle 1|\mathscr{L}^{\alpha}_{i} |\Phi \rangle & = -i \epsilon^{\alpha \beta \gamma} 
\int \prod_{j=1}^{N} {\rm d}^{2}\theta_{j}\frac{\pa}{\pa 
\theta_{i}^{\nu}}[\sqrt{g}(\btheta_{i}) \\
& \times g^{\mu \nu}(\btheta_{i}) S^{\beta}_{i} u_{\mu}^{\gamma} \Phi(\btheta_{1}, .., 
\btheta_{N})] = 0~,
\end{split}
\end{equation}
which vanishes because it is a total divergence, and the configuration space has no 
boundary.
The relation $\langle 1|\mathscr{L}^{\alpha}_{i}$ implies that $\langle 1|\mathscr{H} = 
0$, that is, $\langle 1|$ is a left eigenvector with zero eigenvalue.
This implies in particular that $\langle 1|{\rm e}^{-\mathscr{H}t} = \langle 1|$, which 
is 
the conservation of probability: $\langle 1|\rho(t) \rangle = \langle 1| {\rm 
e}^{-\mathscr{H}t}|\rho\rangle = \langle 1|\rho\rangle = 1$.

Similarly, the Gibbs distribution $|\rho_{\rm G}\rangle$ is a right-eigenvector with zero 
eigenvalue: $\mathscr{H} |\rho_{\rm G}\rangle = 0$.
In addition for all $i$, $(i \epsilon^{\alpha \beta \gamma}S^{\beta}_{i} N^{\gamma}_{i} + 
k_{\rm B}T \mathscr{L}_{i}^{\alpha}) |\rho_{\rm G} \rangle = 0$.
These relations allow to write the correlations in the form~\cite{zinn-justin_qft}
\begin{equation}
\langle S_{i_{1}}^{\alpha_{1}}(t_{1})  ... S^{\alpha_{l}}_{i_{l}}(t_{l})\rangle = \lt 
\langle 1\lt|T \{\mathscr{S}^{\alpha_{1}}_{i_{1}}(t_{1}) ... 
\mathscr{S}^{\alpha_{l}}_{i_{l}}(t_{l})\}\rt|\rho_{\rm G} \rt \rangle~,
\end{equation}
where $\mathscr{S}^{\alpha}_{i}(t) = {\rm e}^{\mathscr{H}t} S_{i}^{\alpha} {\rm 
e}^{-\mathscr{H}t}$ and $T$ is the time-ordering operator.

In this representation, the FDT is encoded by the fact that the time-derivative of the 
correlation function, in the region $t > t'$, is
\begin{equation}
\begin{split}
&\frac{{\rm d}}{{\rm d}t} \langle S^{\alpha}_{i}(t) S^{\beta}_{j}(t') 
\rangle  = \big \langle 1 \big|[\mathscr{H}, \mathscr{S}^{\alpha}_{i}(t)] 
\mathscr{S}^{\beta}_{j}(t') \big|\rho_{\rm G}\big\rangle \\
& = -\big \langle 1\big| \mathscr{S}_{i}^{\alpha}(t-t') [\mathscr{H}, S_{j}^{\beta}]\big| 
\rho_{\rm G} \big\rangle\\
& = i k_{\rm B}T \epsilon^{\beta \gamma \delta} \big \langle 1\big| 
\mathscr{S}_{i}^{\alpha}(t-t') \mathscr{L}^{\gamma}_{j}(0) 
\mathscr{S}_{j}^{\delta}(0)\big| \rho_{\rm G} \big\rangle~,
\end{split}
\end{equation}
and is proportional to the Kubo formula for the linear response function $G^{\alpha 
\beta}_{ij} = \langle \delta S^{\alpha}_{i}(t)/\delta \nu_{j}^{\beta}(t')\rangle$:
\begin{equation}\label{Kubo}
\begin{split}
G^{\alpha \beta}_{ij}(t-t') & = -i \Theta(t-t') \times \\
& \epsilon^{\beta \gamma \delta} \big \langle 1 \big|\mathscr{S}^{\alpha}_{i}(t-t') 
\mathscr{L}_{j}^{\gamma}(0) \mathscr{S}_{j}^{\delta}(0)\big|\rho_{\rm G} \big \rangle\\
& = - \beta \Theta(t-t') \frac{{\rm d}}{{\rm d}t} \langle S^{\alpha}_{i}(t) 
S^{\beta}_{j}(t') \rangle~.
\end{split}
\end{equation}

Eq.~\eqref{Kubo} follows from the general formulas of linear response theory.
It can be derived by using that in presence of a small external field $\delta 
\nu_{j}^{\alpha}(t)$ the FP evolution equations become $\pa \rho/\pa t + \mathscr{H} \rho 
+ \mathscr{H}_{1}(t) \rho = 0$ with $\mathscr{H}_{1}(t) = i\epsilon^{\alpha \beta \gamma} 
\sum_{i=1}^{N} \mathscr{L}_{i}^{\alpha} S_{i}^{\beta} \delta \nu_{i}^{\gamma}(t)$.
The perturbation $\mathscr{H}_{1}(t)$ describes the effect on the motion of the field 
$\delta \nu_{i}^{\alpha}(t)$ which modifies $\bN_{i} \to \bN_{i} + \bm{\delta} \bnu_{i}$.

In analogy with other purely-dissipative equations~\cite{zinn-justin_qft, 
kurchan_jpf_1992}, it is also useful to discuss the FP equation by performing the 
transformation $\rho \to \tilde{\rho}$, $\rho = {\rm e}^{-\beta H/2} \tilde{\rho}$.
The transformed operators are:
\begin{equation}\label{transformed-operators}
\begin{gathered}
\tilde{\mathscr{L}}^{\alpha}_{i} = {\rm e}^{\beta H/2} \mathscr{L}^{\alpha}_{i} {\rm 
e}^{-\beta H/2} = \mathscr{L}^{\alpha}_{i} - \frac{i}{2}\beta \epsilon^{\alpha \beta 
\gamma} S_{i}^{\beta} N_{i}^{\gamma} ~,\\
\tilde{S}^{\alpha}_{i} = {\rm e}^{\beta H/2} \mathscr{L}^{\alpha}_{i} {\rm 
e}^{-\beta H/2} =  S_{i}^{\alpha}~,\\
\tilde{\mathscr{H}} = {\rm e}^{\beta H/2} \mathscr{H} {\rm e}^{-\beta H/2} = 
k_{\rm B}T \sum_{i=1}^{N} \tilde{\mathscr{L}}_{i}^{\alpha} 
\tilde{\mathscr{L}}_{i}^{\alpha +}~.
\end{gathered}
\end{equation}
Here $\tilde{\mathscr{L}}^{\alpha+}_{i} = \mathscr{L}^{\alpha}_{i} + i 
\beta\epsilon^{\alpha \beta \gamma} S_{i}^{\beta} N_{i}^{\gamma}/2$ is the adjoint 
of $\tilde{\mathscr{L}}^{\alpha}_{i}$ ($\mathscr{L}^{\alpha}_{i}$ and $S_{i}^{\alpha}$ are 
self-adjoint).
In this representation $\tilde{\mathscr{H}}$ is Hermitian, and has the same left and 
right eigenvectors.
By contrast the evolution operator $\mathscr{H}$ before the transformation is not 
Hermitian~\cite{zinn-justin_qft}.
Eq.~\eqref{transformed-operators} also shows that $\tilde{\mathscr{H}}$ is positive 
semidefinite: all eigenvalues of $\tilde{\mathscr{H}}$ are $\geq 0$.
The Gibbs distribution $|\tilde{\rho}_{\rm G} \rangle = {\rm e}^{\beta H/2} |\rho_{\rm G} 
\rangle$ is a ground state and satisfies $\tilde{\mathscr{L}}^{\alpha +}_{i} 
|\tilde{\rho}_{\rm G}\rangle = 0$.

All results presented in this appendix, clearly, remain valid when considering the equilibrium
dynamics of a cavity system.

\section{Dynamical two-point correlations}
\label{a:dynamical_two_point_functions}

The derivation in Sec.~\ref{s:self_consistency_equations} implies that any dynamical 
correlation of $\bS_{i}$, $\bb_{i}$, $\bS_{j}$, $\bb_{j}$ can be calculated replacing the 
definitions $b^{\alpha}_{i}(t) = \sum_{k} J_{ik}^{\alpha \beta} S^{\beta}_{k}(t)$, 
$b^{\alpha}_{j}(t) = \sum_{k} J_{ik}^{\alpha \beta} S^{\beta}_{k}(t)$, with 
Eqs.~\eqref{b_ij}.
In more detail, we can calculate the second-order correlations as:
\begin{equation}
\begin{split}
C^{\alpha \beta}_{ij}(t-t') & = \langle S^{\alpha}_{i}(t) 
S^{\beta}_{i}(t')\rangle_{\bnu_{i}, \bnu_{j}, \bzeta_{i}, \bzeta_{j}, \bS_{i0}, 
\bS_{j0}}~,\\
A^{\alpha \beta}_{ij}(t-t') & = \langle S^{\alpha}_{i}(t) 
b^{\beta}_{i}(t')\rangle_{\bnu_{i}, \bnu_{j}, \bzeta_{i}, \bzeta_{j}, \bS_{i0}, 
\bS_{j0}}~,\\
\Lambda^{\alpha \beta}_{ij}(t-t') & = \langle b^{\alpha}_{i}(t) 
b^{\beta}_{i}(t')\rangle_{\bnu_{i}, \bnu_{j}, \bzeta_{i}, \bzeta_{j}, \bS_{i0}, 
\bS_{j0}}~,\\
\end{split}
\end{equation}
where $\bS_{i}(t)$, $\bS_{j}(t)$, $\bb_{i}(t)$, $\bb_{j}(t)$ are solutions of the two-site 
Langevin equations $\dot{\bS}_{i} = - \bS_{i} \times (\bS_{i} \times (\bF(\bS_{i}) + 
\bb_{i} + \bnu_{i}))$, $\dot{\bS}_{j} = - \bS_{j} \times (\bS_{j} \times (\bF(\bS_{j}) + 
\bb_{j} + \bnu_{j}))$, with the fields $\bb_{i}$ and $\bb_{j}$ in Eqs.~\eqref{b_ij}.
The correlations are determined by averaging the solutions over the noises $\bnu_{i}$, 
$\bnu_{j}$, $\bzeta_{i}$, $\bzeta_{j}$, and over the initial conditions $\bS_{i0}$, 
$\bS_{j0}$ of the two spins.

Let us first analyze the correlation $C^{\alpha \beta}_{ij}(t-t')$.
Since $r_{ij}$, $R_{ij}$, and $J_{ij}$ are small ($\approx d^{-\ell_{ij}/2}$), the spins 
are only weakly coupled.
For any fixed realization of the noises and the initial conditions the trajectories can 
be expanded to first order as:
\begin{equation}\label{Si_Sj}
\begin{split}
& S^{\alpha}_{i}(t)  = \tilde{S}^{\alpha}_{i}(t) + \int_{0}^{t} {\rm d}t''~ 
\tilde{G}_{i}^{\alpha, \gamma}(t, t'') \Delta b_{i}^{\gamma}(t'')~,\\
& S^{\alpha}_{j}(t)  = \tilde{S}^{\alpha}_{j}(t) + \int_{0}^{t} {\rm d}t''~ 
\tilde{G}_{j}^{\alpha, \gamma}(t, t'') \Delta b_{j}^{\gamma}(t'')~,\\
&\Delta b^{\gamma}_{i}(t'') = J_{ij}^{\gamma \delta} \tilde{S}_{j}^{\delta}(t'') + \beta 
r^{\gamma \delta}_{ij}(t'') S^{\delta}_{j0} \\
& \qquad + \int_{0}^{t''}{\rm d}t'''~R^{\gamma \delta}_{ij}(t''-t''') 
\tilde{S}_{j}^{\delta}(t''')~,\\
& \Delta b^{\gamma}_{j}(t'') = J_{ji}^{\gamma \delta} \tilde{S}_{i}^{\delta}(t'') + \beta 
r^{\gamma \delta}_{ji}(t'') S^{\delta}_{i0}\\
& \qquad + \int_{0}^{t''}{\rm d}t'''~R^{\gamma \delta}_{ji}(t''-t''') 
\tilde{S}_{i}^{\delta}(t''')~.\\
\end{split}
\end{equation}
Here $\tilde{\bS}_{i}(t)$, $\tilde{\bS}_{j}(t)$ are solutions of the single-site Langevin 
equations
\begin{equation}
\begin{split}
\dot{\tilde{\bS}}_{i} & = - \tilde{\bS}_{i} \times (\tilde{\bS}_{i} \times 
(\bF(\tilde{\bS_{i}}) + \tilde{\bb}_{i}(t) + \bnu_{i}))~,\\
\dot{\tilde{\bS}}_{j} & = - \tilde{\bS}_{j} \times (\tilde{\bS}_{j} \times 
(\bF(\tilde{\bS}_{j}) + \tilde{\bb}_{j}(t) + \bnu_{j}))~,\\
\tilde{b}^{\alpha}_{i}(t) & = \bzeta_{i}(t) + \beta l^{\alpha \beta} S^{\beta}_{i0} + 
\int_{0}^{t''}~K^{\alpha \gamma}(t-t'') \tilde{S}^{\beta}_{i}(t'')~,\\
\tilde{b}^{\alpha}_{j}(t) & = \bzeta_{j}(t) + \beta l^{\alpha \beta} S^{\beta}_{j0} + 
\int_{0}^{t''}~K^{\alpha \gamma}(t-t'') \tilde{S}^{\beta}_{j}(t'')~,
\end{split}
\end{equation}
and $\tilde{G}^{\alpha \beta}_{i}(t, t')$, $\tilde{G}^{\alpha \beta}_{i}(t, t')$ are the 
corresponding response functions.

To find $C^{\alpha \beta}_{ij}(t-t')$ we need to discuss the averaging over initial 
conditions and over the noise.
Since we are studying equilibrium correlations, the initial conditions must be weighted 
with the joint probability $P_{2}(\bS_{i0}, \bS_{j0})$ to extract given values $\bS_{i0}$ 
and $\bS_{j0}$ of two spins in the Gibbs distribution.
This probability can be calculated as:
\begin{equation}
P_{2}(\bS_{i0}, \bS_{j0}) = \int {\rm d}^{3} \bar{\bb}_{i} \int {\rm d}^{3} \bar{\bb}_{j} 
P(\bS_{i0}, \bar{\bb}_{i}; \bS_{j0}, \bar{\bb}_{j})
\end{equation}
where $P(\bS_{i}, \bb_{i}; \bS_{j}, \bb_{j})$ is the distribution found in the static 
computation [Eqs.~\eqref{P_SB_2c} and~\eqref{p2_result}].
At leading order for $d \to \infty$, it is sufficient to expand the 
distribution~\eqref{p2_result} at first order in $J^{\alpha \beta}_{ij}$ and $M^{\alpha 
\beta}_{ij}$.
After integration over the magnetic fields we find:
\begin{equation}
\begin{split}
P_{2}(\bS_{i0}, \bS_{j0})& = P_{1{\rm }}(\bS_{i0}) P_{1{\rm eq}}(\bS_{j0}) (1 + \beta 
J^{\alpha \beta}_{ij} S^{\alpha}_{i0} S^{\beta}_{j0} \\
& - \beta^{2} L^{\alpha \gamma} M_{ij}^{\gamma \delta} L^{\delta \beta} S^{\alpha}_{i0} 
S^{\beta}_{j0})~.
\end{split}
\end{equation}

The static equation~\eqref{p2_result}, on the other hand implies that the fields 
$X^{\alpha}_{1} = \sum_{k \neq i, j} J^{\alpha \beta}_{ik} S^{\beta}_{k}$, $X^{\alpha}_{2} 
= \sum_{k \neq i, j} J^{\alpha \beta}_{jk} S^{\beta}_{k}$ have a correlation $\llangle 
X^{\alpha}_{1} X^{\alpha}_{2} \rrangle^{\prime \prime (ij)} \approx -L^{\alpha \gamma} 
M_{ij}^{\gamma \delta} L^{\delta \beta}$ when they are averaged with the statistical 
distribution of the cavity system.
$X^{\alpha}_{1}$ and $X^{\alpha}_{2}$ play the same role of $\zeta_{i}^{\alpha}$ and 
$\zeta_{j}^{\alpha}$, the Gaussian noises which enter the dynamical 
equations~\eqref{b_ij}.
We can thus identify $\llangle X^{\alpha}_{1} X^{\alpha}_{2} \rrangle^{\prime \prime 
(ij)}$ with the equal-time element $r^{\alpha \beta}_{ij}(0)$ of the correlation 
$r_{ij}^{\alpha \beta}(t-t')$.
As a result we can write the equilibrium distribution of the initial conditions as: 
$P_{2}(\bS_{i0}, \bS_{j0}) = P_{1{\rm eq}}(\bS_{i0}) P_{1{\rm eq}}(\bS_{j0}) (1 + \beta 
J^{\alpha \beta}_{ij} S^{\alpha}_{i0} S^{\beta}_{j0} + \beta^{2} r_{ij}^{\alpha \beta}(0) 
S^{\alpha}_{i0} S^{\beta}_{j0})$, where $P_{1{\rm eq}}(\bS_{i0})$ is the single-site 
probability [Eq.~\eqref{P1}].

Consider now the distribution of the noise.
We can formally write the Gaussian probability $P_{2{\rm noise}}[\bzeta_{i}, \bzeta_{j}]$ 
of a given realization $\bzeta_{i}(t)$, $\bzeta_{j}(t)$ in the form:
\begin{equation}
\begin{split}
& P_{2{\rm noise}}[\bzeta_{i}, \bzeta_{j}] = {\cal N}_{2} \exp \Big[-\frac{1}{2} 
{\textstyle \int}_{0}^{\infty} {\textstyle \int}_{0}^{\infty} {\rm d}t {\rm d}t'~\\
& \qquad  {\textstyle \sum}_{a = i, j} {\textstyle \sum}_{b = i, j}~l^{-1\alpha 
\beta}_{ab}(t-t') \zeta^{\alpha}_{a}(t) \zeta^{\beta}_{b}(t') \Big]~,
\end{split}
\end{equation}
where $l^{-1\alpha \beta}_{ab}(t-t')$ is the inverse of the matrix
\begin{equation}
\begin{split}
l^{\alpha \beta}_{ab}(t-t') & =
\begin{Vmatrix}
l^{\alpha \beta}_{ii}(t-t') & l_{ij}^{\alpha \beta}(t-t') \\
l_{ji}^{\alpha \beta}(t-t') & l_{jj}^{\alpha \beta}(t-t')
\end{Vmatrix}\\
& = \begin{Vmatrix}
l^{\alpha \beta}(t-t') & r_{ij}^{\alpha \beta}(t-t') \\
r_{ji}^{\alpha \beta}(t-t') & l^{\alpha \beta}(t-t')
    \end{Vmatrix}~,
\end{split}
\end{equation}
and ${\cal N}_{2}$ is a normalization constant.
The inverse is intended in the sense of kernels: $\sum_{c} \int_{0}^{\infty}{\rm d}t''~ 
l^{-1\alpha \gamma}_{ac}(t-t'') l^{\gamma \beta}_{cb}(t''-t') = \delta^{\alpha \beta} 
\delta_{ab} \delta(t-t')$.

Expanding to first order in $r_{ij}$ we find:
\begin{equation}
\begin{split}
& P_{2{\rm noise}}[\bzeta_{i}, \bzeta_{j}] \simeq P_{1 {\rm noise}}[\bzeta_{1}] P_{1 {\rm 
noise}}[\bzeta_{1}] \\
& \times \Big[1 + {\textstyle \int}_{0}^{\infty} {\rm d}t {\textstyle \int}_{0}^{\infty} 
{\rm d}t'' {\textstyle \int}_{0}^{\infty} {\rm d}t_{1} {\textstyle \int}_{0}^{\infty} 
{\rm d}t_{2} \big(l^{-1\alpha \gamma}(t-t_{1}) \\
& \times r_{ij}^{\gamma \delta}(t_{1} - t_{2}) l^{-1\delta \beta}(t_{2} - t') 
\zeta^{\alpha}_{i}(t) \zeta^{\beta}_{j}(t') \big) \Big]~,
\end{split}
\end{equation}
where 
\begin{equation}
\begin{split}
P_{1{\rm noise}}[\bzeta] & = {\cal N}_{1} \exp \Big[-\frac{1}{2} {\textstyle 
\int}_{0}^{\infty}{\rm d}t {\textstyle \int}_{0}^{\infty} {\rm d}t' \\
& l^{-1 \alpha \beta}(t-t') \zeta^{\alpha}(t) \zeta^{\beta}(t')\Big]
\end{split}
\end{equation}
is the noise distribution of the single-site problem.

We can now calculate the correlation.
We get three terms at leading order.
The first, which we denote as $C_{1ij}^{\alpha \beta}(t-t')$, comes from the correction 
of the trajectories [Eq.~\eqref{Si_Sj}] calculated in the unperturbed ensemble.
Using that $\langle G^{(n)\alpha \beta}_{i}(t, t')\rangle = \langle G^{(n)\alpha 
\beta}(t, t')\rangle = g^{\alpha \beta}(t-t') = -\beta \dot{c}^{\alpha \beta}_{+}(t-t')$ 
is the single-site response function we get:
\begin{equation}\label{C1}
\begin{split}
& C_{1ij}^{\alpha \beta}(t-t')  = \int_{0}^{t} {\rm d}t''~g^{\alpha \gamma}(t-t'') 
\langle \Delta b^{\gamma}_{i}(t'') S_{j}^{\beta}(t')\rangle \\
& + \int_{0}^{t'} {\rm d}t''~g^{\beta \gamma}(t'-t'') \langle \Delta 
b_{j}^{\gamma}(t'') S_{i}^{\alpha}(t) \rangle~.
\end{split}
\end{equation}

A second term, $C_{2}^{\alpha \beta}(t-t')$, comes from the distribution of initial 
conditions:
\begin{equation}
\begin{split}
& C_{2}^{\alpha \beta}(t-t')= \beta [J^{\gamma \delta}_{ij} + \beta r_{ij}^{\gamma 
\delta}(0)] \langle S^{\alpha}_{i}(t) S^{\gamma}_{i0}\rangle \\
& \times \langle S^{\beta}_{j}(t') S^{\delta}_{j0}\rangle  = \beta [J^{\gamma 
\delta}_{ij} + \beta r_{ij}^{\gamma \delta}(0)] c^{\alpha \gamma}(t) c^{\beta 
\delta}(t')~.
\end{split}
\end{equation}

Finally we have a third term arising from the correlation of the noise:
\begin{equation}
\begin{split} 
C_{3ij}^{\alpha \beta}(t-t')&  = {\textstyle \int}_{0}^{\infty} {\rm d}t_{1} {\textstyle 
\int}_{0}^{\infty} {\rm d}t_{2} {\textstyle \int}_{0}^{\infty} {\rm d}t_{3} {\textstyle 
\int}_{0}^{\infty} {\rm d}t_{4} \\
& l^{-1\gamma \mu}(t_{1}-t_{3}) r_{ij}^{\mu \nu}(t_{3} - t_{4}) l^{-1\nu 
\delta}(t_{4} - t_{2})\\
& \langle S^{\alpha}_{i}(t)  \zeta^{\gamma}_{i}(t_{1}) \rangle \langle 
S_{j}^{\beta}(t') \zeta^{\delta}_{j}(t_{2}) \rangle~,
\end{split} 
\end{equation}
which, using the properties of Gaussian averages~\cite{zinn-justin_qft}, can be rewritten 
as:
\begin{equation}
\begin{split}
C_{3ij}^{\alpha \beta}(t-t') & = {\textstyle \int}_{0}^{\infty} {\rm d}t'' {\textstyle 
\int}_{0}^{\infty} {\rm d}t'''~ \big[g^{\alpha \gamma}(t-t'') \\
& \times g^{\beta \delta}(t'-t''') r_{ij}^{\gamma \delta}(t''-t''')\big]~.
\end{split}
\end{equation}

Collecting all terms $C_{ij}^{\alpha \beta}= C_{1ij}^{\alpha \beta} + C_{2ij}^{\alpha 
\beta} + C_{3ij}^{\alpha \beta}$, and using the FDT, we find and expression manifestly 
consistent with TTI:
\begin{equation}\label{Cij_2}
\begin{split}
& C_{ij}^{\alpha \beta}(t-t') = \Theta(t-t') \Big\{ (J^{\gamma \delta}_{ij} + \beta 
r^{\gamma \delta}_{ij}(0))\\
& \times \Big[\beta c^{\beta \delta}(0) c^{\alpha \gamma}(t-t')+ \int_{t'}^{t} {\rm 
d}t''~g^{\alpha \gamma}(t-t'') c^{\beta \delta}(t'-t'')\Big] \\
& + \int_{t'}^{t} {\rm d}t'' \int_{t'}^{t''} {\rm d}t'''~g^{\alpha \gamma}(t-t'') 
r_{ij}^{\gamma \delta}(t''-t''') g^{\delta \beta}(t'''-t')\Big\}\\
& + \Theta(t'-t) \Big\{ (J^{\gamma \delta}_{ij} + \beta r_{ij}^{\gamma \delta}(0))  
\Big[\beta c^{\alpha \gamma}(0) c^{\beta \delta}(t'-t) \\
& + \int_{t}^{t'}{\rm d}t''~c^{\alpha \gamma}(t-t'') g^{\beta \delta}(t'-t'') \Big] \\
& + \int_{t}^{t'}{\rm d}t''' \int_{t}^{t'''} {\rm d}t''  g^{\gamma \alpha}(t''-t) 
r_{ij}^{\gamma \delta}(t''-t''') g^{\beta \delta}(t'-t''')  \Big\}
\end{split}
\end{equation}

Finally, differentiating over time we find
\begin{equation}
\begin{split}
& \dot{C}_{+ij}^{\alpha \beta}(t-t') = - \beta \Theta(t-t') \Big\{ \int_{t'}^{t}{\rm 
d}t''~\dot{c}_{+}^{\alpha \gamma}(t-t'') \\
& \times \Big[ J_{ij}^{\gamma \delta} \dot{c}_{+}^{\delta \beta}(t''-t') + 
\int_{t'}^{t''} {\rm d}t'''~\big(\dot{c}_{+}^{\alpha \gamma}(t-t'') \\
& \times R_{ij}^{\gamma \delta}(t''-t''') \dot{c}_{+}^{\delta 
\beta}(t'''-t')\big)\Big]\Big\}~,
\end{split}
\end{equation}
an expression which in frequency space becomes Eq.~\eqref{C_ij_plus} in the main text.

Note in passing that for equal times $t = t'$, Eq.~\eqref{Cij_2} reduces to 
$C_{ij}^{\alpha \beta} = \beta c^{\alpha \gamma} (J^{\gamma \delta}_{ij} + \beta 
r_{ij}^{\gamma \delta}(0)) c^{\delta \beta}$, which is equivalent to the first of the 
static equations~\eqref{C_A_Lambda_1st_order} after the identification $r_{ij}^{\alpha 
\beta}(0) = - L^{\alpha \gamma} M_{ij}^{\gamma \delta} L^{\delta \beta}$.

We now consider $A_{ij}^{\alpha \beta}(t-t')$ and $\Lambda^{\alpha \beta}_{ij}(t-t')$.
It is not necessary to go through a detailed calculation, as we did for $C_{ij}^{\alpha 
\beta}(t-t')$, because equations of motion analogue to Eqs.~\eqref{dynamic_eom} relate $A$ 
and $\Lambda$ to $C$.

Writing compactly the fields as:
\begin{equation}
\begin{split}
b_{a}^{\alpha}(t) & = \zeta_{a}^{\alpha}(t) + \sum_{b=i, j} \Big[\beta l^{\alpha 
\gamma}_{ab}(t) S^{\gamma}_{b0} + J_{ab}^{\alpha \beta} S^{\beta}_{b}(t) \\
& + \int_{0}^{t}{\rm d}t''~K_{ab}^{\alpha \gamma}(t-t') S^{\gamma}_{b}(t') \Big]~,
\end{split}
\end{equation}
and introducing
\begin{equation}
\begin{split}
K^{\alpha \beta}_{ab}(t-t') & = -\beta \Theta(t-t')\frac{{\rm d}}{{\rm d}t} 
l_{ab}^{\alpha \beta}(t-t') \\
& = \begin{Vmatrix}
     K^{\alpha \beta}(t-t') & R_{ij}^{\alpha \beta}(t-t') \\
     R_{ji}^{\alpha \beta}(t-t') & K^{\alpha \beta}(t-t')
    \end{Vmatrix}~,
\end{split}
\end{equation}
and
\begin{equation}
\begin{split}
\dot{C}^{\alpha\beta}_{+ab}(t-t') & = \begin{Vmatrix}
                       \dot{C}_{+ii}^{\alpha \beta}(t-t') &\dot{C}_{+ij}^{\alpha 
\beta}(t-t') \\
                       \dot{C}_{+ji}^{\alpha \beta}(t-t') & \dot{C}_{+jj}^{\alpha 
\beta}(t-t')
                      \end{Vmatrix}\\
& = \begin{Vmatrix}
     \dot{c}_{+}^{\alpha \beta}(t-t') &\dot{C}_{+ij}^{\alpha 
\beta}(t-t') \\ 
\dot{C}_{+ji}^{\alpha \beta}(t-t') & \dot{c}^{\alpha \beta}_{+}(t-t')
    \end{Vmatrix}
\end{split}
\end{equation}
we find, using the properties of Gaussian averages and the FDT:
\begin{equation}\label{dot_A_plus_1}
\begin{split}
& \dot{A}_{+ab}^{\alpha \beta}(t-t') = \sum_{c=i, j} \dot{C}^{\alpha 
\gamma}_{+ac}(t-t') J_{cb}^{\gamma \beta}\\
& \qquad +\sum_{c=i, j}\int_{t'}^{t} {\rm d}t''~\dot{C}^{\alpha \gamma}_{+ac}(t-t'') 
K^{\gamma \beta}_{cb}(t''-t')~,
\end{split}
\end{equation}

\begin{equation}\label{dot_Lambda_plus_1}
\begin{split}
& \dot{\Lambda}_{+ab}^{\alpha \beta}(t-t') = -\frac{1}{\beta} K_{ab}^{\alpha \beta}(t-t') 
+ \sum_{c, d=i, j} \Big\{\\
& \qquad \int_{t'}^{t} {\rm d}t'' \int_{t'}^{t''} {\rm d}t''' \big[K^{\alpha 
\gamma}_{ac}(t-t'') \dot{C}_{+cd}^{\gamma \delta}(t''-t''') \\
& \qquad \times K^{\delta \beta}_{db}(t'''-t')\big]  + J^{\alpha \gamma}_{ac} 
\dot{C}_{+cd}^{\gamma \delta}(t-t') J_{db}^{\delta \beta} \\
& \qquad + J_{ac}^{\alpha \gamma} \int_{t'}^{t}{\rm d}t'' \dot{C}_{+cd}^{\gamma 
\delta}(t-t'') K_{db}^{\delta \beta}(t''-t')\\
& \qquad + J_{db}^{\delta\beta}\int_{t'}^{t}{\rm d}t'' K^{\alpha \gamma}_{ac}(t-t'') 
\dot{C}_{+cd}^{\gamma \delta}(t''-t')\Big\}~.
\end{split}
\end{equation}

$\dot{A}_{+ab}^{\alpha \beta}(t-t')$ and $\dot{\Lambda}_{+ab}^{\alpha \beta}(t-t')$ are 
given by convolutions of retarded response functions, acting in sequence.
All time integrals can be extended to $-\infty$ to $+\infty$ because the $\Theta(t-t')$ 
factors in the definition of $\dot{C}_{+}$ and $K$ automatically restrict the integration 
to the causal region.

This allows to write Eqs.~\eqref{dot_A_plus_1},~\eqref{dot_Lambda_plus_1} in frequency 
space as:
\begin{equation}
\begin{split}
&\dot{A}_{+ab}^{\alpha \beta}(\omega) = \sum_{c = i, j} \dot{C}^{\alpha 
\gamma}_{+ac}(\omega) (J_{cb}^{\gamma \beta} + K_{cb}^{\gamma \beta}(\omega))~,\\
&\dot{\Lambda}^{\alpha \beta}_{+ab}(\omega) = - \frac{1}{\beta} K_{ab}^{\alpha 
\beta}(\omega) +  \sum_{c, d = i, j} \big[ (J_{ac}^{\alpha \gamma} + K_{ac}^{\alpha 
\gamma}(\omega)) \\
& \qquad \times \dot{C}_{+cd}^{\gamma \delta}(\omega) (J_{db}^{\delta \beta} + 
K_{db}^{\delta \beta}(\omega))]~.
\end{split}
\end{equation}

We now take the $(ij)$ matrix element of the equations and keep only terms of linear 
order in $J_{ij}$, $r_{ij}$, $R_{ij}$.
Using that $J_{ii} = J_{jj} = 0$ we find:
\begin{equation}
\begin{split}
\dot{A}_{+ij}^{\alpha \beta}(\omega) & = \dot{c}^{\alpha \gamma}_{+}(\omega) (J^{\gamma 
\beta}_{ij} + R_{ij}^{\gamma \beta}(\omega)) \\
& + \dot{C}_{+ij}^{\alpha \gamma}(\omega) K^{\gamma \beta}(\omega)~,
\end{split}
\end{equation}
\begin{equation}
\begin{split}
\dot{\Lambda}^{\alpha \beta}_{+ij}(\omega) & = -\frac{1}{\beta} R_{ij}^{\alpha 
\beta}(\omega) + K^{\alpha \gamma}(\omega) \dot{C}^{\gamma \delta}_{+ij}(\omega) K^{\delta 
\beta}(\omega)\\
& + (J_{ij}^{\alpha \gamma} + R_{ij}^{\alpha \gamma}(\omega)) 
\dot{c}_{+}^{\gamma \delta}(\omega) K^{\delta \beta}(\omega)\\
& + K^{\alpha \gamma}(\omega) \dot{c}_{+}^{\gamma \delta}(\omega) (J_{ij}^{\delta \beta} 
+ R_{ij}^{\delta \beta}(\omega))~,
\end{split}
\end{equation}
which are Eqs.~\eqref{A_ij_plus_Lambda_ij_plus} in the main text.

\footnotetext[1]{Since $[\hat{t}^{2n}]_{ii}$ is of order $d^{n}$ for $d$
large~\cite{parisi_jpa_1994}, the matrix elements $[\hat{t}^{2n}/(2d)^{n}]$ are all of 
order O$(1)$ and the condition $J_{ii}^{\alpha \beta} = 0$ can be consistently imposed 
while keeping finite all coefficients $J_{n}^{\alpha \beta}$, including $J_{0}^{\alpha 
\beta}$.
If the interaction had been scaled as $\hat{J} = f(\hat{t}/d^{\gamma})$ with $\gamma < 
1/2$, the condition $J_{ii}^{\alpha \beta} = 0$ would have required a divergent $f^{\alpha 
\beta}(0) = J_{0}^{\alpha \beta}$ to cancel the divergent on-site matrix elements
$[(\hat{t}/d^{\gamma})^{2n}] = {\rm O} (d^{n(1-2\gamma)})$}

\footnotetext[2]{More precisely, we can distinguish two cases.
If $R$ is even and $J^{\alpha \beta}_{R}$ is negative definite (fully antiferromagnetic),
then the eigenvalues of $f^{\alpha \beta}$ are always bounded from above.
If $R$ is even and $J^{\alpha \beta}_{R}$ is negative semi-definite (with one or two zero
eigenvalues), the boundedness of the eigenvalues requires additional conditions on the
coefficients $J^{\alpha \beta}_{m}$.
}

\footnotetext[3]{In Ref.~\cite{mezard_spin-glass} the cavity equations were derived
by more general arguments, based on a counting of configurations within a given state.
Here we use a simpler approach and assume directly a canonical Gibbs distribution.
}

\footnotetext[4]{For $d \to \infty$, broken-symmetry states could be studied by finding
self-consistent solutions with nonzero space-dependent mean-fields (breaking spin-inversion
symmetry), and with space-dependent self-energies $\sigma^{\alpha \beta}_{i}$ and fluctuation
matrices $L^{\alpha \beta}_{i}$ (due to the breaking of translation invariance).
This approach would allow to describe ordered states and to amorphous metastable states,
as minima of a free energy analogue to the TAP free-energy~\cite{de-dominicis_pr_1980}.}

\end{document}